\documentclass[a4paper]{article}
\usepackage{jheppub}
\usepackage[T1]{fontenc}
\usepackage{graphicx,subcaption}
\usepackage{braket}
\usepackage{bm}
\usepackage{xcolor}

\definecolor{purple}{rgb}{0.4,0,0.7}

\usepackage{physics}
\usepackage{mathtools, paralist}
\usepackage{enumerate, enumitem}
\usepackage{multirow, tabularx, array}
\usepackage[hang,flushmargin]{footmisc}

\newcommand{\nn}{\nonumber \\}
\newcommand{\e}{\mathrm{e}}

\newcommand{\figref}[1]{FIG.~\ref{#1}}
\newcommand{\tabref}[1]{TABLE~\ref{#1}}
\newcommand{\secref}[1]{Sec.~\ref{#1}}
\newcommand{\appref}[1]{Appendix~\ref{#1}}

\newcommand{\re}[1]{\mathrm{Re}\qty[#1]}
\newcommand{\im}[1]{\mathrm{Im}\qty[#1]}
\newcommand{\argu}[1]{\mathrm{Arg}\qty[#1]}

\usepackage{acro}
\DeclareAcronym{gr}{
    short=GR ,
    long=general relativity
}
\DeclareAcronym{ds}{
    short=dS ,
    long=de Sitter
}
\DeclareAcronym{ads}{
    short=AdS ,
    long=anti-de Sitter
}
\DeclareAcronym{qft}{
    short=QFT ,
    long={quantum field theory}
}
\DeclareAcronym{jt}{
    short=JT ,
    long={Jackiw-Teitelboim}
}

\DeclareAcronym{hl}{
    short=HL ,
    long=Ho\v{r}ava-Lifshitz
}

\DeclareAcronym{uv}{
    short=UV ,
    long=ultraviolet
}
\DeclareAcronym{ir}{
    short=IR ,
    long=infrared
}
\DeclareAcronym{wdw}{
    short=WDW ,
    long=Wheeler-DeWitt
}
\DeclareAcronym{rg}{
    short=RG ,
    long=renormalization group
}
\DeclareAcronym{adm}{
    short=ADM ,
    long={Arnowitt, Deser, and Misner}
}

\title{\bf Dimensional (In)dependence of Lorentzian Quantum Cosmology}

\author{\large Masazumi Honda${}^{a,b}$, Hiroki Matsui${}^{c,d,e}$, Kota Numajiri${}^{e}$, \\ Kazumasa Okabayashi${}^{c,e}$, Takahiro Terada${}^{f}$}%\medskip\\ 

\emailAdd{masazumi.honda@riken.jp} 
\emailAdd{hiroki.matsui@yukawa.kyoto-u.ac.jp}
\emailAdd{kota.numajiri@yukawa.kyoto-u.ac.jp} 
\emailAdd{kazumasa.okabayashi@yukawa.kyoto-u.ac.jp}
\emailAdd{terada@eken.phys.nagoya-u.ac.jp}

\affiliation{

${}^a$RIKEN Center for Interdisciplinary Theoretical and Mathematical Sciences (iTHEMS), RIKEN, Wako 351-0198, Japan \medskip\\
${}^b$Graduate School of Science and Engineering, Saitama University, 255 Shimo-Okubo,
Sakura-ku, Saitama 338-8570, Japan\medskip\\
${}^c$Osaka Central Advanced Mathematical Institute (OCAMI), Osaka Metropolitan
University, 3-3-138 Sugimoto, Sumiyoshi, Osaka 558-8585, Japan \medskip\\
${}^d$Department of Physics, College of Humanities and Sciences, Nihon University, Sakurajosui, Tokyo 156-8550, Japan \medskip\\
${}^e$Center for Gravitational Physics and Quantum Information, Yukawa Institute for Theoretical Physics, Kyoto University, Kitashirakawa Oiwakecho, Sakyo-ku, Kyoto 606-8502, Japan \medskip\\
${}^f$Kobayashi-Maskawa Institute for the Origin of Particles and the Universe, Nagoya University, Tokai National Higher Education and Research System, Furo-cho, Chikusa-ku, Nagoya 464-8602, Japan
}

\abstract{
We study transition amplitudes of the quantum universe in $D$-dimensional spacetime within the de Sitter minisuperspace approximation in Einstein gravity. Using the Lorentzian path integral combined with Picard–Lefschetz theory, we identify the saddle points that contribute to the integral and derive a well-defined transition amplitude under Dirichlet boundary conditions. We further show numerically that a resurgence analysis based on Borel-Pad\'e resummation is effective in resolving the residual Stokes ambiguities that arise in certain cases. Analyzing three-, four-, and five-dimensional cases, as well as the large-$D$ limit, we find that the tunneling proposal is commonly realized for the creation of the universe from nothing. 
This result implies a universal feature
of quantum cosmology based on Einstein gravity across dimensions.
}

\begin{document}
\begin{flushright}
    \small
RIKEN-iTHEMS-Report-26, STUPP-26-304, YITP-26-126, OCU-PHYS-631, AP-GR-215
\end{flushright}
\maketitle
\flushbottom

%%%%%%%%%%%%%%%%%%%%%%%%%%%%%%%%%%%%%%%
%%%%%%%%%%%%%%%%%%%%%%%%%%%%%%%%%%%%%%%
\section{Introduction}
\label{sec:introduction}
%%%%%%%%%%%%%%%%%%%%%%%%%%%%%%%%%%%%%%%
%%%%%%%%%%%%%%%%%%%%%%%%%%%%%%%%%%%%%%%

Constructing a quantum theory of gravity remains a fundamental open problem in theoretical physics. A central feature of gravity is that spacetime geometry itself is dynamical, so its quantum description requires quantum fluctuations of the geometry rather than of fields propagating on a fixed spacetime. In quantum cosmology, this is naturally formulated in terms of transition amplitudes between boundary geometries, which can be represented by a gravitational path integral over metrics on a spacetime manifold. In the Lorentzian formulation, the transition amplitude is formally written as
\begin{equation}\label{eq:G-amplitude1}
\mathcal{A}[g_f;g_i] = \int_{\mathcal{M}} \frac{\mathcal{D}g_{\mu\nu}}{\mathrm{Vol}(\mathrm{Diff})}
  \exp\left(\frac{i}{\hbar} S[g_{\mu\nu}] \right),
\end{equation}
where $S[g_{\mu\nu}]$ is the Einstein--Hilbert action supplemented by appropriate boundary terms and, if present, matter contributions. The integration is taken over metrics on $\mathcal{M}$ that induce the prescribed metrics $g_i$ and $g_f$ on the initial and final boundaries, respectively, and the division by $\mathrm{Vol}(\mathrm{Diff})$ removes the overcounting of diffeomorphism-equivalent configurations. Historically, applications of the gravitational path integral to quantum cosmology have largely relied on the Euclidean formulation, most notably the Hartle--Hawking no-boundary proposal~\cite{Hartle:1983ai}. However, the Euclidean action is unbounded from below, because the conformal factor of the metric has a kinetic term of the wrong sign~\cite{Gibbons:1978ac}. As a result, the integral over Euclidean metrics does not converge.

Moreover, there is ambiguity as to whether one should adopt the Wick rotation or the anti-Wick rotation to obtain a Euclidean theory, corresponding to different choices of the contour in the gravitational path integral. 
Different choices of the metric's contour can lead to distinct physical predictions. A well-known example is the tension between the no-boundary proposal~\cite{Hartle:1983ai} and Vilenkin's tunneling proposal~\cite{Vilenkin:1984wp} as well as related early proposals~\cite{Linde:1983cm, Linde:1984ir, Rubakov:1984bh, Zeldovich:1984vk}. They show opposite exponential behaviors in the wave function of the universe and thus lead to radically different cosmological predictions~\cite{Vilenkin:1986cy, Vilenkin:1987kf}. The no-boundary proposal defines the initial wave function through a regular Euclidean geometry, which gives a superposition of expanding and contracting branches through analytic continuation. On the other hand, in the tunneling proposal, the requirement that only the expanding branch remains in the final state leads to a tunneling-like exponential factor in the transition amplitude.
This ambiguity further motivates a formulation that treats Lorentzian metrics more directly.

In recent years, the Lorentzian path integral with Picard-Lefschetz theory \cite{Pham,Berry:1991,Howls,Witten:2010cx}, initiated in Ref.~\cite{Feldbrugge:2017kzv}, has gained renewed attention as a rigorous alternative to the Euclidean approach. 
The central idea is to complexify the metric degrees of freedom
within the \ac{adm} formalism.
The original integration contour is then analytically deformed onto the Lefschetz thimbles, namely the steepest-descent paths associated with saddles, along which the integral becomes absolutely convergent and non-oscillatory.
Picard-Lefschetz theory provides powerful tools to identify contributing saddles and associated thimbles
under explicitly specified boundary conditions, and provides a basis for a well-defined prescription for computing the wave function of the universe~\cite{Feldbrugge:2017kzv, DiazDorronsoro:2017hti,Feldbrugge:2017mbc,DiazDorronsoro:2018wro,Feldbrugge:2018gin,DiTucci:2018fdg,DiTucci:2019dji,Janssen:2019sex,DiTucci:2019bui,Narain:2021bff,Lehners:2021jmv,Narain:2022msz,Matsui:2023hei,Lehners:2023yrj,Ailiga:2023wzl,Matsui:2023tkw,Lehners:2024kus,Ailiga:2024mmt,Ailiga:2024wdx,Honda:2024aro,Honda:2024hdr}.\footnote{Lefschetz-thimble techniques have also been applied to a variety of quantum and statistical systems to control highly oscillatory integrals and sign problems; see, for example, Refs.~\cite{Mou:2019tck,Mou:2019gyl,Millington:2020vkg,Matsui:2021oio,Rajeev:2021zae,
Hayashi:2021kro,Feldbrugge:2022idb,Nishimura:2023dky,Feldbrugge:2023frq,Feldbrugge:2023mhn,Saito:2024acm}.}

In certain cases, as shown in Ref.~\cite{Honda:2024aro},
the so-called Stokes phenomenon \cite{Stokes1858} can occur:  as parameters are varied, the asymptotic expansion discontinuously changes its form across some particular regions of the parameters called Stokes lines.
Namely, when crossing the Stokes lines, the way saddle points contribute can change discontinuously, and the thimble decomposition becomes ambiguous on the Stokes lines.
This makes it much more subtle to identify the contributing saddles, or equivalently, the corresponding cosmological scenario.
For four-dimensional Einstein gravity, Ref.~\cite{Honda:2024aro} shows that the ambiguity in the thimble decomposition cancels against ambiguities in the resummation of perturbative series, leaving the full result continuous across Stokes lines, as we will review.
It is conceptually important to note that on the Stokes lines, the set of contributing saddles may not be unique and may depend on details of how the perturbative series is resummed.
Therefore, simply asking which saddles contribute is not a well-defined question when the Stokes lines are present in the parameter region under consideration, and we need to be careful also about how to resum the perturbative series.
The above discussion is in line with a typical successful scenario of resurgence theory \cite{SC_1977__17_1_A5_0}, which has a long history in applications to quantum mechanics, including exact WKB analyses of the Schr\"{o}dinger equation.\footnote{See, e.g., Refs.~\cite{Costin:1999798,Marino:2012zq,Dorigoni:2014hea,ANICETO20191,2014arXiv1405.0356S} for reviews.}  
There have been applications of the resurgence to various areas of physics from the viewpoint of not only differential equations but also Lefschetz thimble analyses of (path) integrals, including quantum field theory (QFT), string theory~\cite{Marino:2008vx,Garoufalidis:2010ya,Chan:2010rw,Chan:2011dx,Schiappa:2013opa,Marino:2006hs,Marino:2007te,Marino:2008ya,Pasquetti:2009jg,Aniceto:2011nu,Santamaria:2013rua,Couso-Santamaria:2014iia,Grassi:2014cla,Couso-Santamaria:2015wga,Couso-Santamaria:2016vcc,Couso-Santamaria:2016vwq,Kuroki:2019ets,Kuroki:2020rgg,Dorigoni:2022bcx,Baldino:2022aqm,Schiappa:2023ned,Iwaki:2023cek,Alexandrov:2023wdj}, hydrodynamics~\cite{Aniceto:2015mto,Basar:2015ava,Casalderrey-Solana:2017zyh,Behtash:2017wqg,Heller:2018qvh,Heller:2020uuy,Aniceto:2018uik,Behtash:2020vqk}, and Jackiw-Teitelboim gravity \cite{Griguolo:2021wgy,Gregori:2021tvs,Eynard:2023qdr,Honda:2024hdr}.\footnote{In particular, recent applications to QFT span a wide range of theories, including two-dimensional QFTs~\cite{Dunne:2012ae,Dunne:2012zk,Cherman:2013yfa,Cherman:2014ofa,Misumi:2014jua,Behtash:2015kna,Dunne:2015ywa,Buividovich:2015oju,Demulder:2016mja,Okuyama:2018clk,Marino:2019eym,Marino:2019fvu,Abbott:2020qnl,Abbott:2020mba,Marino:2021six,DiPietro:2021yxb,Marino:2021dzn,Marino:2022ykm,Reis:2022tni,Marino:2023epd}, Chern-Simons theory~\cite{Gukov:2016njj,Gang:2017hbs,Wu:2020dhl,Ferrari:2020avq,Gukov:2019mnk,Garoufalidis:2020nut,Fuji:2020ltq,Garoufalidis:2021osl}, the three-dimensional $O(2N)$ model \cite{Dondi:2021buw}, four-dimensional nonsupersymmetric QFTs~\cite{Argyres:2012vv,Dunne:2015eoa,Mera:2018qte,Canfora:2018clt,Unsal:2020yeh}, six-dimensional $\phi^3$ theory~\cite{Borinsky:2021hnd,Borinsky:2022knn}, and supersymmetric gauge theories in various dimensions~\cite{Russo:2012kj,Aniceto:2014hoa,Aniceto:2015rua,Honda:2016mvg,Honda:2016vmv,Honda:2017qdb,Gukov:2017kmk,Dorigoni:2017smz,Honda:2017cnz,Fujimori:2018nvz,Grassi:2019coc,Dorigoni:2019kux,Dorigoni:2021guq,Fujimori:2021oqg,Beccaria:2021ism}.}
Currently there are much fewer applications in the context of cosmology and astrophysics with some exceptions such as stochastic inflation \cite{Honda:2023unh,Honda:2024evc}, particle production \cite{Enomoto:2020xlf,Enomoto:2021hfv,Enomoto:2022mti,Enomoto:2022nuj,Namba:2025ejw}, and quasinormal modes of black holes~\cite{Hatsuda:2021gtn,Hatsuda:2019eoj,Matyjasek:2019eeu,Eniceicu:2019npi,Miyachi:2025ptm,Miyachi:2025dyk,Hatsuda:2026ghx}.

This Lorentzian path-integral approach with resurgence has been successfully applied to four-dimensional Einstein gravity and two-dimensional \ac{jt} gravity~\cite{Jackiw:1984je, Teitelboim:1983ux}, leading to a consistent derivation of the transition amplitude and clarifying whether the no-boundary or tunneling scenario is realized \cite{Honda:2024aro, Honda:2024hdr}.\footnote{Lorentzian quantum cosmology of \ac{jt} gravity and related Kantowski-Sachs models has also been studied in Refs.~\cite{Halliwell:1990tu,Fanaras:2021awm,Fanaras:2022twv,Ghosh:2023njl,Buchmuller:2024ksd}. Numerical implementations using Lefschetz thimbles have been explored in Ref.~\cite{Chou:2024sgk}.}
%%%%%%%%%%%%%%%%%%%%%%%%%%%%%%%%%%
%%%%%%%%%%%
Intriguingly, the conclusions in these two theories are quite different. In four-dimensional Einstein gravity, the tunneling proposal is favored under Dirichlet-type boundary conditions with a vanishing initial scale factor \cite{Honda:2024aro}.
On the other hand, in the two-dimensional \ac{jt} gravity, a solvable toy model of quantum gravity, the no-boundary proposal can be realized by imposing the same boundary conditions on the scale factor together with additional conditions on the dilaton value \cite{Honda:2024hdr, Matsui:2025guo}. 
These results suggest that the favored scenario for the genesis of the universe may depend nontrivially on the spacetime dimensionality.\footnote{
There is also a possibility that the favored scenario is affected by the presence of the dilaton, but here we focus on the dimensionality dependence.
} 
Clarifying this dimensionality dependence may therefore offer valuable insight into the construction of a fully consistent quantum cosmology.

\begin{table}[tb]
    \centering
    \caption{Summary of the results.}
    \renewcommand{\arraystretch}{1.15}
    \begin{tabularx}{\textwidth}{
        >{\raggedright\arraybackslash}p{0.15\textwidth}
        >{\centering\arraybackslash}p{0.11\textwidth}@{\quad/\quad}
        >{\centering\arraybackslash}p{0.11\textwidth}
        >{\raggedright\arraybackslash}p{0.17\textwidth}
        >{\raggedright\arraybackslash}X
    }
        \hline\hline
        Theory &
        \multicolumn{2}{c}{Initial / final spacetime} &
        Favored transition & Note \\
        \hline

        \multirow[t]{2}{*}{2D JT \cite{Honda:2024hdr}}
        & \multicolumn{2}{c}{Lorentzian} & Classical   & \\
        & \multicolumn{2}{c}{Euclidean}   & No-boundary & \\[6pt]

        \multirow[t]{3}{*}{3D GR }
        & Lorentzian & Lorentzian & Classical  & \\
        (\secref{sec:three-dimension})& Euclidean  & Lorentzian & Tunneling  & \\
        & Euclidean  & Euclidean  & Tunneling  &
          w/ saddle ambiguity that resurgence can resolve.\\[6pt]

        \multirow[t]{3}{*}{4D GR }
        & Lorentzian & Lorentzian & Classical  & \\
        (\secref{sec:four-dimension})& Euclidean  & Lorentzian & Tunneling  & \\
        & Euclidean  & Euclidean  & Tunneling  &
          w/ saddle ambiguity that resurgence can resolve.\\[6pt]

        \multirow[t]{3}{*}{5D GR }
        & Lorentzian & Lorentzian & Classical  & \\
        (\secref{sec:five-dimension})& Euclidean  & Lorentzian & Tunneling  & \\
        & Euclidean  & Euclidean  & Tunneling  &
          w/ saddle ambiguity (in some cases) that resurgence can resolve. \\[6pt]

        \multirow[t]{3}{*}{Large-$D$ GR}
        & Euclidean   & Euclidean   & Tunneling &
          w/o parameter rescaling.\\
        (\secref{sec:large-d-limit}) & Lorentzian  & Lorentzian  & Classical &
          w/ parameter rescaling.\\
        & Euclidean   & Euclidean   & Tunneling &
          w/ parameter rescaling.\\

        \hline\hline
        \vspace{1pt}
    \end{tabularx}
    \label{tab:summary}
    \begin{minipage}{0.95\linewidth}
    \textit{Note.} For the $D \geq 3$ setup, the initial and final spacetime geometries are characterized as Lorentzian or Euclidean according to whether the corresponding boundary value of the scale factor lies in the classically allowed or forbidden region, respectively. In the 2D JT case, by contrast, \textit{Lorentzian} and \textit{Euclidean} refer not to the individual boundaries but to whether the initial and final boundary data can be connected by a Lorentzian classical solution; this is determined by the sign of a combination of the differences between the initial and final values of the scale factor and the dilaton.
    The preferred scenarios--Classical/No-boundary/Tunneling--are distinguished by the sign of the imaginary part of the lapse: zero/negative/positive, respectively. In the large-$D$ setup, we compare outcomes with and without a rescaling of the cosmological constant and the lapse, and demonstrate that implementing this rescaling makes the presence of a classical transition more transparent.
\end{minipage}
\end{table}

Motivated by these developments, the goal of this paper is to extend this framework to spacetime dimensions %$D \ge 3$ 
beyond $D=2$ and $4$ (specifically, $D \geq 3$) and to investigate the dimensional dependence of the resulting quantum cosmology. 
We comprehensively analyze the cases with $D=3, 4$, and $5$, as well as the large-$D$ limit in Einstein gravity, and clarify how the saddle structure, thimble contributions, and resulting transition amplitude vary with $D$. 
We begin with the $D$-dimensional de Sitter minisuperspace and construct the Lorentzian path integral for the scale factor and the complex lapse under the Dirichlet-type boundary conditions. Using Picard-Lefschetz theory in each case, we systematically identify the contributing saddles and appropriate thimbles, to reveal the dimensional dependence of the quantum cosmological scenario.

There are two main motivations for considering the large-$D$ limit in the present context. 
The first is to obtain insight into cosmological scenarios based on higher-dimensional theories of gravity. 
The possibility that extra dimensions played an important role in the early universe has long been explored, for instance, in the context of cosmology based on string/M-theory, or in braneworld scenarios motivated by the hierarchy problem among the fundamental interactions~\cite{Arkani-Hamed:1998jmv}. 
Therefore, the large-$D$ limit may provide useful insight into such scenarios.
The second motivation is related to more technical aspects.
It is known that taking the large-$D$ limit in gravity significantly simplifies analyses and provides transparent descriptions in perturbative expansions \cite{Strominger:1981jg,Bjerrum-Bohr:2003veq} and black-hole physics \cite{Emparan:2013moa, Emparan:2013xia, Emparan:2014aba, Emparan:2015hwa, Bhattacharyya:2015dva,Bhattacharyya:2015fdk,Dandekar:2016fvw,Dandekar:2016jrp,Bhattacharyya:2016nhn,Emparan:2020inr, Emparan:2025yfy}. One may therefore expect that a similar large-$D$ analysis in Lorentzian quantum cosmology can lead to a more comprehensive understanding of, for example, the structure of contributing saddles.

In this work, by studying not only specific dimensions such as $D=3, 4$, and $5$ but also the large-$D$ limit, we show that the results obtained in four-dimensional GR \cite{Feldbrugge:2017kzv, Honda:2024aro} are, in fact, robust predictions essentially independent of spacetime dimensionality. Since the tunneling proposal is known to suffer from perturbative instability \cite{Feldbrugge:2017fcc,Feldbrugge:2017mbc,Matsui:2022lfj}, this result suggests an intrinsic limitation of scenarios for the creation of the universe based solely on GR, regardless of dimensionality. All of our results are summarized in \tabref{tab:summary}.

Moreover, as a further application of resurgence to quantum cosmology, we perform a numerical resurgence analysis based on the so-called Borel-Pad\'e resummation
for several specific cases in four and three dimensions. 
Whereas constructing the full Borel transform requires perturbative coefficients to all orders, Borel–Pad\'{e} resummation uses finite coefficients to approximate the Borel transform by a Pad\'{e} approximant.
In our previous study \cite{Honda:2024aro}, the analysis was carried out analytically in the special setup where the initial and final states coincide. In the present work, we generalize this analysis by employing a numerical approach with the Pad\'e approximation, and show that resurgence can resolve the Stokes phenomenon also in more general situations.

The remainder of this paper is organized as follows. In Sec.~\ref{sec:setup}, we introduce the $D$-dimensional minisuperspace model and
canonical variables used in the later sections to evaluate the Lorentzian path integral of the gravitational transition amplitude. We introduce our basic methodology, Picard-Lefschetz theory and resurgence techniques, in Sec.~\ref{sec:four-dimension}, where we review the four-dimensional case. We slightly generalize the analyses in Ref.~\cite{Honda:2024aro} and discuss the Borel-Pad\'e resummation, which is also used in the following section. It is useful when analytic results are difficult to obtain.  The cases $D = 3$ and $D = 5$ are studied in Secs.~\ref{sec:three-dimension} and \ref{sec:five-dimension}, respectively, followed by the large-$D$ analysis in Sec.~\ref{sec:large-d-limit}. Sec.~\ref{sec:conclusion} is devoted to our discussion and conclusion.  Appendix~\ref{sec:5D_fin} is dedicated to cases not covered in Sec.~\ref{sec:five-dimension}.

%%%%%%%%%%%%%%%%%%%%%%%%%%%%%%%%%%%%%%%
%%%%%%%%%%%%%%%%%%%%%%%%%%%%%%%%%%%%%%%
\section{Our Setup}
\label{sec:setup}
%%%%%%%%%%%%%%%%%%%%%%%%%%%%%%%%%%%%%%%
%%%%%%%%%%%%%%%%%%%%%%%%%%%%%%%%%%%%%%%

We work in the minisuperspace approximation, assuming a homogeneous and isotropic $D$-dimensional Friedmann-Lema\^itre-Robertson-Walker (FLRW) ansatz
\begin{align}
    \dd{s}^2 = - N^2 (t) \dd{t}^2 + a^2(t)
    \qty[\frac{\dd{r}^2}{1-kr^2}+r^2 \dd{\Omega}^{2}_{D-2}],
    \label{eq:FLRW}
\end{align}
where $N(t)$ is the lapse function, $a(t)$ is the scale factor, $k\in\{-1,0,1\}$ denotes the constant spatial curvature, and $\dd{\Omega}^{2}_{D-2}$ is the metric of the unit $(D-2)$-sphere. 

On this background, the Einstein-Hilbert action with the cosmological constant $\Lambda$ reduces to
\begin{align}
    S\qty[a, N]
    &=\frac{V_{D-1}}{16 \pi G_D}
    \int \dd{t}
        \frac {a^{ D - 3 }} {N^{2}} 
        \left[(D-1)(D-2) k N^3-2 \Lambda a^2 N^3-2(D-1) a \dot{a} \dot{N}
    \right.
    \nn
    & \hspace{100pt}
    \left.
        +(D-1)(D-2) \dot{a}^2 N+2(D-1) N a \ddot{a}
        \right]
    + \qty({\rm bdy}),
    %\nn
\end{align}
where ``bdy'' collectively denotes boundary contributions including the Gibbons-Hawking-York term, $V_{D-1}$ is the comoving volume of the $(D-1)$-dimensional unit sphere, 
and $G_D$ is the $D$-dimensional Newton constant. After integrating by parts and absorbing total derivatives into the boundary terms, we obtain
\begin{align}\label{eq:eh-action}
    S\qty[a, N]
    &=\frac{V_{D-1}}{16 \pi G_D}
    \int \dd{t}
        \frac {a^{ D - 3 }} {N^{2}} 
        \left[(D-1)(D-2) k N^3-2 \Lambda a^2 N^3
        -(D-1)(D-2) \dot{a}^2 N
        \right]
    + \qty({\rm bdy}).
\end{align}

To carry out the path integral explicitly, it is convenient to introduce ``good'' canonical variables and fix the time-reparametrization gauge.
Since the most useful choice for canonical variables depends on the spacetime dimension, we present two parametrizations that will be particularly useful in this paper.

\bigskip
\noindent\textbf{Option I.} 
We define
\begin{equation}
    \bar{N} \coloneqq N\, a^{D-3}, 
    \quad
    q \coloneqq a^{D-2}\,.
\end{equation}
In terms of these variables, the action \eqref{eq:eh-action} becomes
\begin{align}
    S\qty[q, \bar N]
    &=\frac{V_{D-1}}{16 \pi G_D}
    \int \dd{t}
        \left[(D-1)(D-2) k \bar{N}
        -2 \Lambda q^{\frac{2}{D-2}} \bar{N}
        -\frac{D-1}{D-2} \frac{\dot{q}^2}{\bar{N}} 
    \right]
    + \qty({\rm bdy}).
    \label{eq:d-dim_action1}
\end{align}
Here, we impose the gauge condition $\dot{\bar{N}}=0$. 
For $D \to 4$, this reduces to the standard minisuperspace action used, for example, in Refs.~\cite{Feldbrugge:2017kzv,Honda:2024aro}. Varying the action with respect to $q$ yields
\begin{align}
    \ddot{q} \;=\; \frac{2\,\bar{N}^2\, \Lambda}{D-1}\;
    q^{\frac{4-D}{D-2}},
    \label{eq:d-dimEOM1}
\end{align}
which can be solved in closed form for $D=3$ and $4$ and in the large-$D$ limit, as we will demonstrate below. In this parametrization, the kinetic term is quadratic in $q$, and the effective potential is a simple monomial in $q$, which is advantageous for the path integral analysis.

\bigskip
\noindent\textbf{Option II.}
An alternative parametrization, more convenient for certain dimensions, is obtained by defining
\begin{equation}
    \bar{N} \coloneqq N\, a^{5-D}, 
    \quad
    q \coloneqq a^{2}\,.
\end{equation}
In these variables, the action becomes
\begin{align}
    S[q, \bar N] &=\frac{V_{D-1}}{16 \pi G_D}
    \int \dd{t}
        \left[(D-1)(D-2) k q^{D-4} \bar{N}
        -2 \Lambda q^{D-3} \bar{N}
        -\frac{1}{4}(D-1)(D-2) \frac{\dot{q}^2}{\bar{N}}
        \right]
    + \qty({\rm bdy}),
    \label{eq:d-dim_action2}
\end{align}
with the gauge-fixing $\dot{\bar{N}}=0$.\footnote{
Although the expressions look the same, this gauge fixing is in principle
different from that used in Option~I, due to the different definitions of
$\bar{N}$. Nevertheless, the two coincide for $D=4$.} 
The corresponding equation of motion for $q$ reads
\begin{align}
    \ddot{q} 
    = - 2(D-4) k \bar{N}^2 q^{D-5} 
    + \frac{4\Lambda (D-3) \bar{N}^2}{(D-1)(D-2)} q^{D-4}.
    \label{eq:d-dimEOM2}
\end{align}
This form is particularly convenient in $D=4$ and $5$, where Eq.~\eqref{eq:d-dimEOM2} simplifies and admits analytic control.

In both parametrizations, the kinetic term becomes quadratic in the single minisuperspace degree of freedom $q(t)$, while the potential becomes a sum of powers of $q$, with exponents that depend on $D$. These features are crucial for evaluating the Lorentzian path integral and the associated functional-determinant prefactors in the subsequent sections. 
Although these two choices, which differ in the field redefinition and gauge-fixing, generally lead to different values of the amplitudes, their predictions can be shown to be identical at least in the semiclassical limit \cite{Partouche:2021lyb, Partouche:2021epk, Partouche:2022kfi}. We therefore choose a convenient option in each dimension and focus on the
qualitative behavior of the resulting amplitudes.

In this minisuperspace setup, the full gravitational path integral~\eqref{eq:G-amplitude1}
reduces to a quantum mechanical type path integral~\cite{Halliwell:1988wc,Halliwell:1988ik}:
\begin{equation}\label{eq:G-amplitude2}
\mathcal{A}[q_1; q_0]
= \int \dd{\bar{N}}\,\int \mathcal{D}q\;
\exp\!\left(\frac{i}{\hbar} S[q, \bar{N}]\right),
\end{equation}
where we impose the Dirichlet boundary conditions on the minisuperspace variable,
\begin{equation}\label{eq:Dirichlet-bc}
  q(t_0) = q_0, \quad
  q(t_1) = q_1,
\end{equation}
so that $q_0$ and $q_1$ specify the geometry on the initial and final hypersurfaces at times $t=t_0$ and $t=t_1$, respectively. In this paper, we set $t_0=0, \; t_1=T$ for simplicity. Throughout this paper, we adopt such Dirichlet boundary conditions in the Lorentzian path integral, which provide a natural framework for describing quantum cosmology, including the quantum creation of the universe from nothing. It is also possible to use different boundary conditions, such as Neumann or Robin boundary conditions~\cite{DiTucci:2019dji,DiTucci:2019bui,Narain:2021bff,Narain:2022msz,Ailiga:2023wzl,Ailiga:2024mmt,Ailiga:2024wdx}. In such cases, it is necessary to introduce the appropriate boundary terms into the action. The functional integral $\int \mathcal{D}q$ is taken over all histories $q(t)$ interpolating between $q_0$ and $q_1$, and the remaining integral $\int \dd{\bar N}$ ensures that the amplitude is the Green's function of the Hamiltonian constraint.

Much of the recent work in Lorentzian quantum cosmology~\cite{Feldbrugge:2017kzv,
DiazDorronsoro:2017hti,Feldbrugge:2017mbc,DiazDorronsoro:2018wro,Feldbrugge:2018gin,DiTucci:2018fdg,DiTucci:2019dji,Janssen:2019sex,DiTucci:2019bui,Narain:2021bff,Lehners:2021jmv,Narain:2022msz,Matsui:2023hei,Lehners:2023yrj,Ailiga:2023wzl,Matsui:2023tkw,Lehners:2024kus,Ailiga:2024mmt,Ailiga:2024wdx} has focused on giving a mathematically rigorous definition of the lapse integral $\int \dd{\bar N}$ using Picard-Lefschetz theory, by deforming the integration contour onto appropriate Lefschetz thimbles in the complex $\bar N$-plane. This is crucial for resolving ambiguities between different cosmological proposals, such as the no-boundary proposal~\cite{Hartle:1983ai} and tunneling proposal~\cite{Vilenkin:1984wp}, which correspond to different choices of contour and yield qualitatively distinct wave functions of the universe. However, Picard-Lefschetz theory alone is not sufficient to fully fix the integration contour when parameters lie on Stokes lines. To completely resolve this issue, one must combine a systematic analysis of Lefschetz thimbles with resurgence theory~\cite{Honda:2024aro}. Within this resurgent Picard-Lefschetz framework, the minisuperspace path integral~\eqref{eq:G-amplitude2} can be defined in a self-consistent manner, thereby removing the ambiguity in the wave function of the universe.

%%%%%%%%%%%%%%%%%%%%%%%%%%%%%%%%%%%%%%%
%%%%%%%%%%%%%%%%%%%%%%%%%%%%%%%%%%%%%%%
\section{Review of Four-dimensional Spacetime}
\label{sec:four-dimension}
%%%%%%%%%%%%%%%%%%%%%%%%%%%%%%%%%%%%%%%
%%%%%%%%%%%%%%%%%%%%%%%%%%%%%%%%%%%%%%%

As a warm-up for our methodology, which combines the Lorentzian path integral with the resurgent Picard-Lefschetz framework, we review the four-dimensional case~\cite{Halliwell:1988ik, Feldbrugge:2017kzv, Honda:2024aro}. 
For the four-dimensional case, we adopt the parametrization of Option~I introduced in Sec.~\ref{sec:setup}. With the Dirichlet boundary conditions~\eqref{eq:Dirichlet-bc}, the action $S[q,\bar{N}]$ is quadratic in $q(t)$. Consequently, the corresponding path integral can be evaluated exactly using the standard semiclassical techniques.
We decompose the path as $q(t) = q_{\rm cl}(t) + Q(t)$, 
where $q_{\rm cl}(t)$ is the classical solution satisfying the Dirichlet boundary conditions, and $Q(t)$ denotes the fluctuation with homogeneous Dirichlet boundary conditions,
$Q(0) = Q(T) = 0$.
Inserting this decomposition into the action and using the classical equations of motion to eliminate terms linear in $Q$, the action splits into an on-shell action $S_0$ and a quadratic fluctuation action $S_2$~\cite{Halliwell:1988ik,Feldbrugge:2017kzv}:
\begin{align}
S[q,N] &= S_0[q_{\rm cl},N] + S_2[Q,N], \nn
    S_0[q_{\rm cl},N]
    &= \frac{V_3}{8 \pi G_4}
    \int_{0}^{T} \dd{t}\left[
          3k N- \Lambda N q_{\rm cl}
          - \frac{3}{4}\frac{\dot{q}_{\rm cl}^2}{N}\right], \nn
    S_2[Q,N]
    &= \frac{V_3}{8 \pi G_4}
    \int_{0}^{T} \dd{t}\left[ - \frac{3}{4}\frac{\dot{Q}^2}{N}\right],
    \label{eq:4Daction}
\end{align}
where we drop the bar on the lapse $N$ for simplicity.
The classical solution under the Dirichlet boundary conditions~\eqref{eq:Dirichlet-bc} takes the form
\begin{align}
    q_{\rm cl}(t)
    &= \frac{T-t}{T}q_0 +\frac{t}{T}q_1
    -\frac{1}{3}t(T-t)\Lambda N^2.
\end{align}

It is noteworthy that the action \eqref{eq:4Daction} is nothing but the one for a nonrelativistic particle with a negative kinetic term in a linear potential. Indeed, we can explicitly write down the Hamiltonian:
\begin{align}
    &H_4 = N \qty(-\frac{P^2}{2\mathcal{M}_4} + U_4(q)),
    \quad
    U_4(q) =2\mathcal{M}_4 
    \qty(-k+\frac{1}{3}\Lambda q)
    \nn
    &\mathcal{M}_4 \coloneqq \frac{3V_3}{16 \pi G_4} ,
    \quad P\coloneqq -\frac{\mathcal{M}_4}{N} \dot{q}.
    \label{eq:4DHam}
\end{align}
This Hamiltonian gives a constraint; that is, the expression in parentheses in $H_4$ must vanish.  
Since $P^2 \geq 0$, a classically allowed configuration must satisfy $U_4 \geq 0$, or in other words,
\begin{align}
    q \geq q_{\rm crit} = \frac{3k}{\Lambda}.
\end{align}
This critical value will appear again when we consider the physical interpretation of the saddle points.

The transition amplitude $\mathcal{A}[q_1; q_0]$ is calculated by substituting the classical solution $q_{\rm cl}$ into $S_0$ in Eq.~\eqref{eq:4Daction} and performing the path integral over $Q(t)$. Since $S_2$ is essentially the free-particle action, we can easily perform the Gaussian path integral of it as in Ref.~\cite{Feynman:100771}. Then, we can get the explicit expression 
\begin{align}
    \mathcal{A}[q_1; q_0]
    &=: \sqrt{\frac{3 i V_3}{32\pi^2 T \hbar G_4}}
    \int^\infty_0 \dd{N}
    \exp\qty[F_4(N)]\,,
    \label{eq:4Damp}
\end{align}
with
\begin{align}
    F_4(N) \coloneqq
    -\frac{1}{2}\log N
    +\frac{i V_3}{8\pi \hbar G_4}
        \qty(
            - \frac{3}{4 N T}\,(q_1 - q_0)^2
            + \frac{N T}{2}\,\bigl(6k - \Lambda (q_0+q_1)\bigr)
            + \frac{\Lambda^2}{36}\,N^3 T^3
        ),
    \label{eq:4Dphase}
\end{align}
where the first term is the (logarithm of) prefactor stemming from $Q$-integration, and the remaining is the semiclassical saddle contribution. 
Here we primarily restrict the lapse integral to positive values, $N \in (0,\infty)$, which corresponds to forward evolution in proper time, and this choice of $N$ ensures causality~\cite{Teitelboim:1983fh}. Alternatively, we may extend the integration domain to $N \in (-\infty,\infty)$ 
as considered in Refs.~\cite{DiazDorronsoro:2017hti,Feldbrugge:2017mbc}. In the Lorentzian path integral, however, this extended range leads to a non-unique choice of integration contour in the complexified lapse plane and thereby to an ambiguity between the no-boundary and tunneling proposals.

The main idea of the Lorentzian path-integral method with Picard-Lefschetz theory is the analytical deformation of the original $N$ integration contour onto the appropriate ``steepest descent paths'' within the complex $N$ plane. To find them, we must identify the saddle points, which %that 
satisfy $\dd{F_4}/\dd{N}=0$. They are found analytically in the semiclassical limit $\hbar\rightarrow 0$ as
\begin{align}
    &N_1 = -N_2 = \frac{1}{T\sqrt{\Lambda}}
    \sqrt{3(q_0+q_1-2q_{\rm crit}) + 6 \sqrt{\qty(q_0 - q_{\rm crit})\qty(q_1 - q_{\rm crit})}},
    \nn
    &N_3 = -N_4 = \frac{1}{T\sqrt{\Lambda}}
    \sqrt{3(q_0+q_1-2q_{\rm crit}) - 6 \sqrt{\qty(q_0 - q_{\rm crit})\qty(q_1 - q_{\rm crit})}}.
    \label{eq:4dsaddles}
\end{align}

In the following, we focus on \ac{ds} spacetime with a positive spatial curvature: $k, \, \Lambda >0$ and $0 \leq q_0 \leq q_1$. Then, it is convenient to distinguish three regimes depending on the relative values of $q_0$ and $q_1$ with respect to the critical scale $q_{\rm crit}$:\footnote{Actually, all saddles and the following classification are symmetric under $q_0\leftrightarrow q_1$ since the action \eqref{eq:4Dphase} is. Such solutions correspond to the contracting universe.
}
\footnote{The boundary cases ($q_i=q_{\mathrm{crit}}$) in each region are understood as the limits in which some of the semiclassical saddles for $N$ become degenerate. Notably, the case with $q_0=q_1=q_{\mathrm{crit}}$ is exceptional: all semiclassical saddles merge at $N=0$, and quantum corrections including the log prefactor in Eq.~\eqref{eq:4Dphase} are then crucial for determining the saddle structure. We do not consider these cases in the present work.
}
\begin{enumerate}[label=(\roman*).]
    \item Lorentzian-to-Lorentzian:
    $q_1 \geq q_0 \geq q_{\rm crit} = 3k/\Lambda$ \\
    or equivalently $(3k-\Lambda q_0 )(3k-\Lambda q_1 ) \geq 0$ and $6k \leq \Lambda (q_0 +q_1)$,
    \item Euclidean-to-Lorentzian:
    $q_1 > q_{\rm crit} > q_0 \geq 0$\\
    or equivalently $(3k-\Lambda q_0 )(3k-\Lambda q_1 ) < 0$,
    \item Euclidean-to-Euclidean: 
    $q_{\rm crit} \geq q_1 \geq q_0 \geq 0$ \\
    or equivalently $(3k-\Lambda q_0 )(3k-\Lambda q_1 ) \geq 0$ and $6k \geq \Lambda (q_0 +q_1)$.
\end{enumerate}
In case (i), all saddles become real, and they give the purely imaginary $F_4$ in the $\hbar \rightarrow 0$ limit. 
This indicates the purely Lorentzian evolution of the universe that is already larger than the critical size, and in particular captures the classical evolution of the universe. Cases (ii) and (iii) describe the transition from the classically-forbidden solution to the allowed/forbidden one, respectively. The quantum creation of the universe from nothing, which we are primarily interested in, belongs to these branches. 

The saddles \eqref{eq:4dsaddles} in these cases become complex. Those with $\im{N}<0$ correspond to the usual convention of the Wick rotation, and give $\re{F_4}>0$, implying the no-boundary setting. Those with $\im{N}>0$ mean anti-Wick rotation, and hence correspond to the tunneling scenario with $\re{F_4}<0$. 
The saddle points in the finite $\hbar$ case can be found numerically. Although saddles in such circumstances are generally complex for any boundary conditions due to the $\hbar$ correction, their corresponding scenarios should be identified through the semiclassical ($\hbar\rightarrow 0$) analysis above.

In each of the cases (i)--(iii), the relevant saddle points are determined using the framework of Picard-Lefschetz theory. 
The basic idea is as follows: 
Let us consider a one-dimensional integral along a contour $C$ with the form\footnote{
The contour $C$ is taken such that the integrand vanishes at the end points of $C$ if $C$ is not closed.
} 
\begin{align}
I = \int_C dz \exp\qty[F (z)]\, .
\end{align}
Picard-Lefschetz theory provides a systematic way to rewrite this as a superposition of integrals along steepest-descent paths or Lefschetz thimbles associated with saddle points that is equivalent to the original contour $C$ via Cauchy's integration theorem.
The Lefschetz thimble $\mathcal{J}_\sigma$ associated with the saddle point $\sigma$ is defined as a solution of the following differential equation, called the flow equation
\begin{align}
\left. \frac{dz}{ds} \right|_{\mathcal{J}_\sigma} = -\overline{\frac{\partial F}{\partial z}} \qquad {\rm with}\ \lim_{s\rightarrow -\infty} z(s) =\sigma ,
\end{align}
where $s$ parametrizes the curve $\mathcal{J}_\sigma$ in the complex $z$-plane.
The flow equation implies that the Lefschetz thimble has the following properties
\begin{align}
\left. \frac{d}{ds} {\rm Im}F(z(s)) \right|_{\mathcal{J}_\sigma} = 0 \quad {\rm and} \quad
\left. \frac{d}{ds} {\rm Re}F(z(s)) \right|_{\mathcal{J}_\sigma} \leq 0 .
\end{align}
Namely, the integrand is absolutely convergent and non-oscillatory along the thimbles.
Then one can rewrite the integral as
\begin{align}
I = \sum_{\sigma\in {\rm saddles}} n_\sigma  \int_{\mathcal{J}_\sigma } dz \exp\qty[F (z)]\, ,
\label{eq:thimble-decomposition}
\end{align}
where $n_\sigma$ is an integer called the Stokes multiplier, which determines how each saddle $\sigma$ contributes to the integral.
Topological considerations then show that the contributing saddles and their associated thimbles are precisely those whose steepest ascent paths (dual thimbles), i.e., the steepest upward flows of $F(z)$, intersect the original integration contour.\footnote{
More precisely, the dual thimble $\mathcal{K}_\sigma$ associated with the saddle $\sigma$ is defined as a solution of the differential equation
\[
\left. \frac{dz}{ds} \right|_{\mathcal{K}_\sigma} = +\overline{\frac{\partial F}{\partial z}} \qquad {\rm with}\ \lim_{s\rightarrow -\infty} z(s) =\sigma ,
\]
which has the following properties
\[
\left. \frac{d}{ds} {\rm Im}F(z(s)) \right|_{\mathcal{K}_\sigma} = 0 \quad {\rm and} \quad
\left. \frac{d}{ds} {\rm Re}F(z(s)) \right|_{\mathcal{K}_\sigma} \geq 0 .
\]
}
Thus Picard-Lefschetz theory enables us to systematically rewrite a conditionally convergent integral as a superposition of absolutely convergent integrals.

In general, as the parameters are varied, the integer $n_\sigma$ cannot change continuously; 
rather, it may undergo discontinuous jumps across certain loci in parameter space, referred to as Stokes lines. 
Accordingly, the form of the perturbative expansion can change discontinuously under parameter variation. 
This behavior is known as the Stokes phenomenon.
The Stokes phenomenon may occur when distinct saddle points have the same imaginary part of the exponent appearing in the integrand, namely,
\begin{align}
\textrm{Im}[F(\sigma )]=\textrm{Im}[F(\sigma')],
\end{align}
where $\sigma$ and $\sigma'$ denote different saddle points. 
On a Stokes line, a Lefschetz thimble may pass through multiple saddle points, and consequently the Stokes multiplier $n_\sigma$ is not uniquely defined. 
The thimble decomposition \eqref{eq:thimble-decomposition} is therefore ambiguous on the Stokes lines.
To resolve this ambiguity, it is often useful to deform the parameters slightly away from a Stokes line and study the limiting behavior as we approach the line from different directions. 
In a variety of examples, individual contributions to the thimble decomposition exhibit discontinuous jumps across Stokes lines. The full result, however, remains continuous since these jumps cancel against those arising from the resummation of perturbative series, as we discuss in detail below.

Let us apply Picard-Lefschetz theory to the amplitude \eqref{eq:4Damp} of the four-dimensional minisuperspace model.
After the deformation, the amplitude \eqref{eq:4Damp} is now written as a sum of integrals over the contributing Lefschetz thimbles $\mathcal{J}_\sigma$
%$C_\sigma$
\begin{align}
    \mathcal{A}[q_1; q_0]
    = \sqrt{\frac{3 i V_3}{32\pi^2 \hbar T G_4}}
    \sum_{\sigma} n_\sigma \int_{\mathcal{J}_\sigma} \dd{N}
    \exp\qty[F_4(N)]\,.
\end{align}
Here $n_\sigma=\qty{0, \pm 1}$ denotes the Stokes multiplier, which specifies the oriented contribution of each thimble $\mathcal{J}_\sigma$
%$C_\sigma$ 
to the deformed integration contour. In practice, these coefficients are determined by how the associated steepest ascent path intersects the original contour.

We then show the saddle and thimble analysis for the three cases (i)--(iii) in the following.\footnote{Here we note that $k=\Lambda=T^{-2}$ is assumed in most of the numerical calculations in this paper unless otherwise stated.}

\begin{figure}[tp]%
  \begin{minipage}[t]{0.5\linewidth}%
    \centering%
    \includegraphics[keepaspectratio, width=\linewidth]{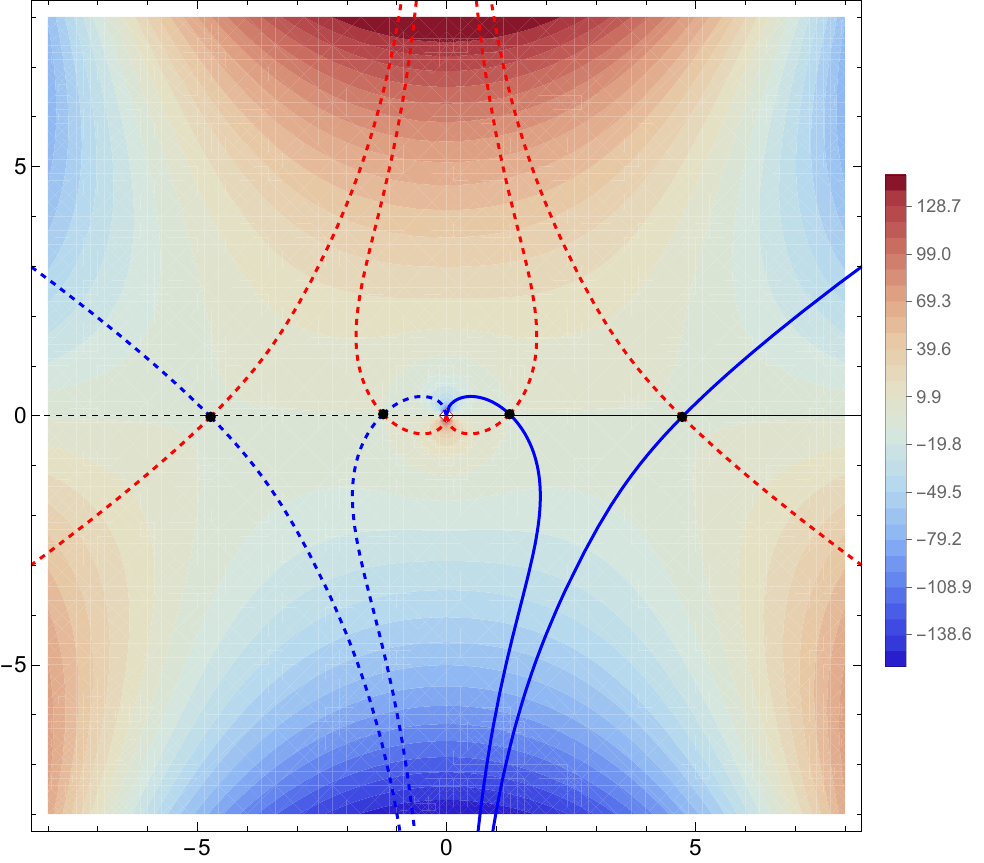}%
    \subcaption{Lorentzian-to-Lorentzian}%
    \label{fig:4dcl_thimbles}%
  \end{minipage}%
  \begin{minipage}[t]{0.5\linewidth}%
    \centering%
    \includegraphics[keepaspectratio, width=\linewidth]{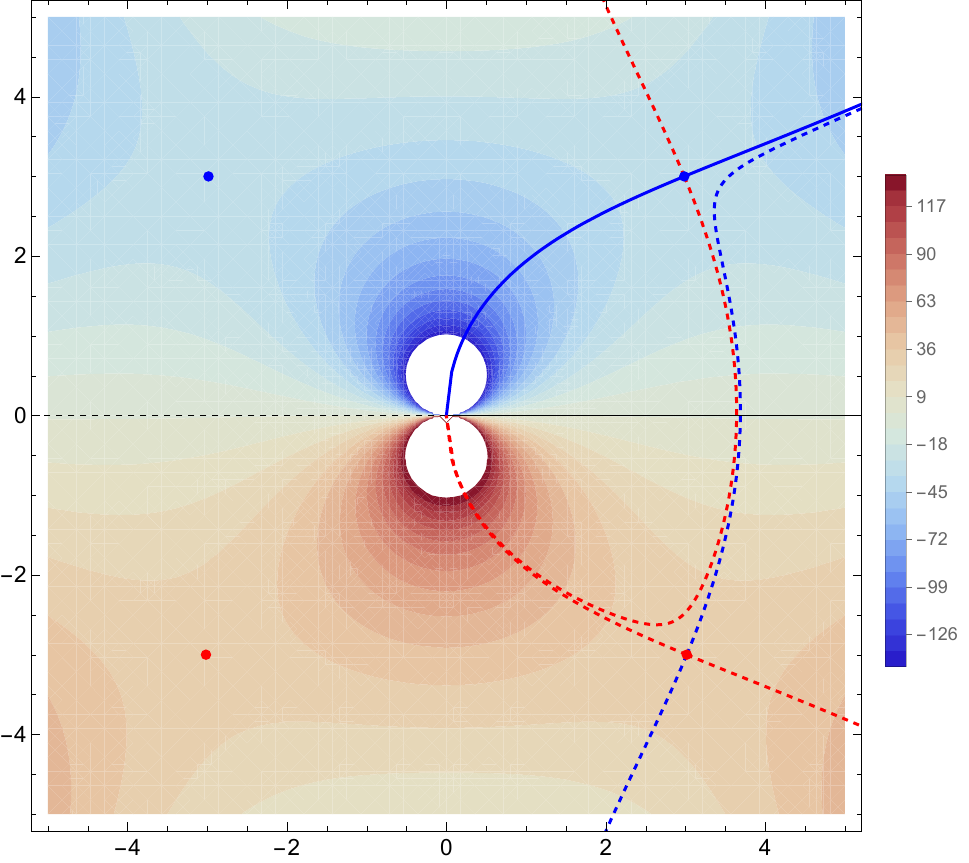}%
    \subcaption{Euclidean-to-Lorentzian}%
    \label{fig:4dinq_thimbles}%
  \end{minipage}%
  \caption{Contour plots of $\re{F_4(N)}$ on the complex $N$ plane show saddle points and associated Lefschetz thimbles in the four-dimensional Lorentzian-to-Lorentzian and Euclidean-to-Lorentzian cases. The black solid line indicates the original integration domain $0 < N < \infty$, while the black dashed lines denote branch cuts originating from the fluctuation prefactor. Black, blue, and red dots mark saddle points associated with classical evolution, tunneling, and no-boundary saddles, respectively. Blue and red dashed lines represent steepest descent and steepest ascent paths, and the blue solid lines indicate the contributing thimbles, which can be obtained by analytic deformation of the original integration contour.
  }
  \label{fig:4dCandINQ}
\end{figure}%

\begin{figure}[tp]%
  \begin{minipage}[t]{0.5\linewidth}%
    \centering%
    \includegraphics[keepaspectratio, width=\linewidth]{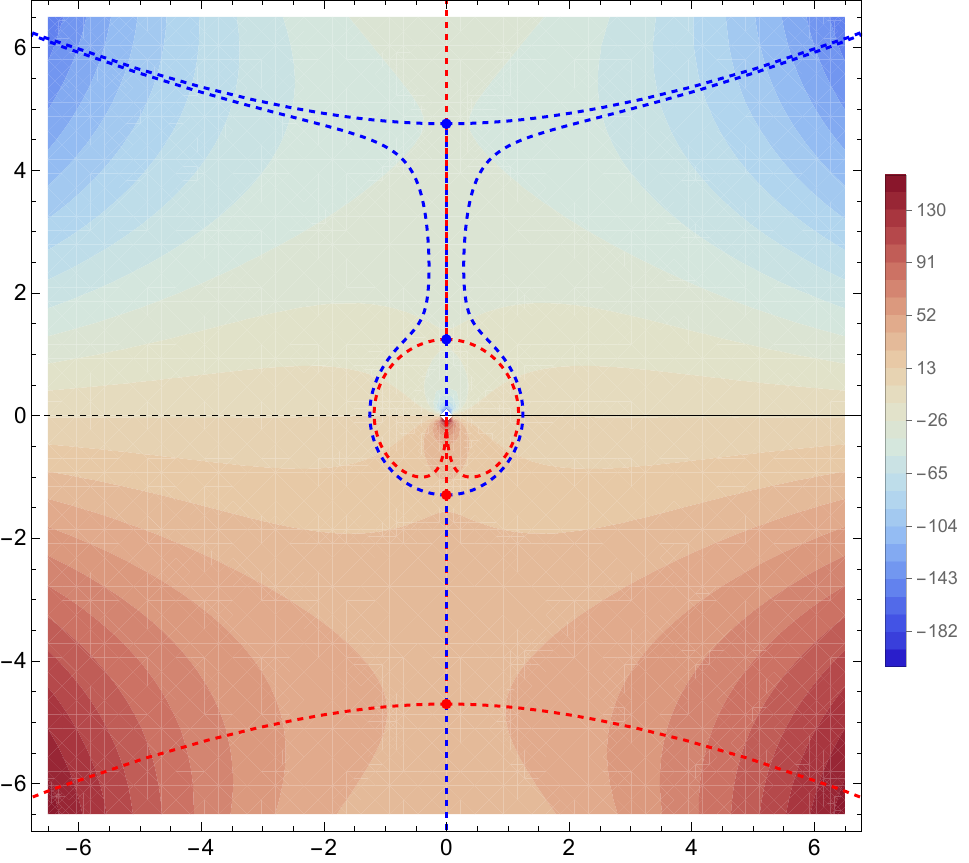}%
    \subcaption{thimbles}%
    \label{fig:4dq_thimbles}%
  \end{minipage}%
  \\[10pt]
  \begin{minipage}[t]{0.5\linewidth}%
    \centering%
    \includegraphics[keepaspectratio, width=\linewidth]{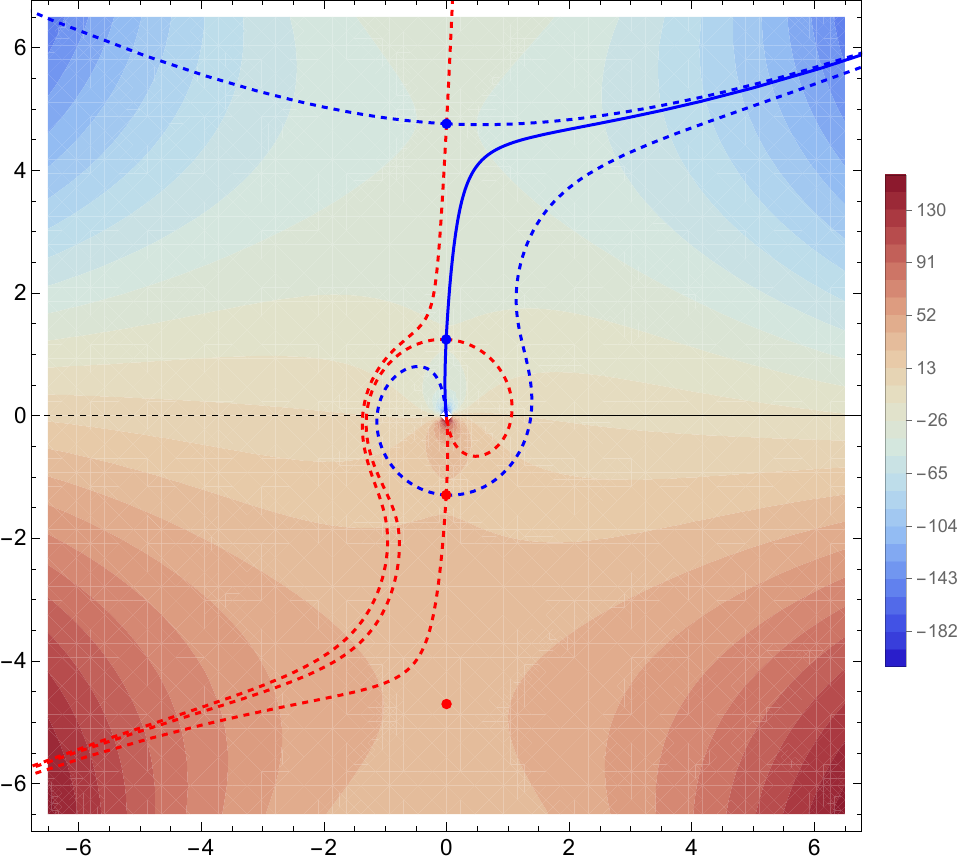}%
    \subcaption{$ \hbar \rightarrow \hbar \e^{-i\pi/30}$}%
    \label{fig:4dq_minrot}%
  \end{minipage}%
  \begin{minipage}[t]{0.5\linewidth}%
    \centering%
    \includegraphics[keepaspectratio, width=\linewidth]{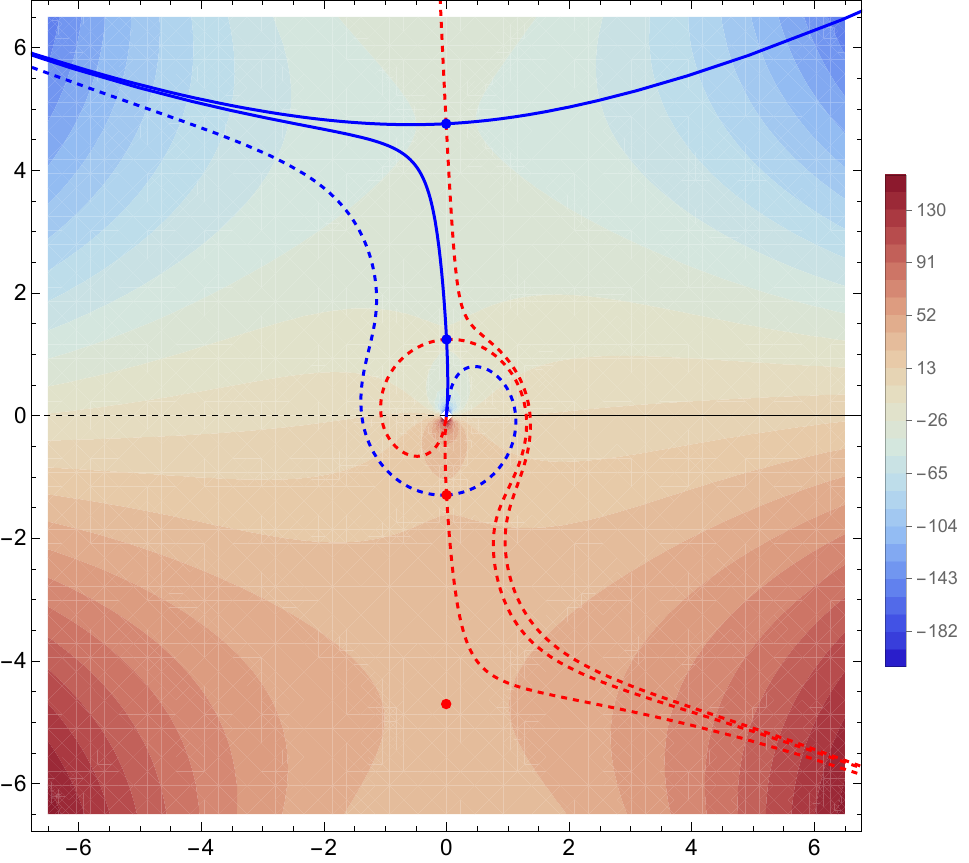}%
    \subcaption{$\hbar \rightarrow \hbar \e^{+i\pi/30}$}%
    \label{fig:4dq_plusrot}%
  \end{minipage}%
  \caption{Contour plot of $\re{F_4(N)}$ with saddle points and Lefschetz thimbles in the four-dimensional Euclidean-to-Euclidean case. Degeneracies among the thimbles are lifted by introducing a small phase rotation of $\hbar$, but an ambiguity in the choice of contributing saddles (Stokes ambiguity) remains.
   }
  \label{fig:4dq}
\end{figure}%

\subsubsection*{(i). Lorentzian-to-Lorentzian 
($q_1 \geq q_0 \geq q_{\rm crit}$)
} 

Contour plots of $\re{F_4(N)}$ for the Lorentzian-to-Lorentzian case with $q_0 = 4 k/\Lambda$ and $q_1 = 6k/\Lambda$ are shown in \figref{fig:4dcl_thimbles}.
Black dots denote the classical saddle points, while blue and red dashed lines represent the steepest descent and ascent paths associated with each saddle, respectively. They are found numerically by solving the stationary condition for $\re{F_4(N)}$ and the corresponding flow equations. One can notice that two of the classical saddles contribute, and the transition amplitude is dominated by the superposition of the corresponding classical Lorentzian histories. The original integration contour, represented as a black solid line, can therefore be deformed into the union of the contributing thimbles indicated by the blue solid lines.

\subsubsection*{(ii). Euclidean-to-Lorentzian 
($q_1 > q_{\rm crit} > q_0 \geq 0$)
}

\figref{fig:4dinq_thimbles} shows the Euclidean-to-Lorentzian case with
$q_0 = 0$ and $q_1 = 6k/\Lambda$, which describes the transition from nothing to a classically allowed configuration. In this regime, the four saddles appear in the complex region: two blue ones for the tunneling scenario, and the other two red ones for the no-boundary scenario. Only one tunneling saddle contributes, characterized by $\re{F_4} < 0$. Consequently, the resulting wave function realizes Vilenkin’s tunneling proposal. Here, we comment that the degeneracy of thimbles shown in Refs.~\cite{Feldbrugge:2017kzv, Honda:2024aro} is resolved in this plot by accounting for the prefactor contribution ($\order{\hbar^0}$ log-term in $F_4$ \eqref{eq:4Dphase}).

\subsubsection*{(iii). Euclidean-to-Euclidean ($q_{\rm crit} \geq q_1 \geq q_0 \geq 0$)
}

The Euclidean-to-Euclidean setup with $q_0 = 0$ and $q_1 = 2k/\Lambda$ is depicted in \figref{fig:4dq}. As illustrated in \figref{fig:4dq_thimbles}, the steepest-descent path from one saddle and the steepest-ascent path from another saddle degenerate,
thereby indicating that the parameters are on a Stokes line. 
As shown in FIGs.~\ref{fig:4dq_minrot} and~\ref{fig:4dq_plusrot}, these degeneracies can be lifted by slightly rotating $\hbar$ in the complex plane, $\hbar \rightarrow \hbar \e^{i\Delta \theta}$.  

Under this deformation, however, the Stokes phenomenon appears: for a small negative phase rotation,
$\hbar \to \hbar e^{-i|\Delta\theta|}$, only one tunneling saddle $N_1$ contributes (FIG.~\ref{fig:4dq_minrot}), while for a small positive phase rotation,
$\hbar \to \hbar e^{+i|\Delta\theta|}$, two tunneling saddles $N_1$ and  $N_3$ contribute (FIG.~\ref{fig:4dq_plusrot}). This abrupt change in the set of contributing thimbles as $\Delta\theta$ passes through zero characterizes the Stokes phenomenon. Since the lower tunneling saddle point $N_1$ (closer to the origin) in the figure contributes irrespective of the sign of $\Delta \theta$, we call such a saddle point an unambiguous saddle point. On the other hand, the upper tunneling saddle point $N_3$ may or may not contribute depending on the sign, so we call such a saddle point an ambiguous saddle point. 

This fact does not mean that the amplitude itself is discontinuous. In fact, we previously showed in Ref.~\cite{Honda:2024aro} through resurgence analysis that this ambiguity of the contributing saddles (the tunneling saddle further from the origin in \figref{fig:4dq}) is completely compensated by that from the Borel resummation (around the closest tunneling saddle), and the resulting amplitude continuously varies with $\Delta\theta$. 

This previous analysis was restricted to the special case \(q_0 = q_1\), where we can perform the calculation analytically. In what follows, we generalize the analysis to the regime \(q_0 \leq q_1 < q_{\rm crit}\) by employing the so-called Borel–Padé resummation. 
This technique combines Borel resummation, which extracts non-perturbative information from a perturbative asymptotic series, with Padé approximation, which constructs a rational 
approximation from a series truncated at finite order.
In practice, this approach is useful to obtain an approximation for the Borel resummation when only a finite number of series coefficients are known, and it has been used in the high-energy physics and cosmology contexts such as quantum field theory \cite{Broadhurst:1999ys,Okuyama:2018clk,Abbott:2020qnl,Fujimori:2021oqg}, quantum gravity \cite{Grassi:2014cla,Baldino:2022aqm,Eynard:2023qdr,Schwick:2026eqp}, black hole quasinormal modes \cite{Hatsuda:2019eoj,Eniceicu:2019npi,Hatsuda:2026ghx} and inflation \cite{Honda:2023unh}.\footnote{For a review of resummation, see for example Ref.~\cite{Dunne:2025mye}.} 
Let us apply this method to show that the ambiguity in the resummation around the unambiguous saddle $N_{1}$ (the closest to the origin) is compensated by another ambiguity from the choice of thimbles in \figref{fig:4dq}.

\begin{figure}[tbp]%
  \begin{minipage}[t]{0.5\linewidth}%
    \centering%
    \includegraphics[keepaspectratio, width=0.96\linewidth]{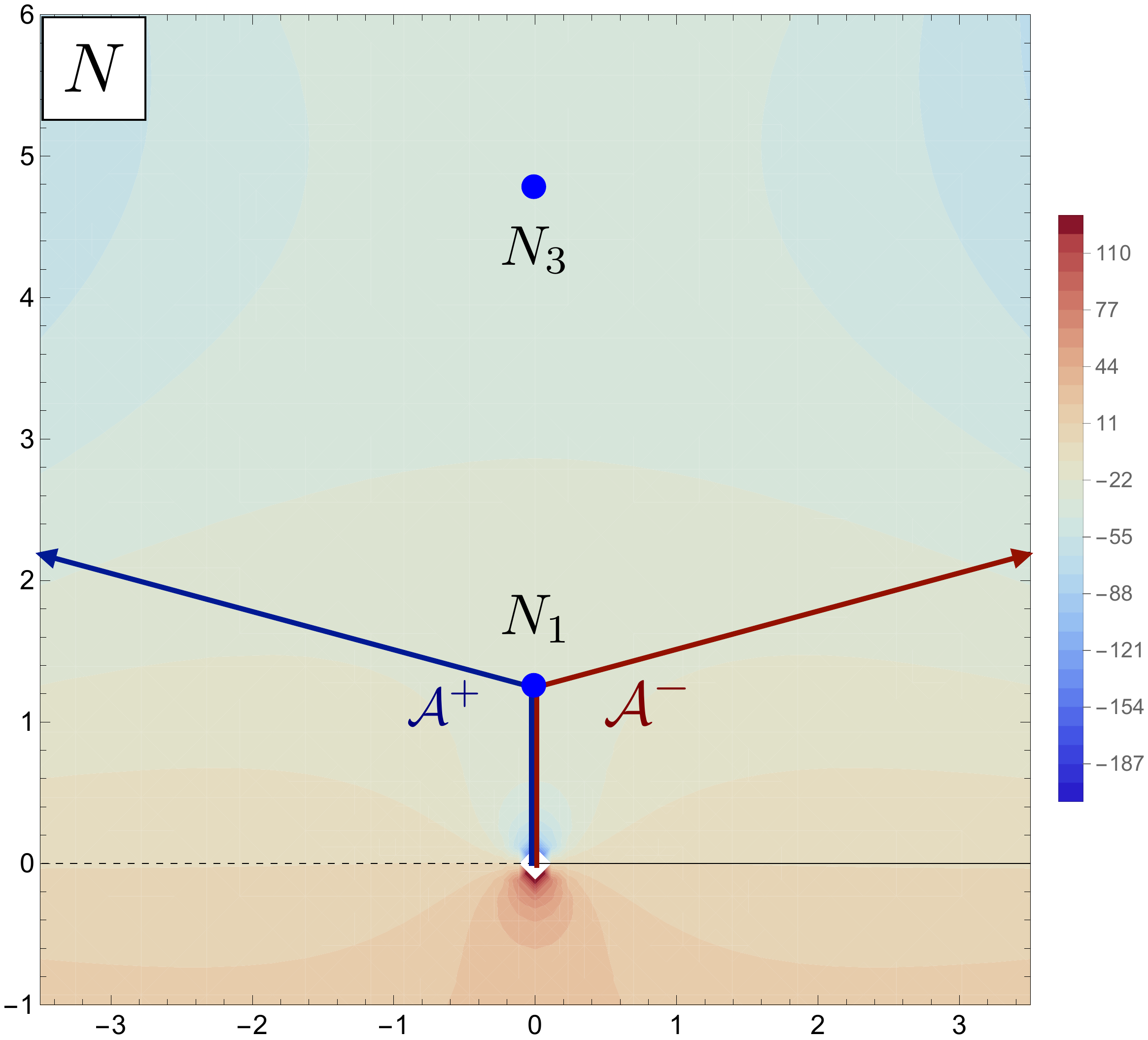}%
    \subcaption{$\re{F_4(N)}$ and integration paths}%
    \label{fig:4dq_path}%
  \end{minipage}%
  \begin{minipage}[t]{0.5\linewidth}%
    \centering%
    \includegraphics[keepaspectratio, width=\linewidth]{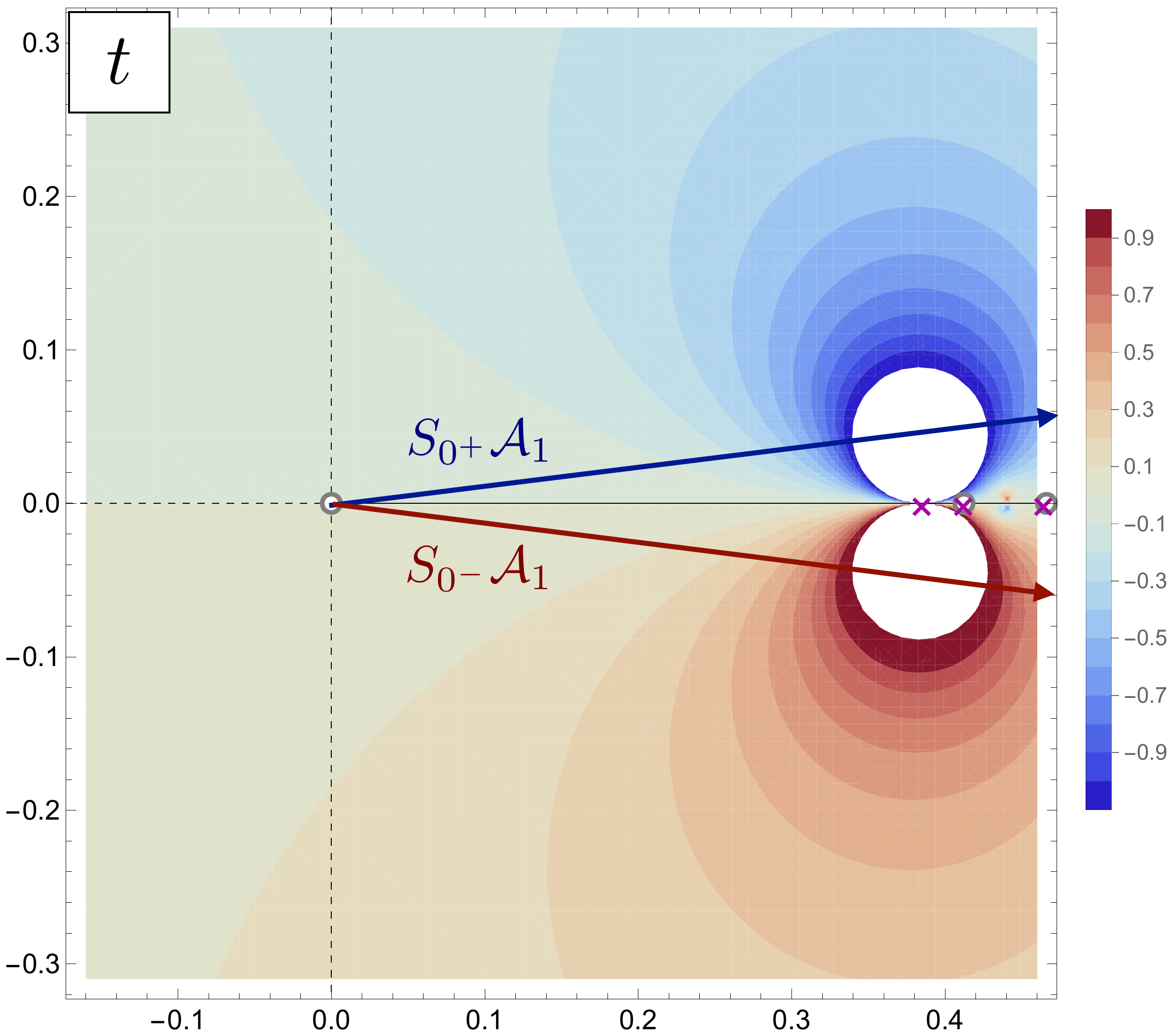}%
    \subcaption{$\re{\mathcal{BP}\bar{\mathcal{A}}(t)}$ and resummation contours}%
    \label{fig:4dq_borel}%
  \end{minipage}%
  \caption{
    (a) Contour plot of $\re{F_4(N)}$ with the integration contours used for numerical evaluation. Blue and red solid lines show the integration paths used for numerical evaluation. (b) Contour plot of $\re{\mathcal{BP}\bar{\mathcal{A}}(t)}$ on the Borel plane and the paths for resummation. Magenta crosses and gray open points are poles and zeros, respectively.
  }
  \label{fig:4dq_Nandt}
\end{figure}

Here we assume 
$q_0=0 < q_1 < q_{\rm crit}$, which leads to the following values of the relevant saddles (see also the focused contour plot \figref{fig:4dq_path}):
\begin{align}
    &N_1 = \frac{i}{T\sqrt{\Lambda}}\sqrt{3(2q_{\rm crit}-q_1) - 6 \sqrt{q_{\rm crit}\qty(q_{\rm crit}-q_1)}},
    \nn
    &N_3 = \frac{i}{T\sqrt{\Lambda}}\sqrt{3(2q_{\rm crit}-q_1) + 6 \sqrt{q_{\rm crit}\qty(q_{\rm crit}-q_1)}},
\end{align}
The saddle $N_1$ contributes unambiguously, while $N_3$ contributes ambiguously depending on $\argu{\hbar}$, as illustrated in \figref{fig:4dq}. 
To evaluate the amplitude \eqref{eq:4Damp}, let us expand it around $N_1$ as\footnote{
This integration contour is a simplified contour to reproduce the same asymptotic expansion as the thimble integral associated with $N_1$. Its direction can be determined by the second derivative of $F_4 (N)$ at $N_1$.  
}
\begin{align}
    \mathcal{A}_{1}[q_1; q_0]
    &= \sqrt{\frac{3 V_3}{32\pi^2 T G_4}}\e^{\frac{3}{4}i\pi}
    \int^\infty_{-\infty} \dd{x}
    \exp\qty[F_4(N_1 + i\sqrt{\hbar}x)]
    \nn
    &= \exp \qty[\frac{\bar{F}_4 (N_{1})}{\hbar}]
     \sum_{n=0}^\infty c_n \hbar^{n}
     \nn
    &=: \exp \qty[\frac{\bar{F}_4 (N_{1})}{\hbar}]
    \qty(c_0 + \bar{\mathcal{A}}_{1}(\hbar)),
    \label{eq:4D_asym_ser}
\end{align}
where $\bar{F}_4 (N_{1})$ is the semiclassical contribution evaluated by the saddle point approximation
\begin{align}
    \bar{F}_4 (N_{1}) 
    = \frac{V_3\sqrt{\Lambda}}{4\sqrt{3}\pi G_4}
    \frac{
        2 q_1 q_{\rm crit}-2 q_{\rm crit}^2-q_1^2
        +\qty(2 q_{\rm crit}-q_1)\sqrt{q_{\rm crit}\left(q_{\rm crit}-q_1\right)}}
        { \sqrt{2 q_{\rm crit}-q_1-2 \sqrt{q_{\rm crit}\left(q_{\rm crit}-q_1\right)}}}.
\end{align}
The coefficients $c_i$ are derived by the Gaussian integration with respect to $x$ as
\begin{align}
    &c_0=\sqrt{\frac{3 V_3}{16\pi T G_4\cdot N_1 \bar{F}^{\prime\prime}_4 (N_{1})} }
    \e^{\frac{3}{4}i\pi}, \nn
    &c_1=\sqrt{\frac{3 V_3}{64\pi T G_4}}
    \e^{\frac{3}{4}i\pi}
    \frac{-9 \bar{F}_4^{\prime \prime}(N_1)^2-5 N_1^2 \bar{F}_4^{(3)}(N_1)^2+3 N_1\bar{F}_4^{\prime \prime}(N_1)\left(-2 \bar{F}_4^{(3)}(N_1)+N_1 \bar{F}_4^{(4)}(N_1)\right)}
    {12N_1^{\frac52}\qty(\bar{F}_4^{\prime\prime} (N_{1}))^{\frac72}}
    , \cdots
\end{align}
where the prime and superscripts on $\bar{F}_4$ denote differentiation.
It should be noted here that, in the current case with $q_1 \neq q_0$, only a finite number of coefficients can be computed in practice, unlike the case with $q_1=q_0$ \cite{Honda:2024aro} where the analytical expression can be obtained. 

These coefficients exhibit factorial growth, indicating that the series in Eq.~\eqref{eq:4D_asym_ser} is an asymptotic series. In such cases, Borel resummation provides a powerful framework for controlling the factorial divergence and extracting non-perturbative information. In the present analysis, however, only a finite number of series coefficients can be computed explicitly. 
The standard Borel resummation, which requires all-order coefficients,
is therefore not directly applicable. 
One might naively think of replacing the Borel transformation with its finite-order counterpart up to the $M$-th order:
\begin{align}
    \mathcal{B}\bar{\mathcal{A}}_{1} (t)
    \coloneqq
    \sum_{n=0}^{M-1} \frac{c_{n+1}}{n!} t^{n},
\label{Borel-finite}
\end{align}
in the standard Borel resummation formula.
However, taking the Laplace transform of Eq.~\eqref{Borel-finite} simply gives the original truncated perturbative series and does not yield anything nontrivial.
Instead, we employ Borel-Pad\'e resummation, where the Pad\'e approximation is used to approximate the Borel transformation by a rational function 
from the finitely many available coefficients:\footnote{
It is known that while the Pad\'e approximation is good at approximating meromorphic functions, it is worse at approximating functions with branch cuts. For cases with branch cuts, the Pad\'e approximation typically has accumulation of poles and zeros around locations of the branch cuts.
Recently, a significant improvement of this approach for such cases has been proposed in Refs.~\cite{Costin:2020pcj,Costin:2021bay}.}
\begin{align}
    \mathcal{BP} \bar{\mathcal{A}}_{1} (t)
    \coloneqq \frac{\sum_{m=0}^p a_m t^m}{\sum_{n=0}^q b_n t^n}.
    \quad \qty(p+q=M-1)
\end{align}
Note that this can have singularities in contrast to Eq.~\eqref{Borel-finite} which partially mimic the Borel singularities.
Then we introduce the (lateral) Borel-Pad\'e resummation of $ {\mathcal{A}}_1(\hbar)$ by the following Laplace transformation
\begin{align}
    \mathcal{S}_{\theta} \mathcal{A}_{1} (\hbar)
    = \exp \qty[\frac{\bar{F}_4 (N_{1})}{\hbar}]
    \qty[
        c_0 + \int^{\infty \e^{i\theta}}_0 \dd{t}
        \e^{-\frac{t}{\hbar}} \,
        \mathcal{BP} \bar{\mathcal{A}}_{1} (t)
        \label{eq:4dq_BPresum}
    ].
\end{align}

%%%
%%%
\begin{figure}[tbp]%
  \begin{minipage}[t]{\linewidth}%
    \centering%
    \includegraphics[keepaspectratio, width=\linewidth]{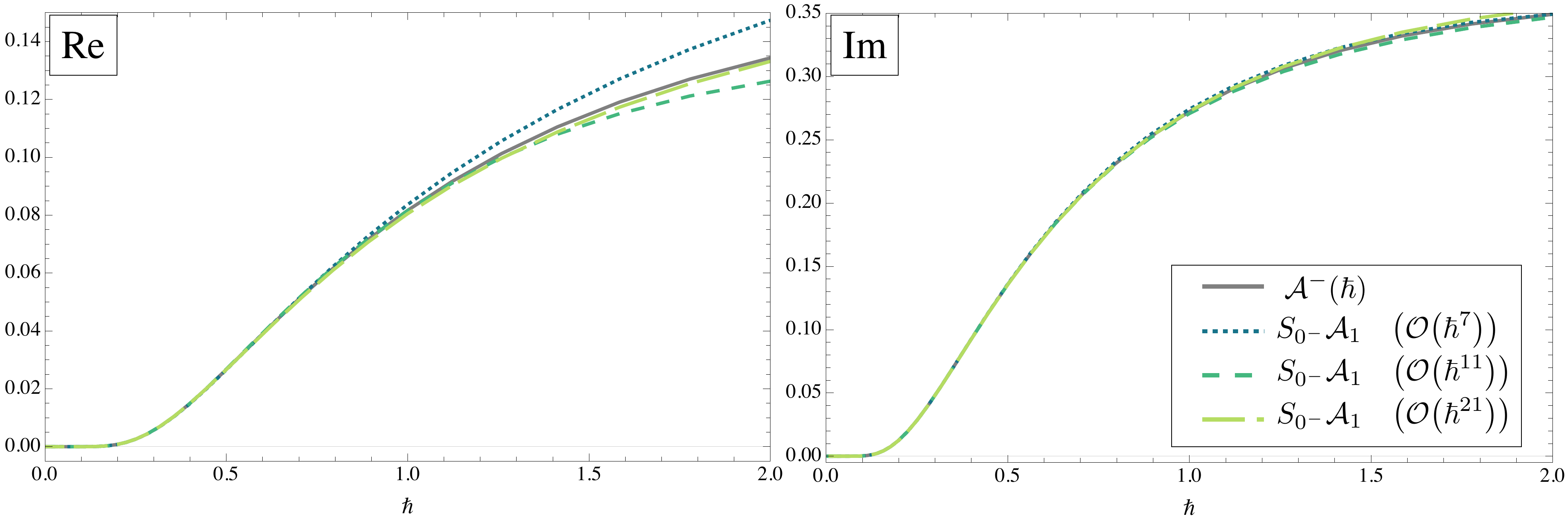}%
    \subcaption{Comparison of the direct numerical result of the amplitude value with the resummed ones.}%
    \label{fig:4dq_amprel}%
  \end{minipage}%
  \\[10pt]
  \begin{minipage}[t]{\linewidth}%
    \centering%
    \includegraphics[keepaspectratio, width=0.8\linewidth]{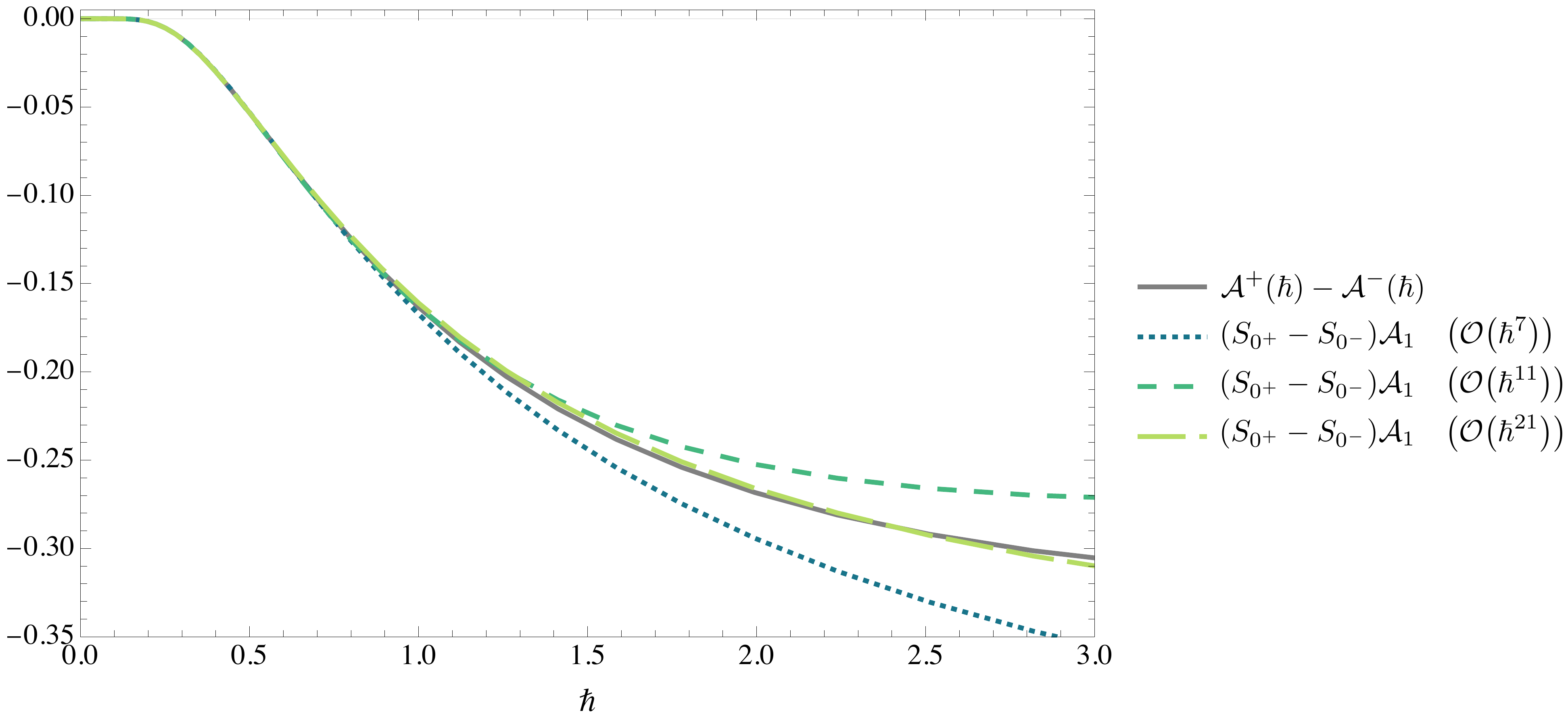}%
    \subcaption{Comparison of the path dependence with the Borel ambiguity.}%
    \label{fig:4dq_resum}%
  \end{minipage}%
  \caption{
    (a) Numerical results of the amplitude. The gray line denotes the direct numerical evaluation of Eq.~\eqref{eq:4Damp}, and colored lines are the Borel-Pad\'{e} resummed ones. 
    (b) Comparison of the dependence on the integration contour with the Borel–Pad\'{e} resummation ambiguity. 
  }
  \label{fig:4dq_resurgence}
\end{figure}%
%%%

If the integrand $\mathcal{BP} \bar{\mathcal{A}}_{1} (t)$ 
did not have singularities in the $\re{t}>0$ region of the complex $t$ plane (called the Borel plane), the integration would be performed unambiguously regardless of the value of $\theta$. 
However, there are singularities on the right half-plane 
in the present case. 
\figref{fig:4dq_borel} shows the contour plot of $\mathcal{BP} \bar{\mathcal{A}}_{1} (t)$ with $q_1 = 2k/\Lambda, M=21, p=q=10$. Magenta crosses and gray open dots denote the poles and zeros of the function, respectively. While the points where the pole and the zero overlap are numerical artifacts due to the Pad\'{e} approximation, the single pole around $t\sim 0.38$ on the real axis corresponds to the stable pole of this approximated function. The existence of this pole leads to the $\theta$-dependence in Eq.~\eqref{eq:4dq_BPresum}, that is, ambiguity of the Borel resummation (the Borel ambiguity). Indeed, we confirmed that the value of this pole position is  $t\sim \bar{F}_4 (N_{1}) - \bar{F}_4 (N_{3})$, the difference between the semiclassical action around $N_1$ and $N_3$. This fact implies that 
\begin{align*}
    (\mathcal{S}_{0^+}-\mathcal{S}_{0^-}) \mathcal{A}_{1} (\hbar)
    \sim \exp \qty[\frac{\bar{F}_4 (N_{3})}{\hbar}],
\end{align*}
and that this Borel ambiguity %approximately 
corresponds to the ambiguous $N_3$ saddle contribution.

We also explicitly confirmed the equivalence of this Borel-Pad\'{e} resummed result and the direct numerical evaluation of the amplitude \eqref{eq:4Damp}. For the latter calculation, we used the two paths as illustrated in \figref{fig:4dq_path} to see the correspondence of the ambiguities we mentioned. The comparison is exhibited in FIGs.~\ref{fig:4dq_amprel} and \ref{fig:4dq_resum}. \figref{fig:4dq_amprel} compares the $\hbar$ dependence of the numerically evaluated value $\mathcal{A}^{-} (\hbar)$ and the Borel-Pad\'{e} resummed ones $\mathcal{S}_{0^-} \mathcal{A}_{1} (\hbar)$. Lines of different colors correspond to different truncation orders $M$.\footnote{Although we present here the numerical results obtained with the diagonal Pad\'{e} approximation $p=q$, we have verified that the relevant pole in the Borel plane (\figref{fig:4dq_borel}) and the resulting conclusions remain unchanged when using other off-diagonal type approximations.}  
The resummed one 
asymptotically approaches the direct calculation result more accurately as the truncation order $M$ increases. This demonstrates the effectiveness of the Borel-Pad\'{e} resummation in evaluating the amplitude by making use of only the finite terms of the asymptotic expansion.

\figref{fig:4dq_resum} compares the integration-path dependence, $\mathcal{A}^{+}(\hbar)-\mathcal{A}^{-}(\hbar)$, with the Borel ambiguity $(\mathcal{S}_{0+}-\mathcal{S}_{0-})\mathcal{A}_{1}(\hbar)$. The figure shows that the Borel ambiguity value approaches the numerically obtained $\mathcal{A}^{+}(\hbar)-\mathcal{A}^{-}(\hbar)$. The latter corresponds to the (minus-signed) contribution from the ambiguous saddle $N_3$, i.e., the Stokes ambiguity, as understood from \figref{fig:4dq_path}. Therefore, this agreement of ambiguities indicates that the Stokes ambiguity is canceled by the Borel ambiguity arising in the resummation, so that the net value of the amplitude is unambiguous.

%%%%%%%%%%%%%%%%%%%%%%%%%%%%%%%%%%%%%%%
%%%%%%%%%%%%%%%%%%%%%%%%%%%%%%%%%%%%%%%
% \clearpage
\section{Three-dimensional Spacetime}
\label{sec:three-dimension}
%%%%%%%%%%%%%%%%%%%%%%%%%%%%%%%%%%%%%%%
%%%%%%%%%%%%%%%%%%%%%%%%%%%%%%%%%%%%%%%

In three dimensions, \ac{gr} has no local degrees of freedom (gravitational waves), but it still exhibits nontrivial global and cosmological dynamics. In the minisuperspace context, this cosmological dynamics is captured by a single scale factor, and the formalism of Sec.~\ref{sec:setup} can be specialized to the case $D=3$.

\begin{figure}[tp]%
  \begin{minipage}[t]{0.5\linewidth}%
    \centering%
    \includegraphics[keepaspectratio, width=\linewidth]{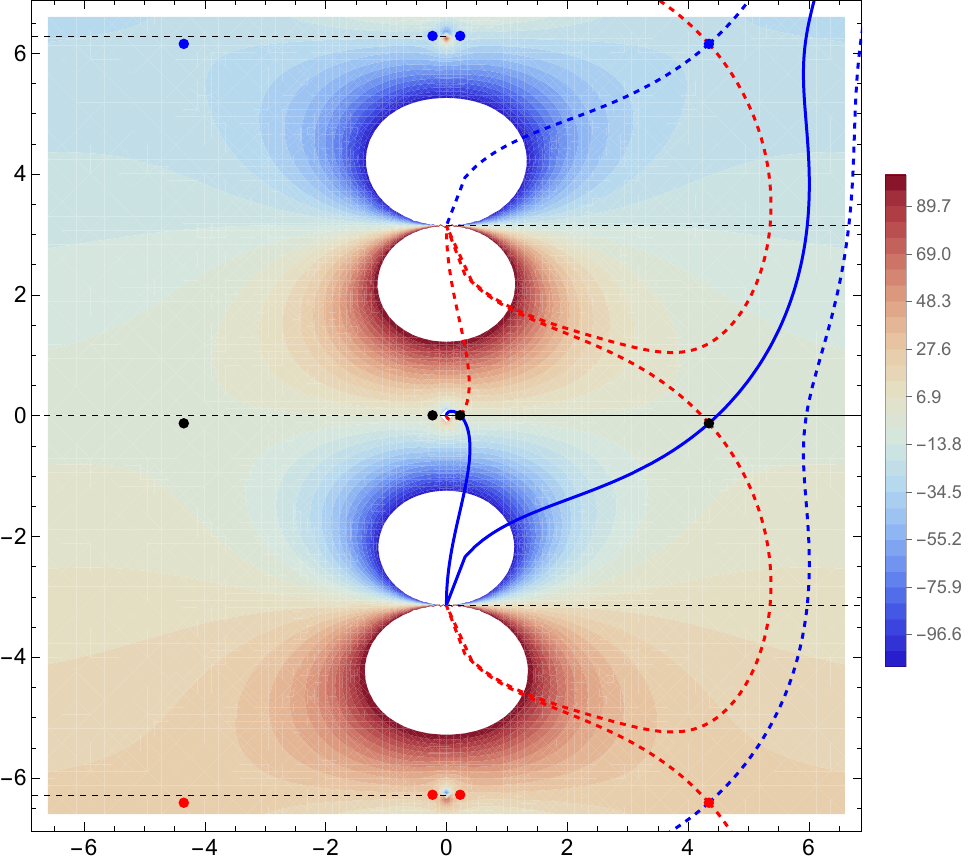}%
    \subcaption{Lorentzian-to-Lorentzian}%
    \label{fig:3dcl_thimbles}%
  \end{minipage}%
  \begin{minipage}[t]{0.5\linewidth}%
    \centering%
    \includegraphics[keepaspectratio, width=\linewidth]{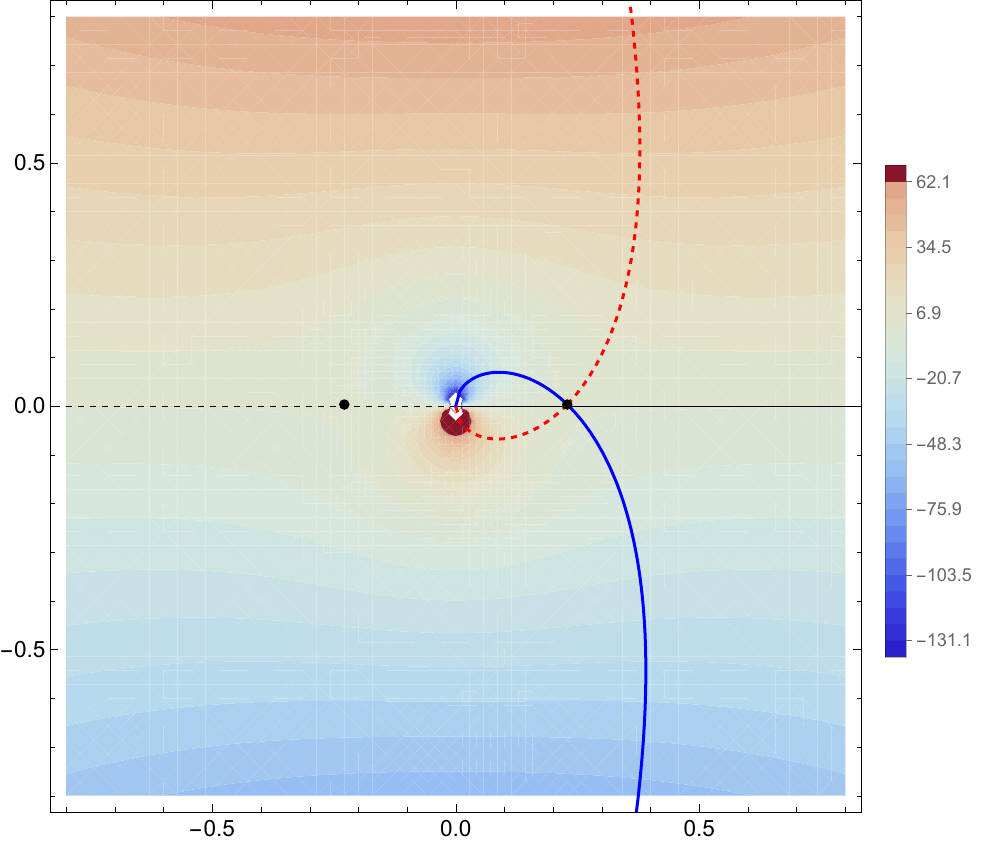}%
    \subcaption{Lorentzian-to-Lorentzian (near the origin)
    }%
    \label{fig:3dcl_center}%
  \end{minipage}%
  \\[10pt]
  \begin{minipage}[t]{0.5\linewidth}%
    \centering%
    \includegraphics[keepaspectratio, width=\linewidth]{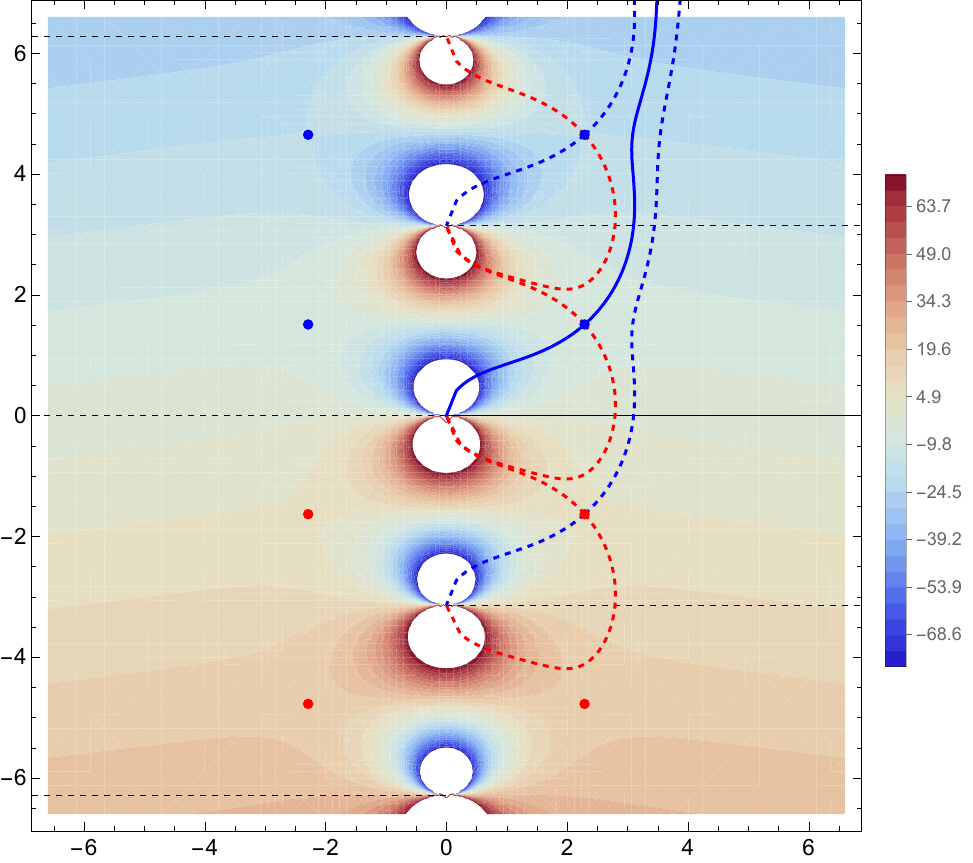}%
    \subcaption{Euclidean-to-Lorentzian}%
    \label{fig:3dinq_thimbles}%
  \end{minipage}%
  \caption{Contour plots of $\re{F_3(N)}$ with saddle points and associated Lefschetz thimbles in the three-dimensional Lorentzian-to-Lorentzian and Euclidean-to-Lorentzian cases.
  The black dashed lines denote the branch cuts.
  }
  \label{fig:3dCandINQ}
\end{figure}

\begin{figure}[tp]%
  \begin{minipage}[t]{0.5\linewidth}%
    \centering%
    \includegraphics[keepaspectratio, width=\linewidth]{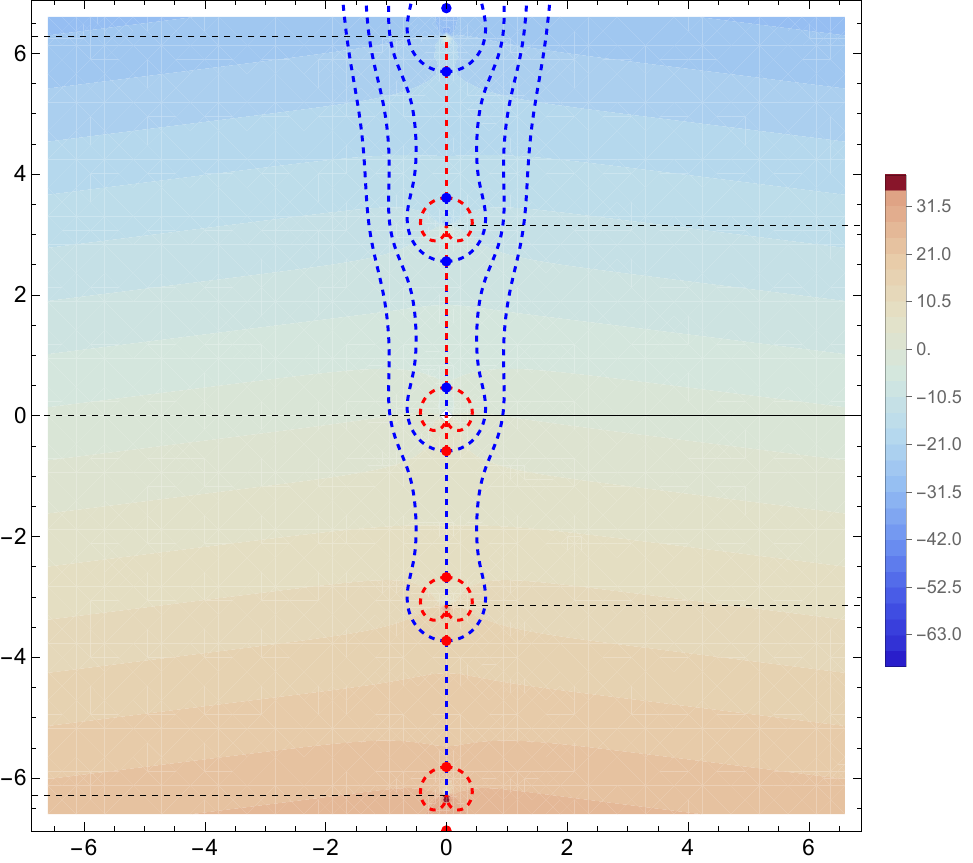}%
    \subcaption{Thimbles}%
    \label{fig:3dq_thimbles}%
  \end{minipage}%
  \\[10pt]
  \begin{minipage}[t]{0.5\linewidth}%
    \centering%
    \includegraphics[keepaspectratio, width=\linewidth]{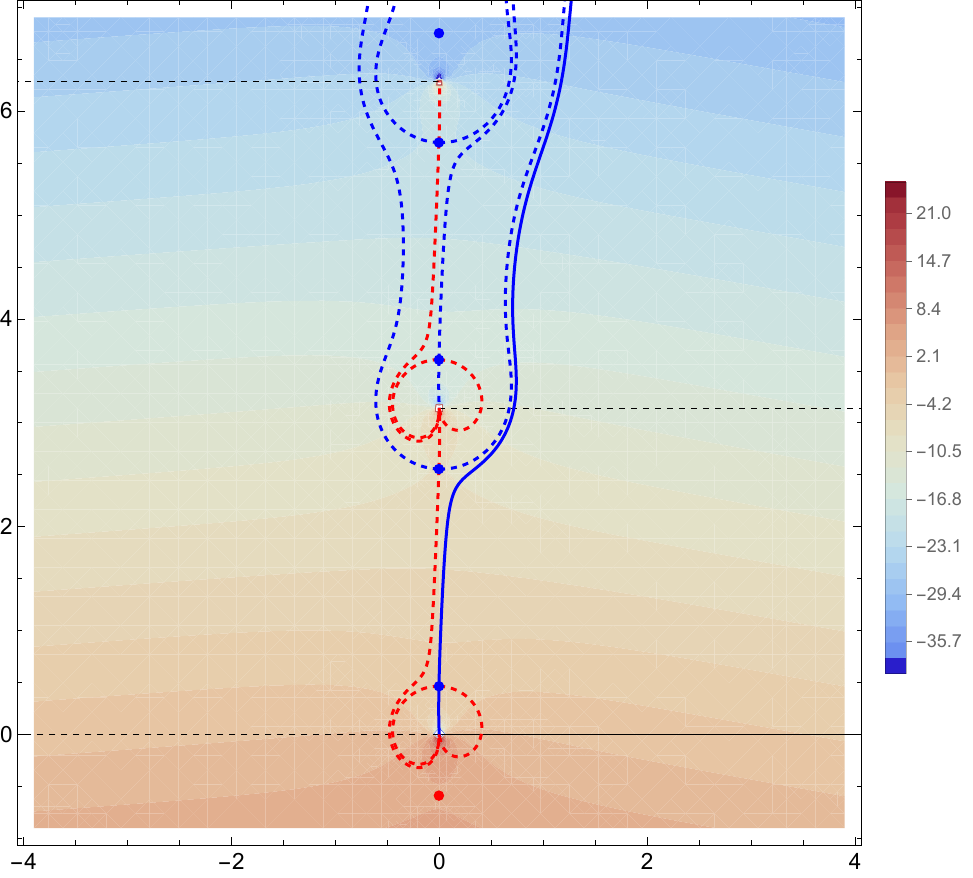}%
    \subcaption{$ \hbar \rightarrow \hbar \e^{-i\pi/60}$}%
    \label{fig:3dq_minrot}%
  \end{minipage}%
  \begin{minipage}[t]{0.5\linewidth}%
    \centering%
    \includegraphics[keepaspectratio, width=\linewidth]{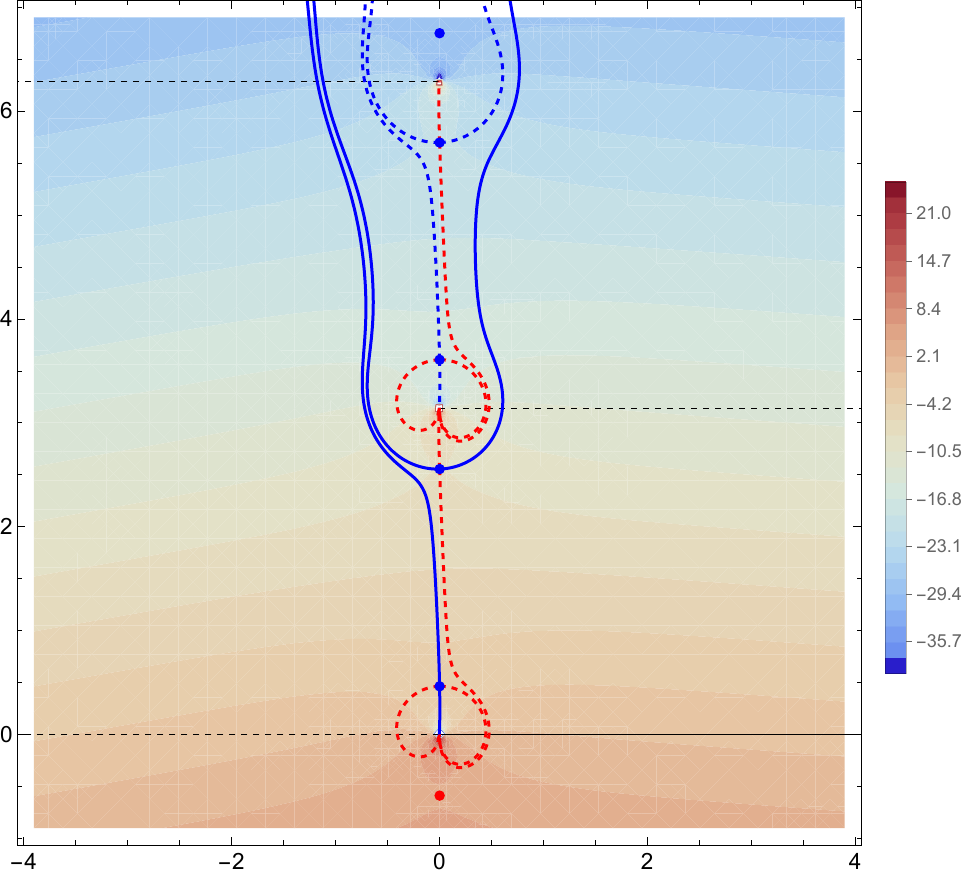}%
    \subcaption{$\hbar \rightarrow \hbar \e^{+i\pi/60}$}%
    \label{fig:3dq_plusrot}%
  \end{minipage}%
  \caption{$\re{F_3(N)}$ with saddles and thimbles in the three-dimensional Euclidean-to-Euclidean case. Thimble degeneracy and the Stokes phenomenon appear as in the four-dimensional case.
  }
  \label{fig:3dq}
\end{figure}

We adopt the parametrization of Option~I introduced in Sec.~\ref{sec:setup}, 
and the action and the field equation are explicitly written as
\begin{align}
    S\qty[q, N]
    &=\frac{V_{2}}{8 \pi G_3}
    \int \dd{t}
        \left[k N
        - \Lambda q^{2} N
        - \frac{\dot{q}^2}{N} 
    \right]
    + \qty({\rm bdy})\,,
    \label{eq:3Daction}
\end{align}
and
\begin{align}
    \ddot{q} = N^2 \Lambda q\,,
    \label{eq:3DEOM1}
\end{align}
where we again omit the bar on the lapse for simplicity. 
The classical solution of the field equation \eqref{eq:3DEOM1} under the Dirichlet condition~\eqref{eq:Dirichlet-bc} is obtained as
\begin{align}
    q_{\rm cl}(t)
    &= \csch\qty(\sqrt{\Lambda}NT)
    \qty[
        q_0\sinh\qty(
        \sqrt{\Lambda}N(T-t))
        +q_1\sinh\qty( \sqrt{\Lambda}Nt)].
\end{align}
The action~\eqref{eq:3Daction} has the same functional form as that of a one-dimensional harmonic oscillator with a negative kinetic term, whose Hamiltonian is
\begin{align}
    &H_3 =    N \qty(-\frac{P^2}{2\mathcal{M}_3} + U_3(q)), 
    \quad
    U_3(q) = \frac{\mathcal{M}_3}{2} 
    \qty(\Lambda q^2-k)
    \nn
    &\mathcal{M}_3 \coloneqq \frac{V_2}{4\pi G_3} ,
    \quad P\coloneqq -\frac{\mathcal{M}_3}{N} \dot{q}.
    \label{eq:Ham3d}
\end{align}
Thus, there are possible classical configurations for $q\geq q_{\rm crit} = \sqrt{k/\Lambda}$.
Indeed, with this choice of gauge and variables, we can compute the transition amplitude in close analogy with the familiar textbook treatment of the harmonic oscillator \cite{Feynman:100771, Mukhanov:2007zz}, yielding
\begin{align}
    &\mathcal{A}[q_1; q_0]
    =
    \qty[\frac{i\Lambda^{\frac{1}{2}} V_2}{8\pi^2 \hbar G_3}]^\frac{1}{2}
    \int^\infty_0 \dd{N} \exp \qty[F_3 (N)],
    \label{eq:3Damp}
\end{align}
with
\begin{align}
    &F_3(N)
    = - \frac{1}{2} \log \qty[\sinh (\sqrt{\Lambda} N T)]
    \nn
    & \hspace{50pt}
    + \frac{iV_{2}}{8 \pi \hbar G_3}
    \qty(k N T 
        - \sqrt{\Lambda}
        \csch\qty[\sqrt{\Lambda}NT]
        \qty((q_0^2+q_1^2)\cosh\qty[\sqrt{\Lambda}NT]
        -2q_0q_1)
    ).
    \label{eq:3dphase}
\end{align}
The first term in Eq.~\eqref{eq:3dphase} arises from the prefactor, whereas the remaining terms arise from the semiclassical action evaluated at $q_{\rm cl}$. This prefactor term introduces branch points at $N=i\pi n/\sqrt{\Lambda}T$ ($n \in \mathbb{Z}$)
together with associated branch cuts in the complex $N$-plane, as  illustrated in FIGs.~\ref{fig:3dCandINQ} and~\ref{fig:3dq}.%%%%%%%%%%%%%%%%%%%%%%%%%%%%%%
\footnote{The Lorentzian path integral of three-dimensional spacetime has already been discussed in Refs.~\cite{Chen:2024vpa,Chen:2024qmn}, where the integration contour for $N$ is fixed by the holographic correspondence. In our setup, however, the lapse integration is originally defined along the real axis, so the relevant contour differs from the one used in those references.}
%%%%%%%%%%%%%%%%%%%%%%%%%%%%%%

The locations of the saddle points of $F_3(N)$ in the semiclassical limit $\hbar \to 0$ can be obtained analytically as
\begin{align}
    &N_{1, n} = \frac{1}{T\sqrt{\Lambda}} 
    \Bigl[
        2i\pi n  
        + \log \Bigl[ \mathcal{N}_{3A} - \mathcal{N}_{3B} - \sqrt{(\mathcal{N}_{3A} - \mathcal{N}_{3B})^2 - 1}
        \Bigr]
    \Bigr],
    \nn
    &N_{2, n} = 
    \frac{1}{T\sqrt{\Lambda}}\Bigl[
        2i\pi n 
        + \log \Bigl[ \mathcal{N}_{3A} - \mathcal{N}_{3B} + \sqrt{(\mathcal{N}_{3A} - \mathcal{N}_{3B})^2 -1}
        \Bigr]
    \Bigr],
    \nn
    &N_{3, n} =
    \frac{1}{T\sqrt{\Lambda}}\Bigl[
        2i\pi n  
        + \log \Bigl[ \mathcal{N}_{3A} + \mathcal{N}_{3B} - \sqrt{(\mathcal{N}_{3A} + \mathcal{N}_{3B})^2 - 1}
        \Bigr]
    \Bigr],
    \nn
    &N_{4, n} = 
    \frac{1}{T\sqrt{\Lambda}}\Bigl[
        2i\pi n 
        + \log \Bigl[ \mathcal{N}_{3A} + \mathcal{N}_{3B} + \sqrt{(\mathcal{N}_{3A} + \mathcal{N}_{3B})^2 - 1}
        \Bigr]
    \Bigr],
\end{align}
where $n \in \mathbb{Z}$ and
\begin{align}
    &\mathcal{N}_{3A} = \frac{q_0 q_1 \Lambda}{k},
    \quad
    \mathcal{N}_{3B} = \frac{\Lambda}{k} \sqrt{(q_0^2 - q_{\rm crit}^2 )(q_1^2 - q_{\rm crit}^2)},
    \quad
    q_{\rm crit}^2 = \frac{k}{\Lambda}.
\end{align}
Here, we focus on \ac{ds} spacetime with a positive curvature: $k, \, \Lambda >0$ and $0 \leq q_0 \leq q_1$. In particular, if $q_0, q_1 > q_{\rm crit}$, it follows that $\mathcal{N}_{3A} > \mathcal{N}_{3B}$, and that all $N_{i,0}$ become real. These physically correspond to the classical saddle points of the lapse integral. The saddle points in the nonzero $\hbar$ case can be found numerically. 

As in the 4D case, we distinguish three cases depending on the relative values of $q_0$ and $q_1$ with respect to the critical scale $q_{\rm crit}$:
\begin{enumerate}[label=(\roman*)]
    \item Lorentzian-to-Lorentzian:
    $q_1 \geq q_0 \geq q_{\rm crit} = \sqrt{k/\Lambda}$ \\
    or equivalently $(k-\Lambda q_0^2 )(k-\Lambda q_1^2 ) \geq 0$ and $2k \leq \Lambda (q_0^2 +q_1^2)$,
    \item Euclidean-to-Lorentzian: 
    $q_1 > q_{\rm crit} > q_0 \geq 0$ \\
    or equivalently $(k-\Lambda q_0^2 )(k-\Lambda q_1^2 ) < 0$,
    \item Euclidean-to-Euclidean:
    $q_{\rm crit} \geq q_1 \geq q_0 \geq 0$ \\
    or equivalently $(k-\Lambda q_0^2 )(k-\Lambda q_1^2 ) \geq 0$ and $2k \geq \Lambda (q_0^2 +q_1^2)$.
\end{enumerate}
Case (i) describes the transition between classically allowed solutions, as expected from the Hamiltonian and the potential \eqref{eq:Ham3d}. The cases (ii) and (iii) give quantum transitions from classically forbidden solutions to classically allowed and forbidden solutions, respectively.

\subsubsection*{(i). Lorentzian-to-Lorentzian 
($q_1 \geq q_0 \geq q_{\rm crit}$)
}

Contour plots of $\re{F_3(N)}$ for the Lorentzian-to-Lorentzian case are shown in FIGs.~\ref{fig:3dcl_thimbles} and \ref{fig:3dcl_center}, which display the saddle points and associated thimbles for $q_0 = 4 \sqrt{k/\Lambda}$
and $q_1 = 5 \sqrt{k/\Lambda}$.
\figref{fig:3dcl_center} is a magnified view of the region near the origin. All saddle points and thimbles are obtained numerically by solving the stationary condition for $\re{F_3(N)}$ and the corresponding flow equations. 
As in the four-dimensional case, there are two classical saddles for which the phase $F_3$ of the exponent is almost purely imaginary, $\re{F_3} \simeq 0$. These saddles contribute to the path integral, implying the superposition of the classical transitions. The original integration contour can therefore be deformed into the union of the contributing thimbles indicated by the blue solid lines in FIGs.~\ref{fig:3dcl_thimbles} and~\ref{fig:3dcl_center}.

\subsubsection*{(ii). Euclidean-to-Lorentzian 
($q_1 > q_{\rm crit} > q_0 \geq 0$)
}

\figref{fig:3dinq_thimbles} shows the Euclidean-to-Lorentzian case with
$q_0 = 0$ and $q_1 = 5 \sqrt{k/\Lambda}$, which describes the transition from nothing to a classically allowed Lorentzian configuration. In this case, the relevant contributing saddle is characterized by $\re{F_3} < 0$. Consequently, the resulting transition amplitude follows Vilenkin’s tunneling proposal in three dimensions, in complete analogy with the four-dimensional analysis~\cite{Feldbrugge:2017kzv,Honda:2024aro}.\footnote{
This thimble structure has also been derived in a different context in the very recent paper \cite{Honda:2026joe}, which studies the correspondence between three-dimensional Lorentzian quantum gravity in dS spacetime and Euclidean quantum gravity in AdS space from both bulk and boundary perspectives.
}

\begin{figure}[tp]%
  \begin{minipage}[t]{0.5\linewidth}%
    \centering%
    \includegraphics[keepaspectratio, width=0.96\linewidth]{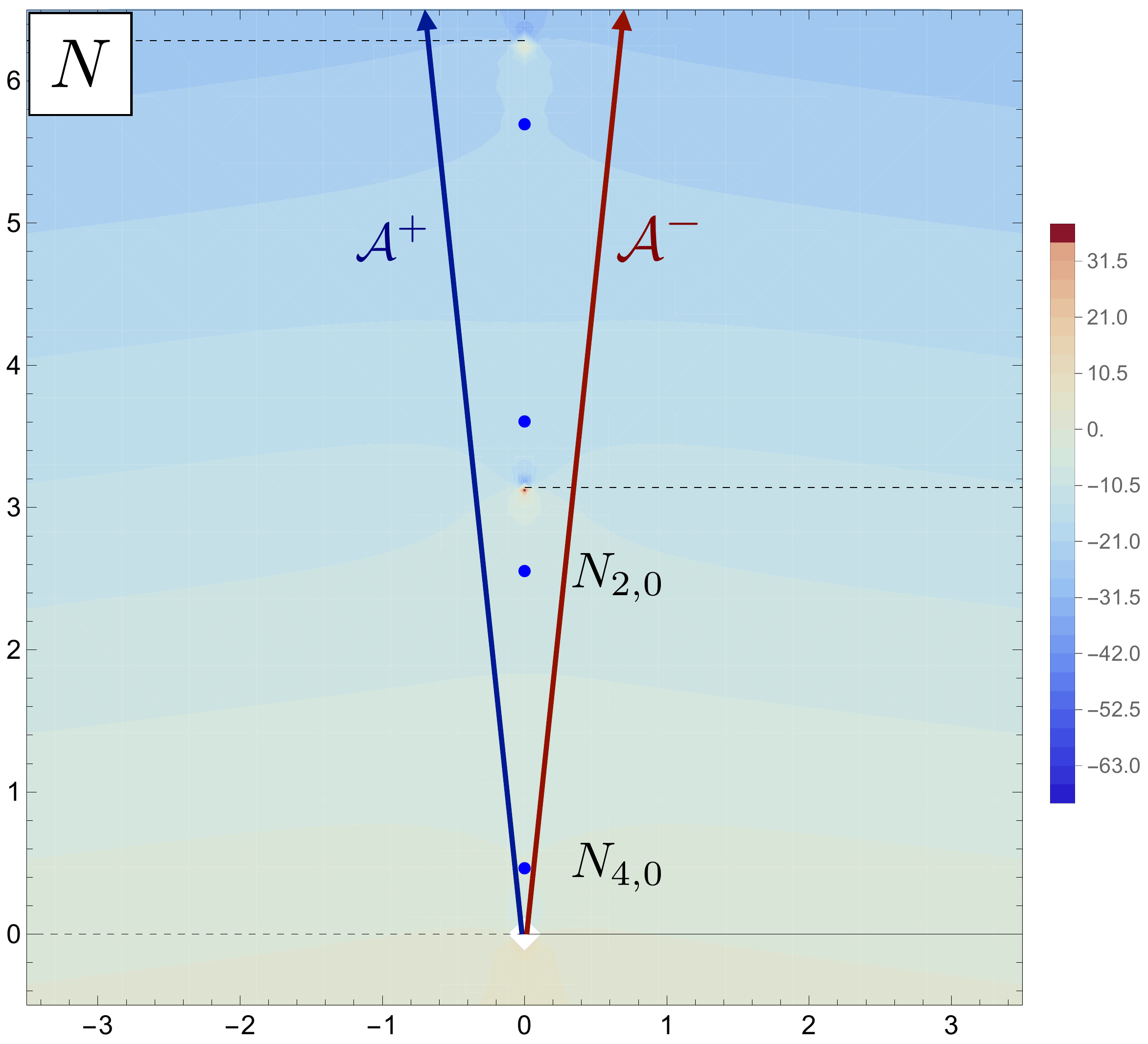}%
    \subcaption{$\re{F_3(N)}$ and integration paths}%
    \label{fig:3dq_path}%
  \end{minipage}%
  \begin{minipage}[t]{0.5\linewidth}%
    \centering%
    \includegraphics[keepaspectratio, width=\linewidth]{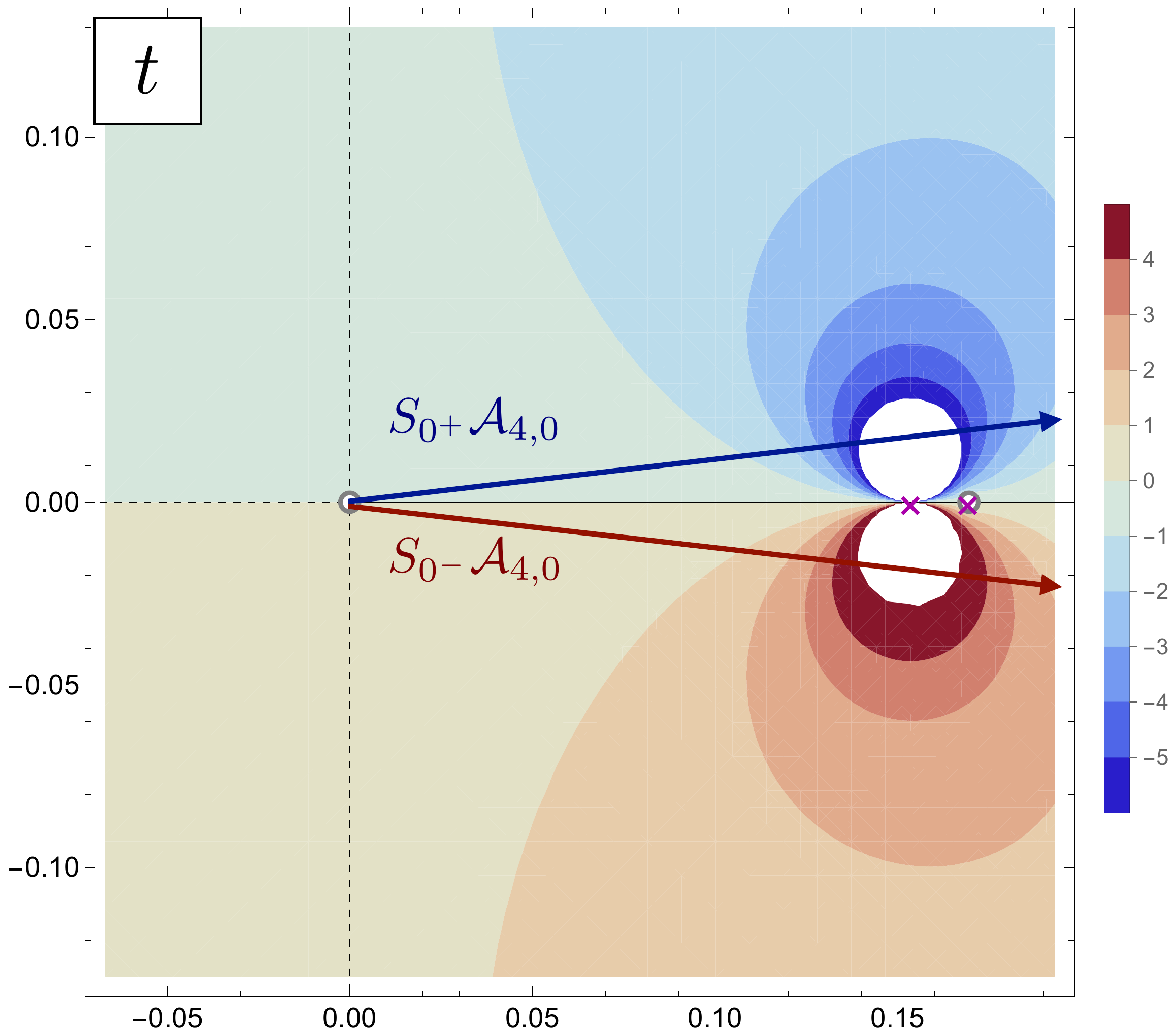}%
    \subcaption{$\re{\mathcal{BP}\bar{\mathcal{A}}(t)}$ and resummation contours}%
    \label{fig:3dq_borel}%
  \end{minipage}%
  \caption{
    (a) Contour plot of $\re{F_3(N)}$ with the contours for numerical evaluation. (b) Contour plot of $\re{\mathcal{BP}\bar{\mathcal{A}}(t)}$ in the Borel plane and paths for resummation.
  }
  \label{fig:3dq_resurgence}
\end{figure}%

\subsubsection*{(iii). Euclidean-to-Euclidean 
($q_{\rm crit} \geq q_1 \geq q_0 \geq 0$)
}

The Euclidean-to-Euclidean case with $q_0 = 0$ and $q_1 = \sqrt{k/4\Lambda}$ is illustrated in \figref{fig:3dq}. As shown in \figref{fig:3dq_thimbles}, the thimbles become degenerate, and several steepest-descent and -ascent paths intersect the original contour. Although we can lift these degeneracies by rotating $\hbar$ toward a complex value $\hbar \rightarrow \hbar \e^{i\Delta \theta}$ as in FIGs.~\ref{fig:3dq_minrot} and \ref{fig:3dq_plusrot}, they show the Stokes phenomenon: for a small negative phase rotation,
$\hbar \to \hbar e^{-i|\Delta\theta|}$, only a single tunneling saddle contributes (\figref{fig:3dq_minrot}), whereas for a small positive phase rotation,
$\hbar \to \hbar e^{+i|\Delta\theta|}$, two tunneling saddles contribute (\figref{fig:3dq_plusrot}). 

As in the four-dimensional case, the resurgence program can solve this problem: this ambiguity of contributing saddles can be compensated for by another ambiguity from the resummation of the asymptotic series expansion of the amplitude \eqref{eq:3Damp}. 
Here we take $q_0 =0$ and $q_1 <q_{\rm crit}$.
\figref{fig:3dq} illustrates that one tunneling saddle contributes to the path integral regardless of $\argu{\hbar}$. Its location in the semiclassical limit is
\begin{align}
    N_{4,0} = \frac{1}{T\sqrt{\Lambda}} \log \qty[\frac{i q_1 + \sqrt{q_{\rm crit}^2-q_1^2}}{q_{\rm crit}}].
\end{align}
On the other hand, there is one ambiguous saddle that contributes in the case of $\argu{\hbar} > 0$
\begin{align}
    N_{2,0} = \frac{1}{T\sqrt{\Lambda}} \log \qty[\frac{i q_1 - \sqrt{q_{\rm crit}^2-q_1^2}}{q_{\rm crit}}].
\end{align}
Then, the contribution from $N_{4,0}$ in the amplitude can be evaluated as 
\begin{align}
    &\mathcal{A}_{4,0}(\hbar)
    =
    \qty[\frac{\Lambda^{\frac{1}{2}} V_2}{8\pi^2 G_3}]^\frac{1}{2}\e^{i\frac{3}{4}\pi}
    \int^{\infty}_{-\infty} \dd{x}\exp \qty[F_3 (N_{4,0}+i\sqrt{\hbar}x)],
    \label{eq:Amp_40}
\end{align}
and expressed as an asymptotic series in $\hbar$ 
\begin{align}
    \mathcal{A}_{4,0}(\hbar)= \exp \qty[\frac{\bar{F}_3 (N_{4,0})}{\hbar}]
    \sum_{n=0}^\infty c_n \hbar^{n}
    =: \exp \qty[\frac{\bar{F}_3 (N_{4,0})}{\hbar}]
    \qty(c_0 + \bar{\mathcal{A}}_{4,0}(\hbar)),
    \label{eq:asymser_40}
\end{align}
where the semiclassical action $\bar{F}_3$ is given by
\begin{align*}
    \bar{F}_3 (N_{4,0}) =
    \frac{iV_{2}\Lambda^{\frac12}}{8 \pi G_3}
    \frac{q_1\left(q_{\rm crit}^2-q_1^2+i q_1 \sqrt{q_{\rm crit}^2-q_1^2}\right)
    +q_{\rm crit}^2\left(q_1-i \sqrt{q_{\rm crit}^2-q_1^2}\right) \log \left[\frac{i q_1+\sqrt{q_{\rm crit}^2-q_1^2}}{q_{\rm crit}}\right]}{q_1 -i \sqrt{q_{\rm crit}^2-q_1^2}}.
\end{align*}
The coefficients $c_i$ in Eq.~\eqref{eq:asymser_40} are evaluated by expanding Eq.~\eqref{eq:Amp_40} with respect to $\hbar$ and performing the Gaussian integral over $x$.
As in the four-dimensional case, a closed-form expression valid for all $c_i$ is not available, so we have to truncate 
the series at $\order{\hbar^{M}}$ practically. Moreover, the coefficients $c_i$ exhibit factorial growth, which is to be handled using Borel resummation.

\begin{figure}[tp]%
  \begin{minipage}[t]{\linewidth}%
    \centering%
    \includegraphics[keepaspectratio, width=\linewidth]{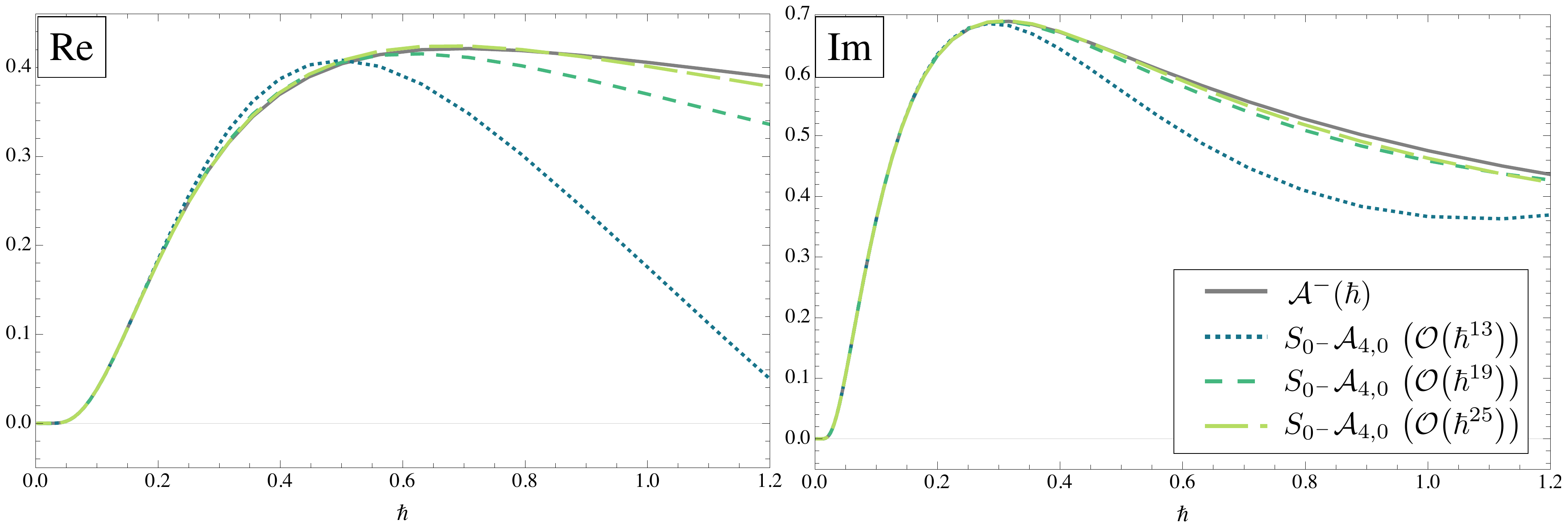}%
    \subcaption{Comparison of the direct numerical result of the amplitude value with the resummed ones.}%
    \label{fig:3dq_amprel}%
  \end{minipage}%
  \\[10pt]
  \begin{minipage}[t]{\linewidth}%
    \centering%
    \includegraphics[keepaspectratio, width=0.8\linewidth]{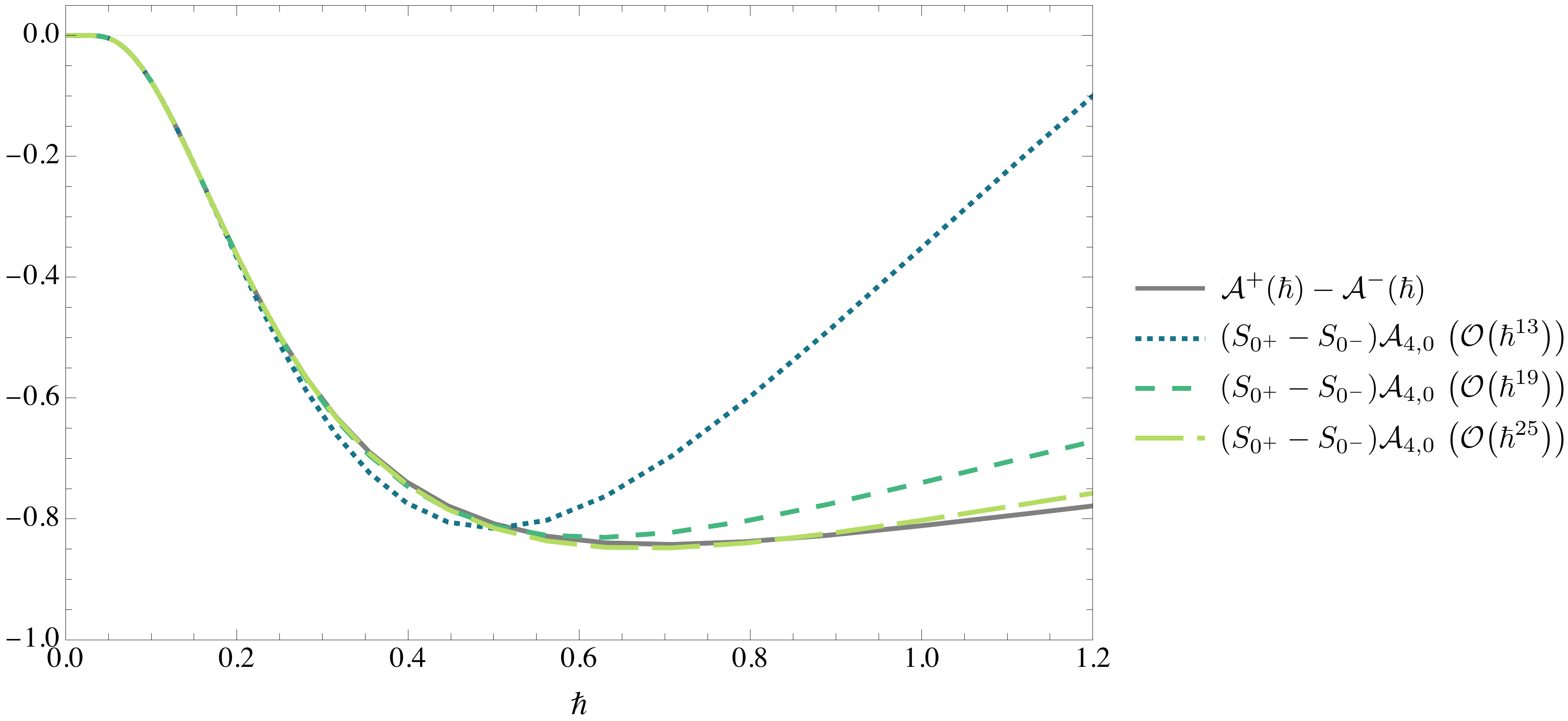}%
    \subcaption{Comparison of the path dependence with the Borel ambiguity.}%
    \label{fig:3dq_resum}%
  \end{minipage}%
  \caption{
    (a) Numerical results of the total amplitude. 
    (b) Comparison of the dependence on the integration contour with the Borel–Pad\'{e} resummation ambiguity. 
  }
  \label{fig:3dq_resurgence_resum}
\end{figure}%

Then we perform the resummation. Since the series expansion \eqref{eq:asymser_40} is truncated, we use the Borel-Pad\'{e} transformation, where the Borel transformation is approximated by the Pad\'{e} approximation. 
We denote the Borel-Pad\'{e} transformed function as $\mathcal{BP} \bar{\mathcal{A}}_{4,0}$. 
\figref{fig:3dq_borel} is the contour plot of $\re{\mathcal{BP} \bar{\mathcal{A}}_{4,0}}$ in the Borel-plane with $q_1 =\sqrt{k/4\Lambda}$ and $M=19$ (or $\order{\hbar^{19}}$ equivalently). One can see that several poles (magenta crosses) lie on the positive $\re{t}$ axis.
In particular, the position of the pole closest to the origin is $t\sim \bar{F}_3 (N_{4,0})-\bar{F}_3 (N_{2,0})$. This means that the Borel ambiguity exists, and that a difference in resummed amplitude values between paths passing above and below the poles is $\sim \exp\qty[\bar{F}_3 (N_{2,0})/\hbar]$, matching the ambiguous saddle contribution. 

We numerically confirm this expectation. In \figref{fig:3dq_amprel}, we compare the amplitude values obtained by direct numerical evaluation of $\mathcal{A}^{-}(\hbar)$ (gray line, see \figref{fig:3dq_path} for the path we take) and by the Borel-Pad\'{e} resummation
\begin{align}
    \mathcal{S}_{\theta} \mathcal{A}_{4,0} (\hbar)
    = \exp \qty[\frac{\bar{F}_3 (N_{4,0})}{\hbar}]
    \qty[
        c_0 + \int^{\infty \e^{i\theta}}_0 \dd{t}
        \e^{-\frac{t}{\hbar}} \,
        \mathcal{BP} \bar{\mathcal{A}}_{4,0} (t)
    ],
\end{align}
with $\theta \rightarrow 0-$ (colored dashed lines). The resummed result clearly converges toward the direct numerical evaluation of Eq.~\eqref{eq:3Damp} as the truncation order $M$ increases.
\figref{fig:3dq_resum} compares the integration-path dependence, $\mathcal{A}^{+}(\hbar)-\mathcal{A}^{-}(\hbar)$,
with the Borel ambiguity, $(\mathcal{S}_{0+}-\mathcal{S}_{0-})\mathcal{A}_{4,0}(\hbar)$.
The figure shows that the latter approaches the numerical value of the former, demonstrating that the resurgence analysis resolves the Stokes ambiguity in the three-dimensional case as well.\footnote{As we show in \appref{sec:5D_fin}, the five-dimensional case with $0<q_0<q_1<q_{\rm crit}$ resembles this three-dimensional setting, and thus this resurgence technique can be utilized to resolve the Stokes phenomenon in almost the same manner.}

% \clearpage
%%%%%%%%%%%%%%%%%%%%%%%%%%%%%%%%%%%%%%%
%%%%%%%%%%%%%%%%%%%%%%%%%%%%%%%%%%%%%%%
\section{Five-dimensional Spacetime}
\label{sec:five-dimension}
%%%%%%%%%%%%%%%%%%%%%%%%%%%%%%%%%%%%%%%
%%%%%%%%%%%%%%%%%%%%%%%%%%%%%%%%%%%%%%%

\begin{figure}[tp]%
  \begin{minipage}[t]{0.5\linewidth}%
    \centering%
    \includegraphics[keepaspectratio, width=\linewidth]{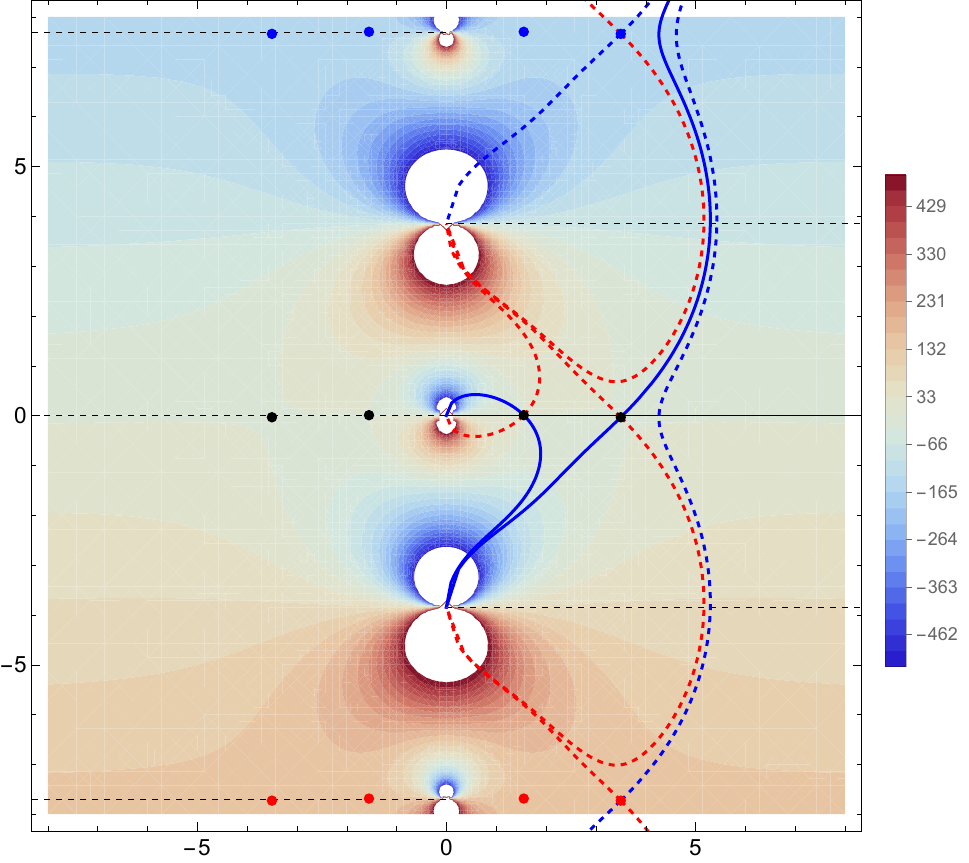}%
    \subcaption{Lorentzian-to-Lorentzian}%
    \label{fig:5dcl_thimbles}%
  \end{minipage}%
  \begin{minipage}[t]{0.5\linewidth}%
    \centering%
    \includegraphics[keepaspectratio, width=\linewidth]{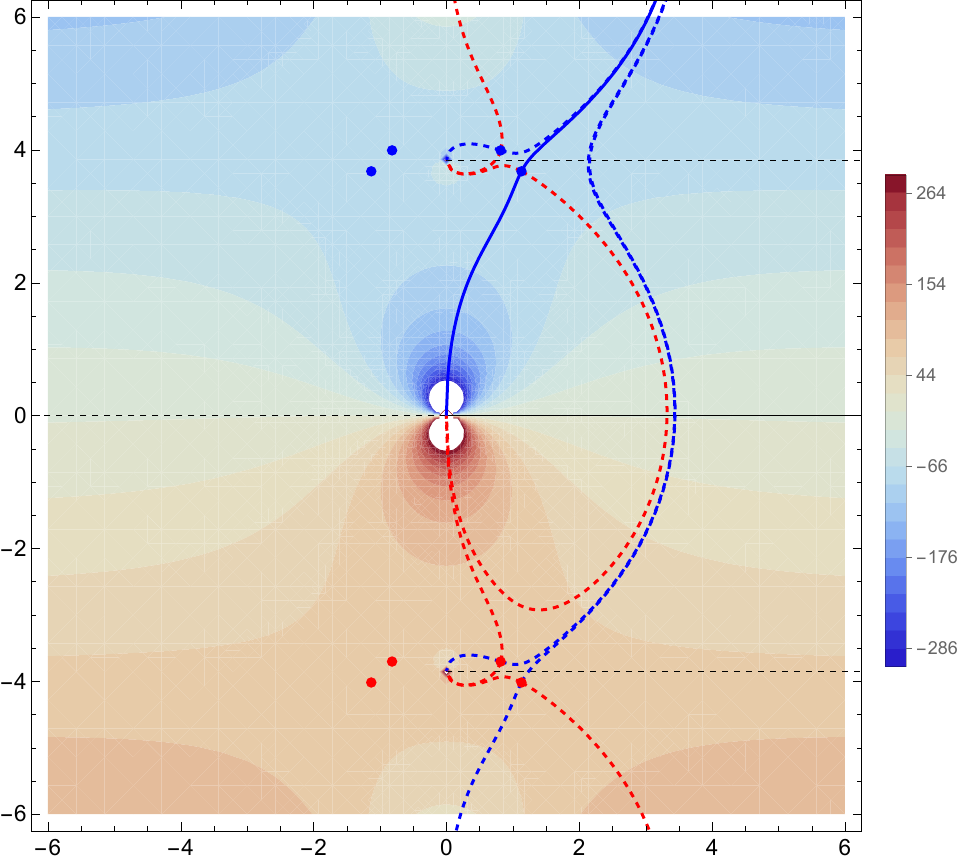}%
    \subcaption{Euclidean-to-Lorentzian ($q_0=0$)
    }%
    \label{fig:5dINQ_thimbles}%
  \end{minipage}%
  \caption{Thimbles in the 5D case with the Lorentzian final spacetime. }
  \label{fig:5dclINQ}
\end{figure}

For the five-dimensional case, it is more convenient to work with Option~II rather than Option~I in Sec.~\ref{sec:setup}.
Setting $D=5$ on the action~\eqref{eq:d-dim_action2}, the gravitational action takes the form
\begin{align}
    &S\qty[q, N]
    =\frac{V_{4}}{16 \pi G_5}
    \int \dd{t}
    \qty[12 k N q
        -2 \Lambda N q^2 
        - \frac{3\dot{q}^2}{N}
    ]
    + \qty({\rm bdy})\,.
    \label{eq:5Daction}
\end{align}
The classical solution with the Dirichlet condition~\eqref{eq:Dirichlet-bc} is obtained as
\begin{align}
    q_{\rm cl}(t) 
    =& 
    \frac{3 k}{\Lambda}+\left(q_0-\frac{3 k}{\Lambda}\right) \frac{\sinh [A_N(T-t)]}{\sinh (A_N T)}+\left(q_1-\frac{3 k}{\Lambda}\right) \frac{\sinh (A_N t)}{\sinh (A_N T)}
    \quad \qty(A_N = \sqrt{\frac{2\Lambda}{3}}N).
\end{align}
As in the three-dimensional case, the action \eqref{eq:5Daction} is of the harmonic oscillator type with the Hamiltonian, 
\begin{align}
    &H_5 = N \qty(-\frac{P^2}{2\mathcal{M}_5} + U_5(q)),
    \quad
    U_5(q) = 
    \frac{\mathcal{M}_5}{3} q\qty(\Lambda q -6k),
    \nn
    &\mathcal{M}_5 \coloneqq \frac{3V_{4}}{8\pi G_5} ,
    \quad P\coloneqq -\frac{\mathcal{M}_5}{N} \dot{q},
    \label{eq:Ham5d}
\end{align}
so the transition amplitude, including the prefactor from fluctuations around $q_{\rm cl}$, can be evaluated exactly by Gaussian path integration. We find
\begin{align}
    &\mathcal{A}[q_1; q_0]
    = \qty[\frac{i\sqrt{6\Lambda} V_4}{16\pi^2 \hbar G_5}]^\frac{1}{2}
    \int^\infty_0 \dd{N}
    \exp \qty[F_5(N)],
    \nonumber
\end{align}
\begin{align}
    &F_5(N) = 
    -\frac{1}{2}\log\qty[\sinh (\sqrt{\frac{2\Lambda}{3}} N T)]
    \nn
    & \hspace{60pt} +
    \frac{iV_{4}}{16 \pi\hbar  G_5}
    \left[
        \frac{18 k^2}{\Lambda} NT
        -\frac{\sqrt{6} \csch\qty(\sqrt{\frac{2\Lambda}{3}} N T)}{\Lambda^{3 / 2}}
        \biggl[
            -2(3 k-q_0 \Lambda)(3 k-q_1 \Lambda)
    \right.
    \nn 
    & \hspace{120pt}
    \left.\left.
            +\qty(18 k^2-6 k(q_0+q_1) \Lambda
            +\left(q_0^2+q_1^2\right) \Lambda^2) 
            \cosh \left(\sqrt{\frac{2\Lambda}{3}} NT\right)
        \right] 
    \right].
    \label{eq:5Dphase}
\end{align}
The first term in $F_5(N)$ arises from the functional determinant associated with fluctuations around the classical solution $q_{\rm cl}(t)$, while the remaining terms come from the semiclassical action evaluated on $q_{\rm cl}(t)$. As in the three-dimensional case, the prefactor term induces nontrivial analytic structure in the complex $N$-plane: in particular, it introduces branch points at $N=i\pi n \cdot \sqrt{3}/\sqrt{2\Lambda}T$ together with associated branch cuts, which are shown as black dotted lines in FIGs.~\ref{fig:5dclINQ} and~\ref{fig:5dq}.

The saddle points in the semiclassical limit $\hbar \to 0$
can be found analytically as
\begin{align}
    &N_{1, n} = \sqrt{\frac{3}{2\Lambda T^2}} \Bigl[
        2i\pi n  
        + \log \Bigl[ \mathcal{N}_{5A} - \mathcal{N}_{5B} - \sqrt{(\mathcal{N}_{5A} - \mathcal{N}_{5B})^2 - 1}
        \Bigr]
    \Bigr],
    \nn
    &N_{2, n} = \sqrt{\frac{3}{2\Lambda T^2}} \Bigl[
        2i\pi n 
        + \log \Bigl[ \mathcal{N}_{5A} - \mathcal{N}_{5B} + \sqrt{(\mathcal{N}_{5A} - \mathcal{N}_{5B})^2 -1}
        \Bigr]
    \Bigr],
    \nn
    &N_{3, n} = \sqrt{\frac{3}{2\Lambda T^2}} \Bigl[
        2i\pi n  
        + \log \Bigl[ \mathcal{N}_{5A} + \mathcal{N}_{5B} - \sqrt{(\mathcal{N}_{5A} + \mathcal{N}_{5B})^2 - 1}
        \Bigr]
    \Bigr],
    \nn
    &N_{4, n} = \sqrt{\frac{3}{2\Lambda T^2}} \Bigl[
        2i\pi n 
        + \log \Bigl[ \mathcal{N}_{5A} + \mathcal{N}_{5B} + \sqrt{(\mathcal{N}_{5A} + \mathcal{N}_{5B})^2 - 1}
        \Bigr]
    \Bigr],
    \label{eq:sad5D}
\end{align}
with $n \in \mathbb{Z}$ and 
\begin{align}
    &\mathcal{N}_{5A} = 1- \frac{\Lambda}{3k}\qty(q_0+ q_1 )
    +\frac{\Lambda^2}{9k^2}q_0 q_1,
    \quad
    \mathcal{N}_{5B} = \frac{\Lambda^2}{9k^2} \sqrt{q_0 q_1 (q_0 - q_{\rm crit} )(q_1 - q_{\rm crit})},
    \quad
    q_{\rm crit} = \frac{6k}{\Lambda}. \label{eq:saddle_constants_5D}
\end{align}
For \ac{ds} spacetime with $k, \, \Lambda >0$ and $0 \leq q_0 \leq q_1$, the boundary conditions $q_0 , q_1>q_{\rm crit}$ lead to $\mathcal{N}_{5A} > \mathcal{N}_{5B}$, and $N_{i,0} \in \mathbb{R}$, implying the existence of the classical saddles. It should be noted that these four types of saddles show degeneracy as $N_{1, n}=N_{3, n},\, N_{2,n}=N_{4,n}$ when $q_0=0$. The saddle locations receive quantum corrections of order $\hbar$ and must be determined numerically. As we will show, this $\hbar$ correction lifts the degeneracy of the saddles in the case of $q_0=0$. 

As in other dimensions, it is useful to classify the physical conditions into three
regimes according to the values of $q_0$ and $q_1$:
\begin{enumerate}[label=(\roman*).]
    \item Lorentzian-to-Lorentzian:
    $q_1 \geq q_0 \geq q_{\rm crit} = 6k/\Lambda$ \\
    or equivalently $(6k-\Lambda q_0 )(6k-\Lambda q_1 ) \geq 0$ and $12k \leq \Lambda (q_0 +q_1)$,
    \item Euclidean-to-Lorentzian:
    $q_1 > q_{\rm crit} > q_0 \geq 0$ \\
    or equivalently $(6k-\Lambda q_0 )(6k-\Lambda q_1 ) < 0$,
    \item : Euclidean-to-Euclidean:
    $q_{\rm crit} \geq q_1 \geq q_0 \geq 0$ \\
    or equivalently $(6k-\Lambda q_0 )(6k-\Lambda q_1 ) \geq 0$ and $12k \geq \Lambda (q_0 +q_1)$.
\end{enumerate}
Most of the qualitative features of the five-dimensional analysis closely parallel those of the three-dimensional case, since the structure of the action \eqref{eq:5Daction} is essentially the same as that of Eq.~\eqref{eq:3Daction}. In the following, we focus on the quantum creation of the universe, the cases (ii) and (iii) with $q_0=0$. In \appref{sec:5D_fin}, we show the second and the third cases with $0<q_0<q_{\rm crit} $, where the semiclassical saddles have no degeneracy. In particular, case (iii) with $q_0\neq 0$ exhibits the Stokes phenomenon and more closely resembles the three-dimensional results. 

\subsubsection*{(i). Lorentzian-to-Lorentzian 
($q_1 \geq q_0 \geq q_{\rm crit}$)
}

\figref{fig:5dcl_thimbles} shows the contour plot %s 
of $\re{F_5(N)}$, displaying the saddle points and associated thimbles for $q_0 = 7k/\Lambda$
and $q_1 = 15k/\Lambda$,
which lie in the Lorentzian-to-Lorentzian regime. In this case, two real (classical) saddles contribute, characterized by an almost purely imaginary phase $\re{F_5(N)} \simeq 0$. As a result, the transition amplitude is dominated by the superposition of the corresponding classical Lorentzian histories, in close analogy with the three-dimensional analysis.

\subsubsection*{(ii). Euclidean-to-Lorentzian 
($q_1 > q_{\rm crit} > q_0 \geq 0$)
}

The Euclidean-to-Lorentzian case with $q_0 = 0$ and $q_1 = 30k/\Lambda$ is exhibited in \figref{fig:5dINQ_thimbles}. 
As in lower-dimensional examples, the relevant contributing saddle is the tunneling saddle with $\re{F_5(N)} < 0$, and the resulting wave function realizes Vilenkin’s tunneling proposal in five dimensions. 
As in the three-dimensional case, the inclusion of the fluctuation prefactor (or equivalently, performing the thimble analysis at finite $\hbar$) resolves the degeneracy among thimbles that would appear if we considered only the semiclassical action, thereby selecting a unique set of contributing Lefschetz thimbles.

\subsubsection*{(iii). Euclidean-to-Euclidean 
($q_{\rm crit} \geq q_1 \geq q_0 \geq 0$)
}

\begin{figure}[tp]%
  \begin{minipage}[t]{0.5\linewidth}%
    \centering%
    \includegraphics[keepaspectratio, width=\linewidth]{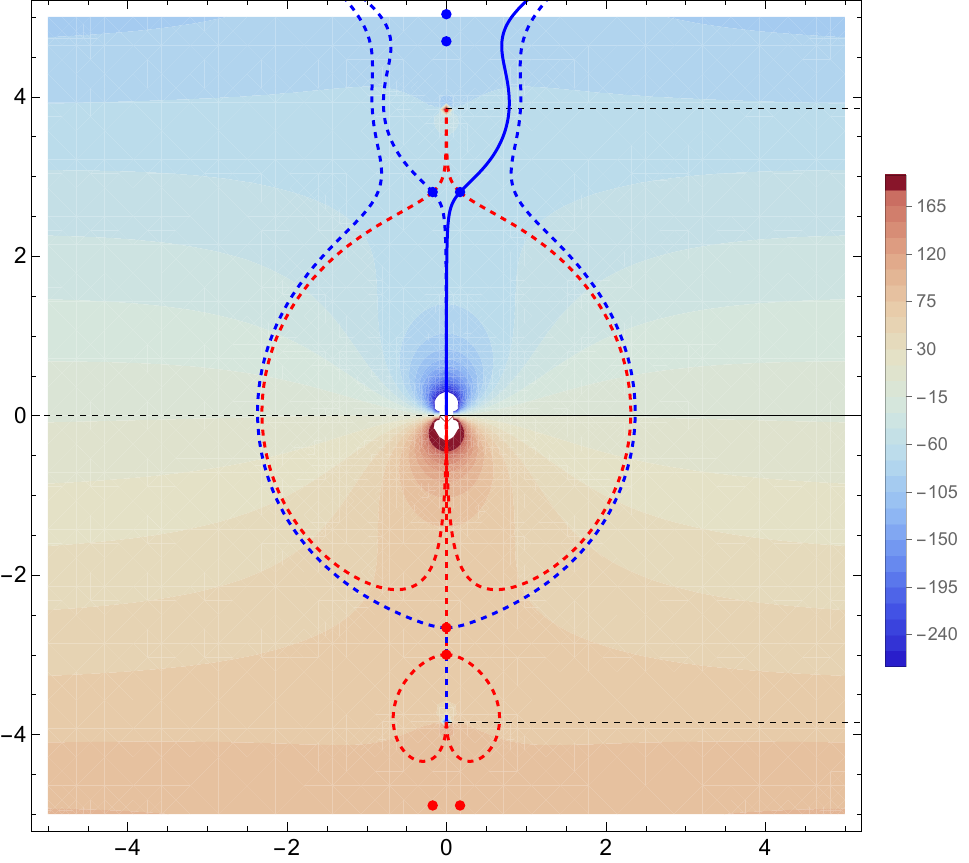}%
    \subcaption{Euclidean-to-Euclidean ($q_0=0, q_1=5k/\Lambda$)}%
    \label{fig:5dq_fromN_thimbles}%
  \end{minipage}%
  \begin{minipage}[t]{0.5\linewidth}%
    \centering%
    \includegraphics[keepaspectratio, width=\linewidth]{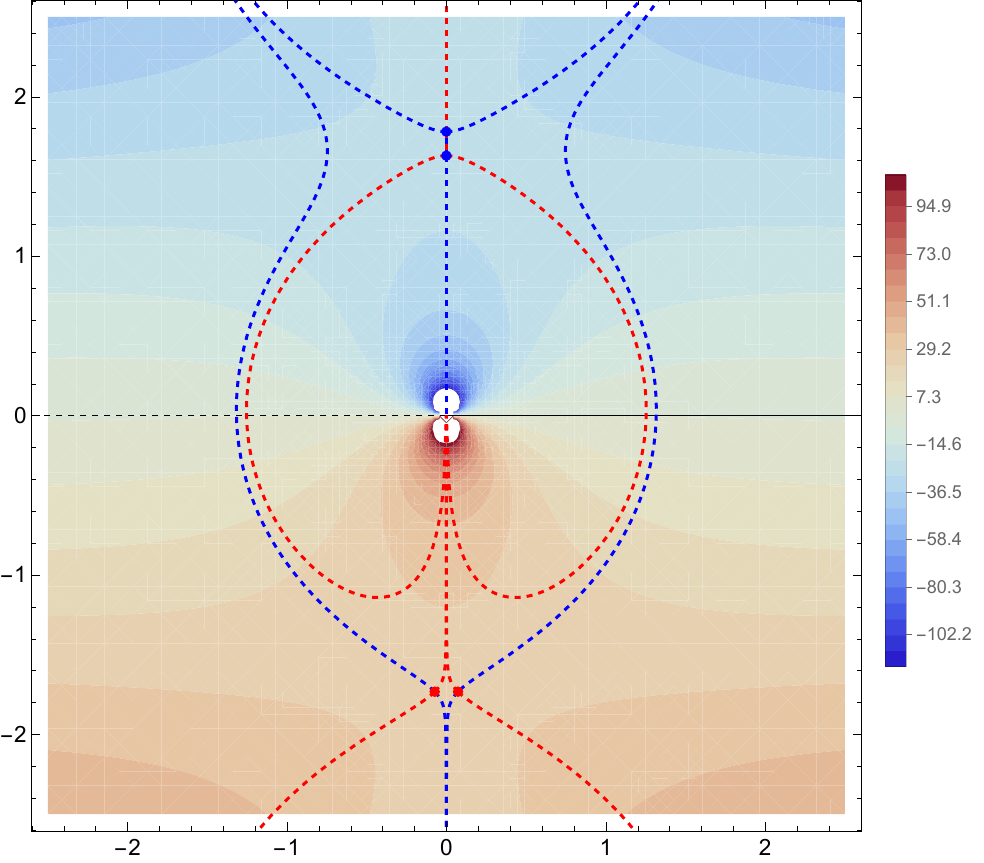}%
    \subcaption{Euclidean-to-Euclidean ($q_0=0, q_1=k/2\Lambda$)}%
    \label{fig:5dq_fromN_lq1_thimbles}%
  \end{minipage}%
  \\[5pt]
  \begin{minipage}[t]{0.5\linewidth}%
    \centering%
    \includegraphics[keepaspectratio, width=\linewidth]{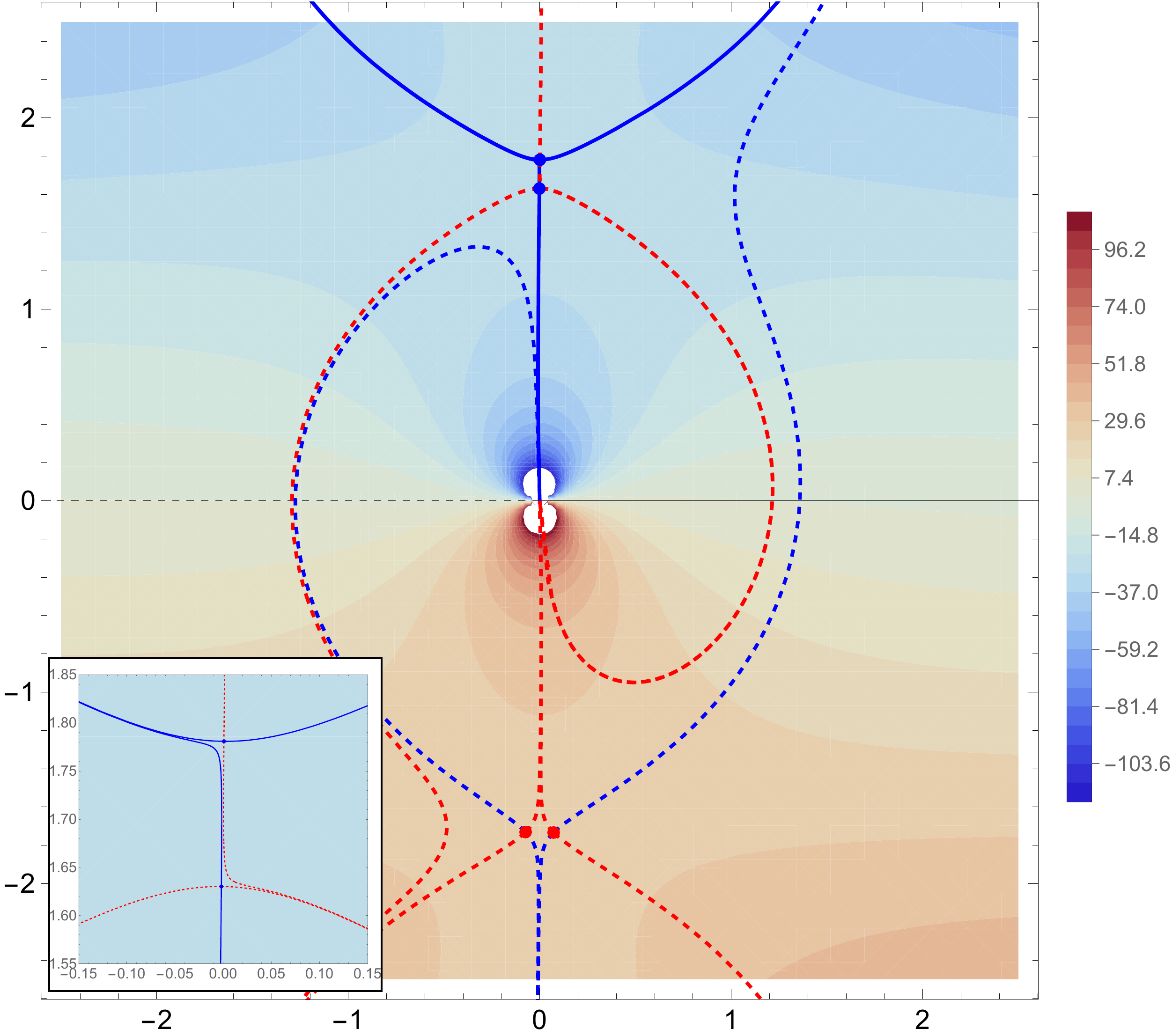}%
    \subcaption{$ \hbar \rightarrow \hbar \e^{-i\pi/60}$ in (b)}%
    \label{fig:5dq_fromN_lq1_thimbles_RM}%
  \end{minipage}%
  \begin{minipage}[t]{0.5\linewidth}%
    \centering%
    \includegraphics[keepaspectratio, width=\linewidth]{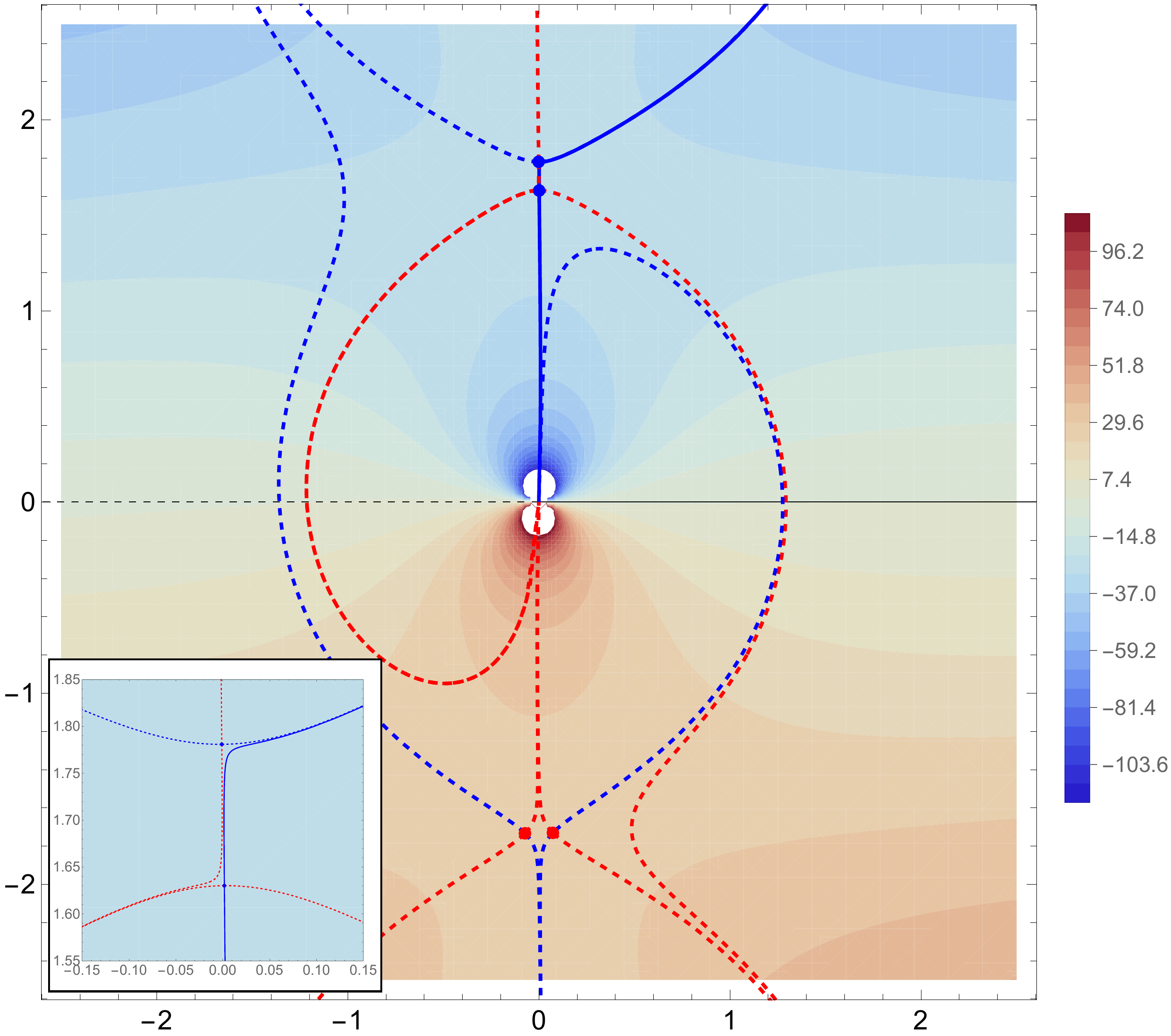}%
    \subcaption{$ \hbar \rightarrow \hbar \e^{+i\pi/60}$ in (b)}%
    \label{fig:5dq_fromN_lq1_thimbles_RP}%
  \end{minipage}%
  \caption{Thimbles in the 5D Euclidean-to-Euclidean case. 
  (a), (b): The saddles and thimbles for the cases with $q_0=0, q_1=5k/\Lambda$ and $q_1=k/2\Lambda$, respectively. (c),(d): Thimbles after $\hbar$ rotation in (b). Each mini panel is an enlarged view around the two tunneling saddles ($N\sim 1.7i$).
  }
  \label{fig:5dq_fromN_lq1}
\end{figure}%

FIG.~\ref{fig:5dq_fromN_thimbles} illustrates the Euclidean-to-Euclidean case with $q_0 = 0$ and $q_1 = 5k/\Lambda$.
Unlike in lower dimensions, there is no degeneracy of saddles and thimbles due to the finite $\hbar$ in such cases with $3k/\Lambda < q_1 <6k/\Lambda$. Hence, we can unambiguously identify the contributing saddle and its associated integration contour, which indicates the tunneling scenario. 

In contrast, for $q_0 = 0, \; 0 < q_1 < 3k / \Lambda$,\footnote{The value $q_1=3k/\Lambda$ corresponds to the minimum of the effective potential $U_5$ in Eq.~\eqref{eq:Ham5d} at the semiclassical limit $\hbar \rightarrow 0$. At precisely this value, the difference between the two saddles becomes higher-order in $\hbar$.}
the split direction of the two closest tunneling saddles becomes vertical.
The system exhibits the Stokes phenomenon: a positive rotation in $\hbar$ yields a contribution from only one tunneling saddle, while a negative rotation leads to contributions from two tunneling saddles, as illustrated in FIGs.~\ref{fig:5dq_fromN_lq1_thimbles}--\ref{fig:5dq_fromN_lq1_thimbles_RP}.\footnote{In \appref{sec:5D_fin}, we show the thimble structure for the case with $q_0 >0$. 
The Stokes phenomenon also appears in such cases.
 Because the thimble configuration closely parallels that of the 3D case, resurgence can resolve the ambiguity about which saddles and thimbles contribute. Most of the computation proceeds in the same way, so we do not repeat the steps here.
}

%%%%%%%%%%%%%%%%%%%%%%%%%%%%%%%%%%%%%%%
%%%%%%%%%%%%%%%%%%%%%%%%%%%%%%%%%%%%%%%
\section{Large-\texorpdfstring{$D$}{D} limit}
\label{sec:large-d-limit}
%%%%%%%%%%%%%%%%%%%%%%%%%%%%%%%%%%%%%%%
%%%%%%%%%%%%%%%%%%%%%%%%%%%%%%%%%%%%%%%

Finally, we consider the large-$D$ limit of Lorentzian quantum cosmology.
In the large-$D$ limit ($D \gg 1$), the gravitational action reduces to the following form under Option~I \eqref{eq:d-dim_action1}:\footnote{
    Note that here we implicitly assume that $|\log q| \ll D$. 
    However, even if we relax this condition, the qualitative results below remain unchanged as long as $q \leq \e^{-D}$, since the constant curvature term still dominates the action \eqref{eq:d-dim_action1} in the large-$D$ limit.     
    Also, the scaling of $V_{D-1}/\hbar G_D T$ is assumed to be fixed.
}
\footnote{
$\order{D^{-1}}$ is neglected in $S_2$, as this term is already higher-order in the small fluctuation $Q$.
}
\begin{align}
    &S = S_0[q_{\rm cl},N] + S_2[Q,N] + \order{Q^3}, \nonumber
\end{align}
\begin{align}
    &S_0[q_{\rm cl},N]
    = \frac{V_{D-1}}{16 \pi G_D}
    \int \dd{t}
    \left[(D-1)(D-2) k N
        -2 \Lambda N
        - \frac{\dot{q}_{\rm cl}^2}{N} 
        - \frac{1}{D}
        \qty(4 \Lambda N \log q_{\rm cl}
        + \frac{\dot{q}_{\rm cl}^2}{N} )
        + \order{D^{-2}}
    \right],
    \nonumber
\end{align}
\begin{align}
    &S_2[Q,N]
    = -\frac{V_{D-1}}{16 \pi G_D}
    \int \dd{t}
    \left[
        \frac{\dot{Q}^2}{N}
        + \order{D^{-1}}
    \right],
    \label{eq:action_LD}
\end{align}
with the quantum fluctuation $Q(t)$ around the classical solution $q_{\rm cl}(t)$. The classical solution $q_{\rm cl}(t)$ with the Dirichlet boundary conditions is obtained perturbatively as
\begin{align}
    q_{\rm cl}(t)
    &= q_\infty (t) +\frac{2N^2 \Lambda T^2}{D(q_1 -q_0 )^2} \Biggl[
    q_\infty (t) \log{q_\infty (t)} 
    -\frac{q_0 (T-t)}{T} \log{q_0} - \frac{q_1 t}{T}\log{q_1} 
    \Biggr]
    +\order{D^{-2}},
\end{align}
where $q_\infty (t)$ denotes the $\order{D^0}$ classical solution
\begin{align}
    q_\infty (t) = q_0 \qty(1-\frac{t}{T}) +q_1 \frac{t}{T} . 
\end{align}
After integrating over $Q$ and changing the variable from $N$ to $x$ via $x^2\coloneqq N$, the amplitude is obtained as 
\begin{align}
    &\mathcal{A}[q_1; q_0]
    \approx \sqrt{\frac{iV_{D-1}}{16\pi^2 T\hbar G_D}}
    \int^{\infty}_{-\infty} \dd{x} \; 
    \exp \qty[F_D(x)],
    \noindent \\ 
    &F_D = 
        -\frac{iV_{D-1}}{16\pi \hbar G_D}
        \left[
             \frac{(D+1)}{DTx^2} (q_1-q_0)^2
        \right.
        \nn
        & \qquad \qquad
        \left.
            + \frac{x^2T}{D}\qty((D-2)(2\Lambda - D(D-1)k)
            +\frac{4\Lambda}{q_1-q_0} (q_1 \log q_1 - q_0 \log q_0 )
            )
        \right].
\end{align}
The phase of the exponent is expanded with a remainder of $\order{D^{-2}}$. The saddles are found as follows:
\begin{align}
    &x_{1} = \e^{i\frac{\pi}{4}}
    \frac{\qty(q_1 - q_0)^{\frac{1}{2}}}{k^{\frac{1}{4}} T^{\frac{1}{2}}} \frac{1}{D^\frac{1}{2}}+\order{D^{-1}}, \quad
    x_{2} = \e^{-\frac{3}{4}i\pi}
    \frac{\qty(q_1 - q_0)^{\frac{1}{2}}}{k^{\frac{1}{4}} T^{\frac{1}{2}}} \frac{1}{D^\frac{1}{2}}+\order{D^{-1}},
    \quad
    x_{3} = x_1^*,
    \quad
    x_4 = x_2^*.
\end{align}
Here we assume that $q_1 \geq q_0$. In terms of $N$, these saddles are expressed as
\begin{align}
    N_{\rm T} \coloneqq x_1^2 = x_2^2 
    = i\frac{\qty(q_1 - q_0)}{k^{\frac{1}{2}} T} \frac{1}{D}+\order{D^{-2}}, 
    \quad
    N_{\rm NB} \coloneqq x_3^2 = x_4^2 
    = -i\frac{\qty(q_1 - q_0)}{k^{\frac{1}{2}} T} \frac{1}{D}+\order{D^{-2}}.
\end{align}
This result implies that only two types of saddles exist: the tunneling one $x_{1,2}$ or, equivalently, $N_{\rm T}$, and the no-boundary one $x_{3,4}$ or $N_{\rm NB}$. There is no classical saddle unless $q_0 = q_1$, or, equivalently, $N_{\rm T}=N_{\rm NB}=0$.

\begin{figure}[tp]%
  \begin{minipage}[t]{0.5\linewidth}%
    \centering%
    \includegraphics[keepaspectratio, width=\linewidth]{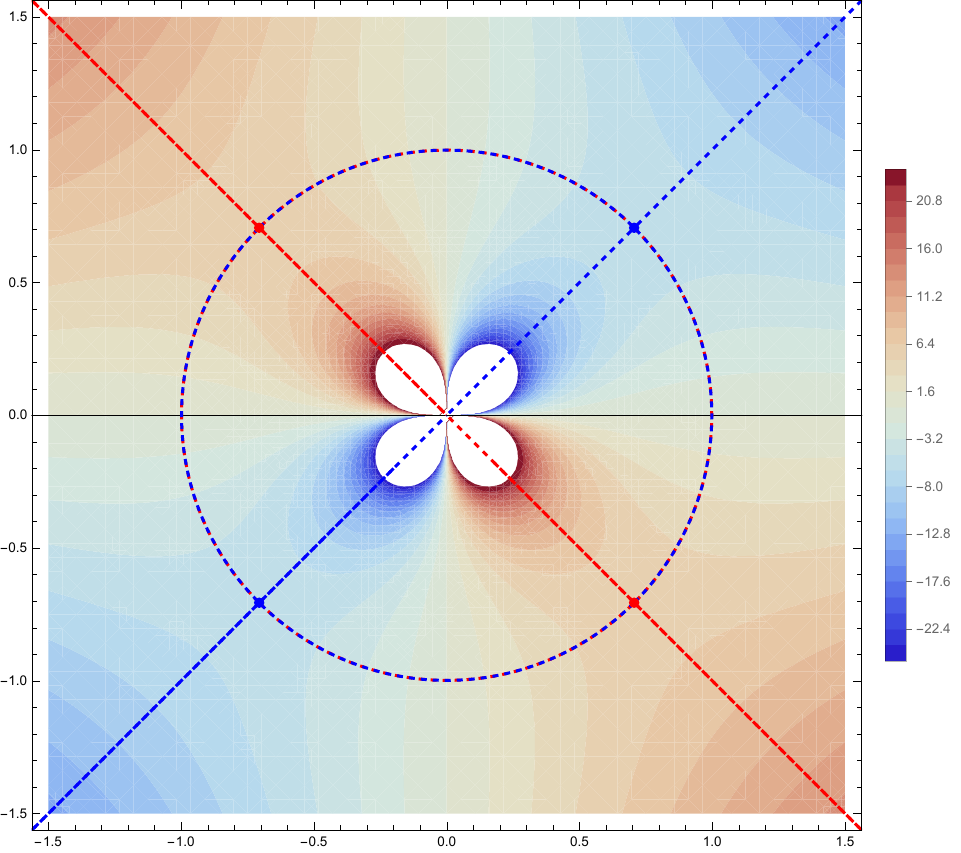}%
    \subcaption{Thimbles ($\argu{\hbar}=0$)}%
    \label{fig:ldq_thimbles}%
  \end{minipage}%
  \\[10pt]
  \begin{minipage}[t]{0.5\linewidth}%
    \centering%
    \includegraphics[keepaspectratio, width=\linewidth]{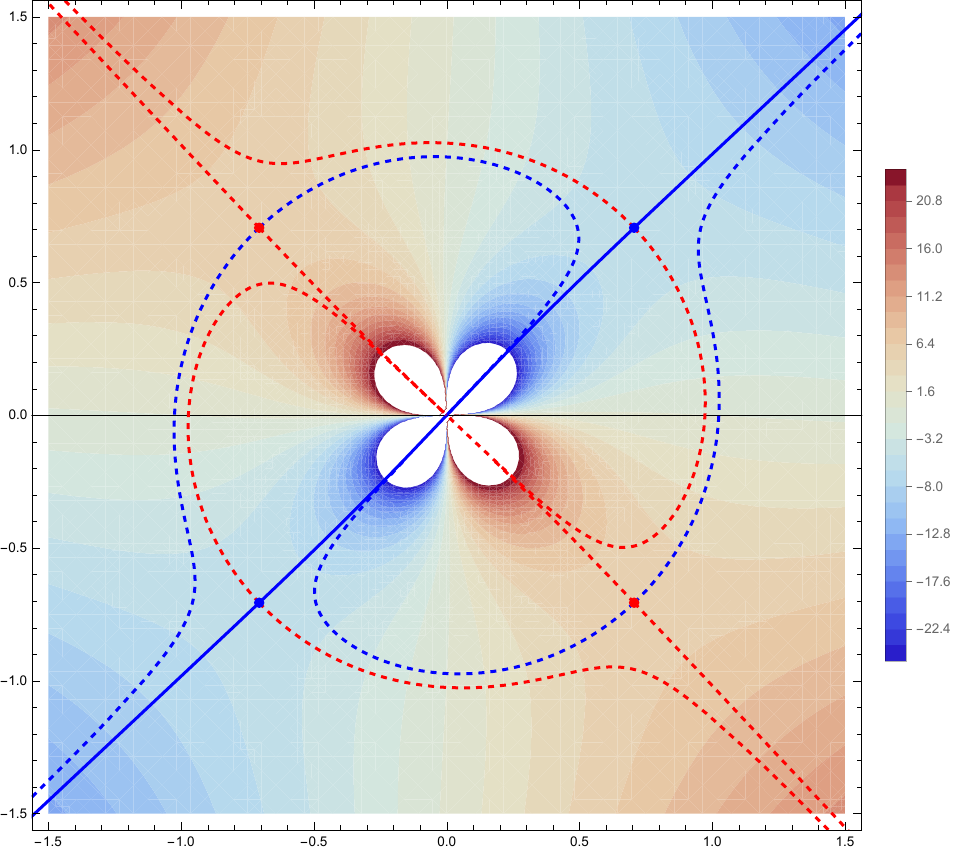}%
    \subcaption{Thimbles ($\argu{\hbar}=-\pi/60$)}%
    \label{fig:ldq_RM}%
  \end{minipage}%
  \begin{minipage}[t]{0.5\linewidth}%
    \centering%
    \includegraphics[keepaspectratio, width=\linewidth]{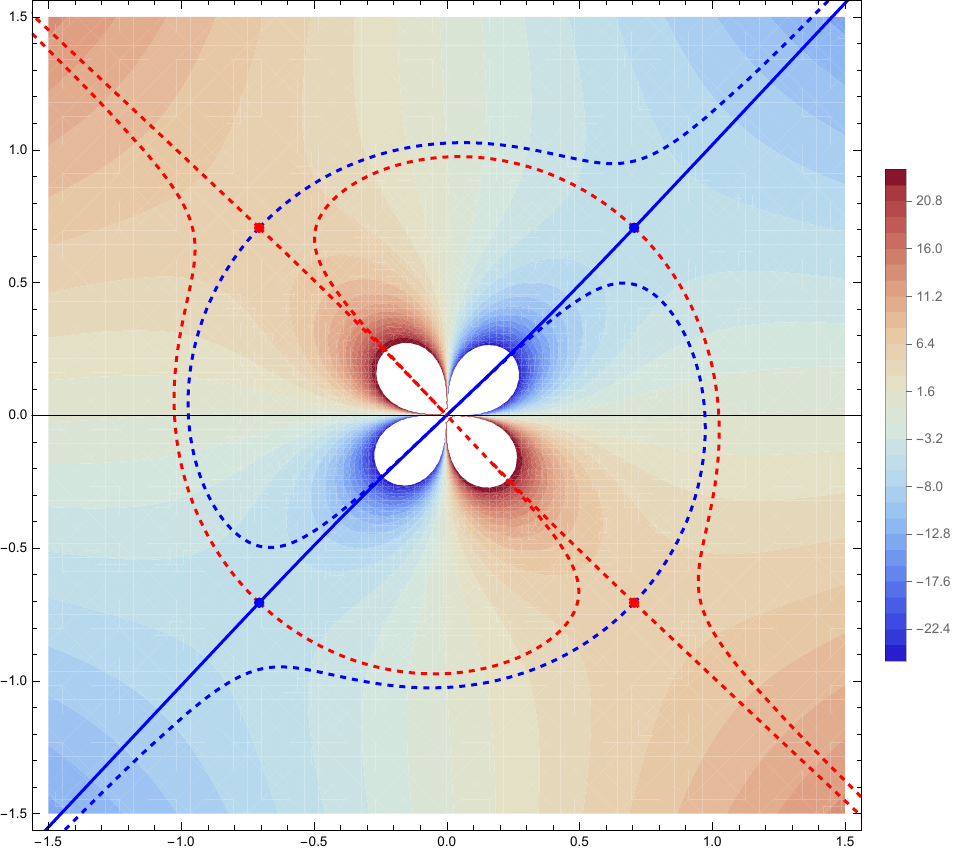}%
    \subcaption{Thimbles ($\argu{\hbar}=+\pi/60$)}%
    \label{fig:ldq_RP}%
  \end{minipage}%
  \caption{Thimbles in the large-$D$ case ($D=7$ as an example). Note that the original integration domain is now the entire real axis, $-\infty < x < \infty$. Two tunneling saddles contribute in all cases.}
  \label{fig:ldq}
\end{figure}%

The contour plot of $\re{F_D(x)}$ with the four saddles and corresponding thimbles is exhibited in \figref{fig:ldq}. The thimbles are degenerate for the case of $\argu{\hbar}=0$ (\figref{fig:ldq_thimbles}). However, this degeneracy can be resolved by considering complex $\hbar$ as shown in FIGs.~\ref{fig:ldq_RM} and \ref{fig:ldq_RP}. They show that the two tunneling saddles $x=x_1$ and $x=x_2$ contribute without ambiguity. That is, only a single tunneling saddle $N=N_{\rm T}$ contributes in terms of the original variable $N$. Unlike the Euclidean-to-Euclidean cases in lower dimensions, this case exhibits no Stokes phenomenon.

Furthermore, we can explicitly evaluate the transition amplitude:
\begin{align}
    \mathcal{A}[q_1; q_0]
    \approx & \,
    \sqrt{\frac{iV_{D-1}}{16\pi^2 T\hbar G_D}}
    \int^{\infty}_{-\infty} \dd{x} \; 
    \exp \qty[i A x^2 - i B \frac{1}{x^2}]
    \nn
    =& \,
    \sqrt{\frac{iV_{D-1}}{16\pi^2 T\hbar G_D} \cdot \frac{\pi}{A}}
    \, \e^{i\frac{\pi}{4}}\,  \e^{-2\sqrt{AB}}
    \nn
    \approx & \, \frac{i}{DT\sqrt{k}}
    \exp\qty[
        -\frac{V_{D-1} D k^{\frac{1}{2}}}{8\pi \hbar G_D}
         (q_1 -q_0)
    ]
    ,
    \label{eq:tamp_ld_exact}
\end{align}
where
\begin{align}
    & A \coloneqq 
    \frac{V_{D-1}}{16\pi \hbar G_D}
    \frac{T}{D}\qty((D-2)(D(D-1)k -2\Lambda)
            -\frac{4\Lambda}{q_1-q_0} (q_1 \log q_1 - q_0 \log q_0 )
    )+\order{D^{-2}},
    \nn
    & B \coloneqq
    \frac{V_{D-1}}{16\pi \hbar G_D}
    \frac{(D+1)}{DT} (q_1-q_0)^2
    +\order{D^{-2}}.
\end{align}
This result indicates that the transition amplitude decays exponentially for higher dimensions, and that the dominant transition becomes the trivial one $q_0=q_1$. 

\begin{figure}[tp]%
  \begin{minipage}[t]{0.5\linewidth}%
    \centering%
    \includegraphics[keepaspectratio, width=\linewidth]{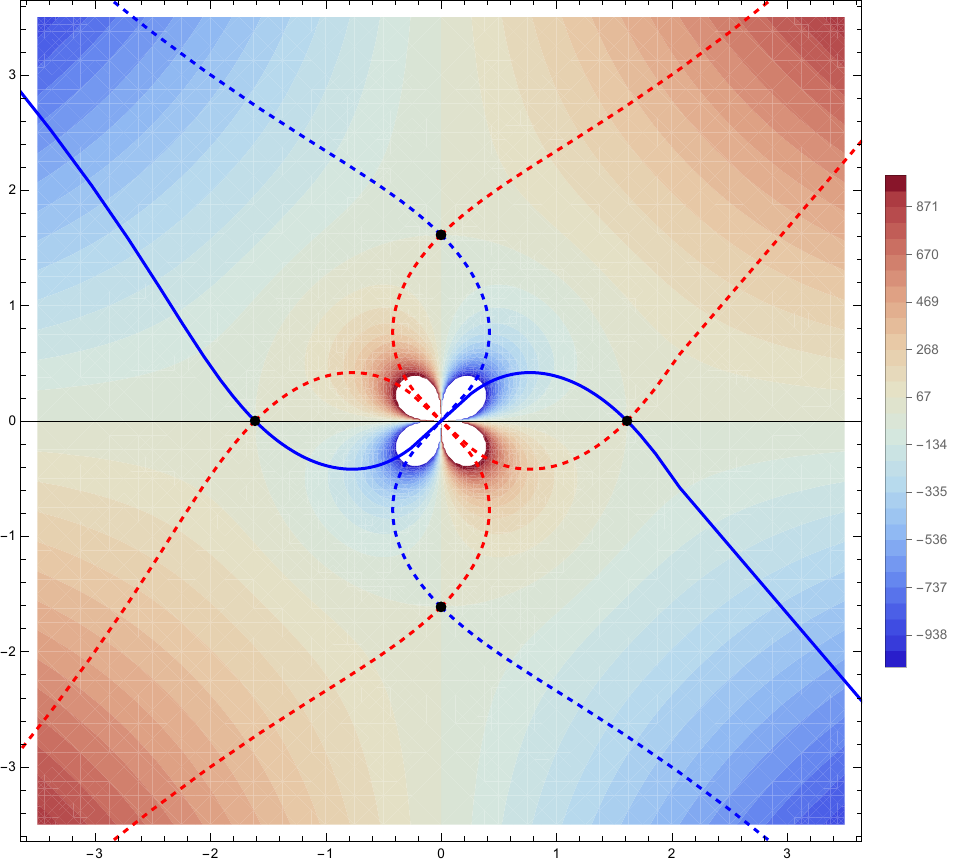}%
    \subcaption{Lorentzian}%
    \label{fig:ldcl_re_thimbles}%
  \end{minipage}%
  \begin{minipage}[t]{0.5\linewidth}%
    \centering%
    \includegraphics[keepaspectratio, width=\linewidth]{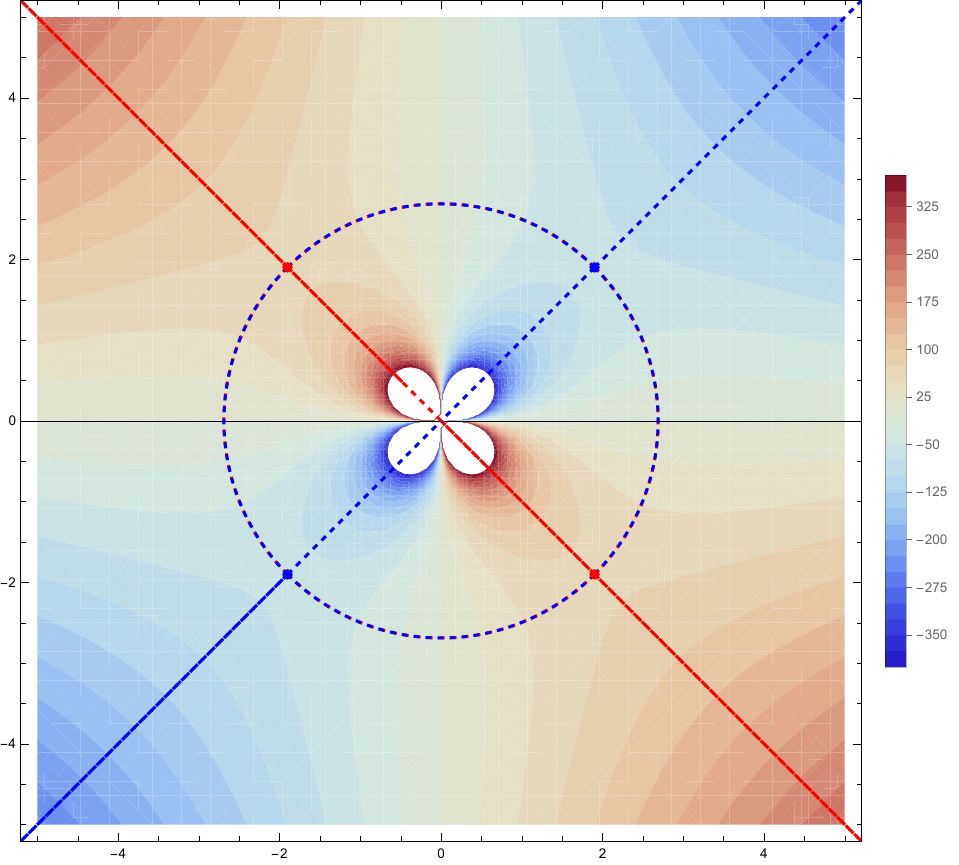}%
    \subcaption{Euclidean ($\argu{\hbar}=0$)}%
    \label{fig:ldq_re_thimbles}%
  \end{minipage}%
  \\[10pt]
  \begin{minipage}[t]{0.5\linewidth}%
    \centering%
    \includegraphics[keepaspectratio, width=\linewidth]{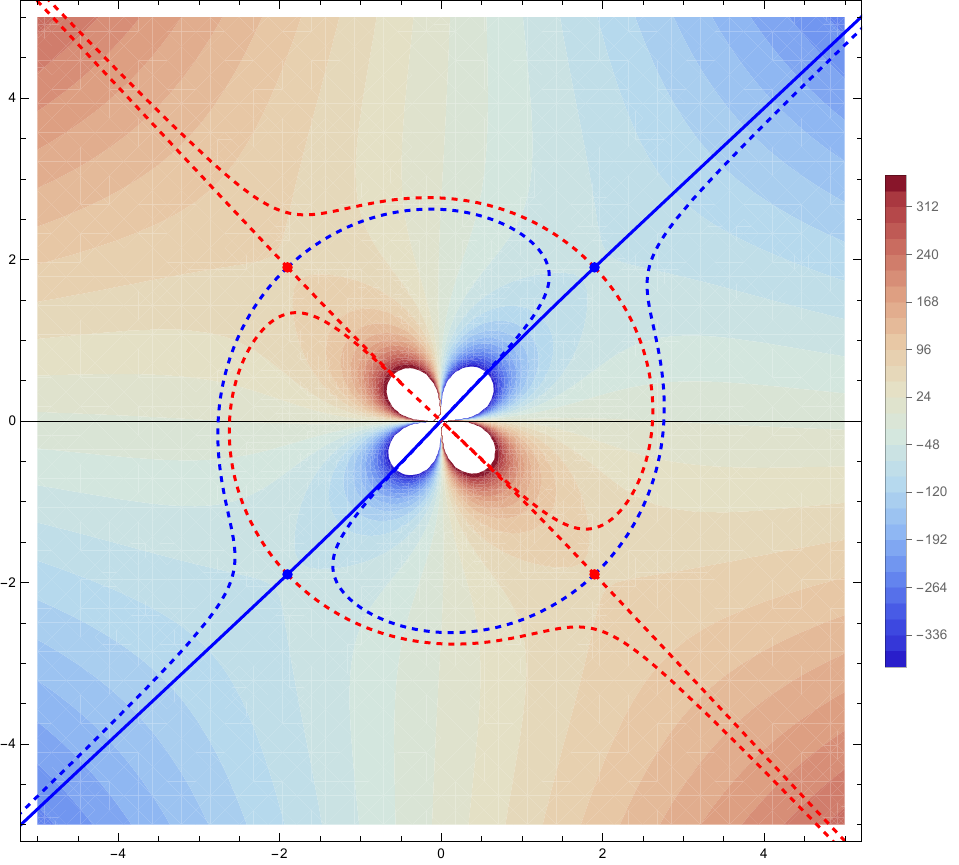}%
    \subcaption{Euclidean ($\argu{\hbar}=-\pi/60$)}%
    \label{fig:ldq_re_RM}%
  \end{minipage}%
  \begin{minipage}[t]{0.5\linewidth}%
    \centering%
    \includegraphics[keepaspectratio, width=\linewidth]{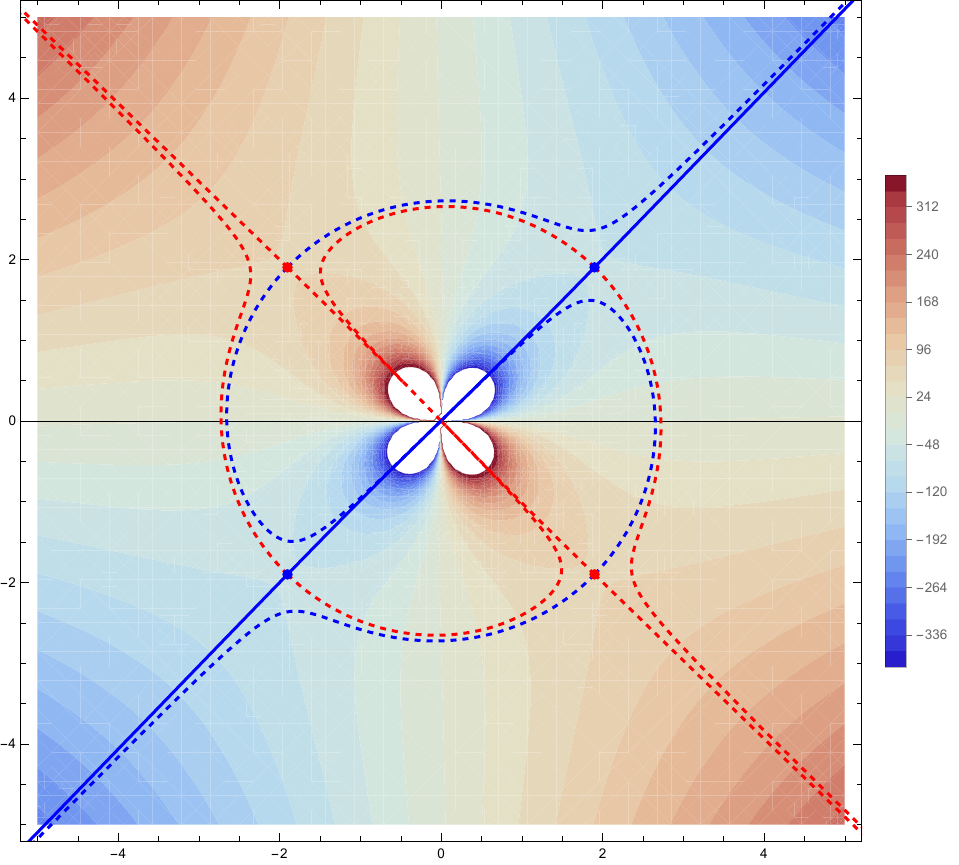}%
    \subcaption{Euclidean ($\argu{\hbar}=+\pi/60$)}%
    \label{fig:ldq_re_RP}%
  \end{minipage}%
  \caption{Thimbles in the large-$D$ limit with the parameter rescalings \eqref{eq:param_res}.}
  \label{fig:ld_re}
\end{figure}%

\subsubsection*{Rescaling the parameters}

One may wonder why there appears to be no classical transition and why only quantum tunneling occurs. The reason lies in the present choice of parameters, for which $kT^2 \sim \Lambda T^2 \sim \order{D^0}$. From the action \eqref{eq:action_LD}, we can derive the Hamiltonian as
\begin{align}
    H_D = N\qty(-\frac{P_D^2}{2\mathcal{M}_D}+U_D)
\end{align}
where the mass parameter, conjugate momentum, and leading-order potential are given by
\begin{align}
    &\mathcal{M}_D = \frac{V_{D-1}}{8\pi G_D} ,
    \quad
    P_D = -\frac{\mathcal{M}_D}{N} \dot{q},
    \quad
    U_D = -\frac{\mathcal{M}_D}{2}
    D^2 k +\order{D^1}.
\end{align}
Since the effective potential has a large negative constant at the leading order, the Hamiltonian constraint forces $P_D^2$ to be negative, so that the classical transition is prohibited.

In fact, classical transitions are found possible when we appropriately rescale the parameters and obtain a positive effective potential. One sensible choice is the following:
\begin{align}
    \Lambda \rightarrow \frac{(D-1)(D-2)}{2\ell^2}, \quad
    N \rightarrow \frac{N}{(D-2)},
    \label{eq:param_res}
\end{align}
The former condition corresponds to fixing the dS radius $\ell$, rather than the cosmological constant, in the large-$D$ limit. The latter condition guarantees that the kinetic term $\dot{q}^2/N$ in Eq.~\eqref{eq:action_LD} dominates over the $q$-dependent potential term proportional to $\Lambda N\log q$ in the potential.
The action \eqref{eq:d-dim_action1} with this choice is rewritten as
\begin{align}
    S\qty[q, N]
    &=\frac{(D-1)V_{D-1}}{16 \pi G_{D}}
    \int \dd{t}
        \left[k N
        -\frac{1}{\ell^2} q^{\frac{2}{D-2}} N
        -\frac{\dot{q}^2}{N} 
    \right]
    \label{eq:LDaction_woLD}
    \\
    &=
    \frac{(D-1)V_{D-1}}{16 \pi G_{D}}
    \int \dd{t}
        \left[\qty(k
        -\frac{1}{\ell^2}) N
        -\frac{\dot{q}^2}{N} 
        -\frac{2N}{D\ell^2} \log q
        +\order{D^{-2}}
    \right],
    \label{eq:d-dim_action1_rep}
\end{align}
whose classical solution is 
\begin{align}
    q_{\rm cl}(t)
    &= q_\infty (t) +\frac{N^2 T^2}{D \ell^2 (q_1 -q_0 )^2} \Biggl[
    q_\infty (t) \log{q_\infty (t)} 
    -\frac{q_0 (T-t)}{T} \log{q_0} - \frac{q_1 t}{T}\log{q_1} 
    \Biggr]
    +\order{D^{-2}}.
    \label{eq:LD_solcl_re}
\end{align}
It is noteworthy that the classical transition can be realized in this parametrization, since the effective potential $V(q) \sim (k-1/\ell^2)N$ can be positive. The corresponding transition amplitude is calculated as\footnote{
    We note here that the $q_0 \rightarrow 0$ limit must be carefully taken, unlike the case without the rescaling. In the preceding discussion, we have implicitly required $|\log q_0|, |\log q_1| \ll D$. 
    In that sense, the universe’s ``sizeless'' limit here corresponds to $q_0 \rightarrow 0$ while still maintaining this condition. In particular, this corresponds to $a_0 \to \text{const.}> 0$.  
    Otherwise, the second term involving $\ell$ in Eq.~\eqref{eq:LDaction_woLD} (equivalently, in $\Tilde{U}_D$) is no longer amenable to a series expansion, and the equation of motion for $q$ must be solved non-perturbatively with respect to $D$. 
    A genuine $a_0 \to 0$ limit would require a separate treatment beyond the present perturbative framework.
}
\footnote{The measure of the $N$-integration is also rescaled here.}
\begin{align}
    &\mathcal{A}[q_1; q_0]
    \approx \sqrt{\frac{iV_{D-1}}{16\pi^2 DT\hbar G_D}}
    \int^{\infty}_{-\infty} \dd{x} \; 
    \exp \qty[F_D(x)] ,
    \nonumber
\end{align}
\begin{align}
    &F_D(x) = \frac{i(D-1)V_{D-1}}{16 \pi \hbar G_{D}}
    \Bigl[\qty(k-\frac{1}{\ell^2}) Tx^2
    - \frac{(q_1-q_0)^2}{Tx^2}
    \nn
    & \hspace{120pt}
    + \frac{2Tx^2}{D\ell^2(q_1-q_0)}
    \qty(
     q_1(1-\log q_1)-q_0(1-\log q_0)
    )
    \Bigr]
    +\order{D^{-1}}.
    \label{eq:LD_phase_wres}
\end{align}
The saddles for $x$ now become
\begin{align}
    &x_1 = -x_2 
    \coloneqq \frac{1}{(1-k\ell^2)^{\frac{1}{4}}} 
    \sqrt{\frac{\ell}{T}(q_1-q_0)}
    \qty[1
    + \frac{
     q_1(1-\log q_1)-q_0(1-\log q_0)
    }{2D(1-k\ell^2)
    (q_1-q_0)}
    ]
    +\order{D^{-2}},
    \nn
    &x_3 = -x_4 \coloneqq i x_1,
    \label{eq:LDsaddles_res}
\end{align}
or equivalently
\begin{align}
    N_1 = - N_2 
    \coloneqq \frac{\ell (q_1-q_0)}{T\sqrt{1-k\ell^2}} 
    \qty[1
    + \frac{
     q_1(1-\log q_1)-q_0(1-\log q_0)
    }{D(1-k\ell^2)
    (q_1-q_0)}
    ]+\order{D^{-2}}.
    \label{eq:LDsaddles_res_N}
\end{align}
These saddles correspond to classical contribution ($N_i \in \mathbb{R}$) if $1 > k\ell^2$, and Euclidean contribution ($\im{N_i} \neq 0$) if $1<k\ell^2$. The saddles and corresponding thimbles in each case are shown in \figref{fig:ld_re}. It is confirmed that the classical saddles contribute in the former case, whereas tunneling saddles contribute in the latter, respectively.

This behavior can be observed more explicitly by evaluating the amplitude exactly. Analogously to the derivation of Eq.~\eqref{eq:tamp_ld_exact}, 
\begin{align}
    \mathcal{A}[q_1; q_0]
    \approx & \,
    \sqrt{\frac{iV_{D-1}}{16\pi^2 DT\hbar G_D}}
    \int^{\infty}_{-\infty} \dd{x} \; 
    \exp \qty[-i A x^2 - i B \frac{1}{x^2}]
    \nonumber \\[5pt]
    & \approx 
    \begin{dcases}
    \frac{\ell}{DT\sqrt{1-k\ell^2}}
    \exp \qty[
        -\frac{iDV_{D-1}}{8\pi \hbar \ell G_D}
        \sqrt{1-k\ell^2} (q_1 - q_0)
    ], & (1>k\ell^2)
    \\
    \frac{i \ell}{DT\sqrt{k\ell^2-1}}
    \exp \qty[
        -\frac{DV_{D-1}}{8\pi \hbar \ell G_D}
        \sqrt{k\ell^2-1} (q_1 - q_0)
    ], & (1<k\ell^2)
    \end{dcases}
    \nn
\end{align}
where
\begin{align}
    & A \coloneqq 
    \frac{(D-1)TV_{D-1}}{16\pi \hbar G_D}
    \qty[
    \qty(\frac{1}{\ell^2}-k) 
    -\frac{2}{D\ell^2(q_1-q_0)}
    \qty(
     q_1(1-\log q_1)-q_0(1-\log q_0)
    )
    ]+\order{D^{-1}},
    \nn
    & B \coloneqq
    \frac{(D-1)V_{D-1}}{16\pi \hbar G_D T}
    (q_1-q_0)^2+\order{D^{-1}}.
\end{align}
This result explicitly shows that not only quantum tunneling but also classical transition can occur, depending on the value of the parameter combination $k\ell^2$. 

The Hamiltonian analysis also confirms this fact. The Hamiltonian in the current parametrization is
\begin{align}
    \Tilde{H}_D = N\qty(-\frac{\Tilde{P}_D^2}{2\Tilde{\mathcal{M}}_D}+\Tilde{U}_D),
\end{align}
with
\begin{align}
    &\Tilde{\mathcal{M}}_D = \frac{(D-1)V_{D-1}}{8\pi G_D} ,
    \quad
    \Tilde{P}_D = -\frac{\Tilde{\mathcal{M}}_D}{N} \dot{q},
    \quad
    \Tilde{U}_D = \frac{\Tilde{\mathcal{M}}_D}{2}
    \qty(\frac{1}{\ell^2}-k) +\order{D^{0}},
\end{align}
at leading order in $D$. We can notice that the classical transition with $\Tilde{P}_D^2>0$ is allowed if $\Tilde{U}_D>0$, or, equivalently, $1>k\ell^2$. We note that dependence on the boundary conditions on $q_0$ and $ q_1$ is effectively negligible as being of higher order in $D^{-1}$, and that only $\ell$ and $k$ govern the qualitative features of the transition. These contributions would, of course, be recovered if terms of order $D^{-2}$ and higher were included in the action \eqref{eq:d-dim_action1_rep}, the classical solution \eqref{eq:LD_solcl_re}, and the subsequent expressions, although it is hard to find a closed-form expression for the solutions.

We conclude this section with a comparison of the saddle-point structures in different spacetime dimensions. In the large-$D$ limit, there are two independent semiclassical saddles with respect to the lapse function $N$, as shown in Eq.~\eqref{eq:LDsaddles_res_N}. This number is clearly smaller than the corresponding numbers in the three- and four-dimensional cases. 
It is therefore natural to ask how the remaining saddles found in lower dimensions disappear from the leading saddle-point structure as the spacetime dimension is increased. To this end, we extend the preceding large-$D$ expansion from first to second order in $1/D$. When terms of order $1/D^2$ are included, the action takes the following schematic form:
\begin{align}
    F_D (x) = \frac{i(D-1)V_{D-1}}{16\pi \hbar G_D}
    \qty[f_{0,2} x^2 + \frac{f_{0,-2}}{x^2} + \frac{f_{1,2}}{D} x^2 +\frac{1}{D^2} \qty(f_{2,2} x^2 + f_{2,6} x^6)
    + \order{D^{-3}}],
    \label{eq:LDaction_higher}
\end{align}
where $f_{i,j}=f_{i,j}(q_0,q_1,T,k,\ell)$ are coefficient functions.\footnote{The first index specifies the order in $1/D$, while the second specifies the power of $x$. The explicit expressions for the first several coefficient functions can be read off from Eq.~\eqref{eq:LD_phase_wres}.}  
Accordingly, we can find more semiclassical saddles than those given in Eq.~\eqref{eq:LDsaddles_res}. These saddles are classified into two groups according to their behavior in the large-$D$ limit. The first group remains at finite values of $x$ as $D\rightarrow \infty$:
\begin{align}
    x^4_{\rm fin} = 
    \frac{f_{0,-2}}{f_{0,2}}
    \qty[ 1
        -\frac{f_{1,2}}{D f_{0,2}}
        +\frac{f_{1,2}^2-f_{0,2}f_{2,2}-3f_{0,-2}f_{2,6}}{D^2 f_{0,2}^2}
        + \order{D^{-3}}
    ],
\end{align}
which are essentially Eq.~\eqref{eq:LDsaddles_res} with additional $1/D^2$ corrections. 
The second group consists of additional saddles that appear when the higher-order correction is included
\begin{align}
    x^4_{\infty} = 
    -\frac{1}{3 f_{2,6}} \qty[
        D^2 f_{0,2}
        + D f_{1,2}
        + f_{2,2} + \frac{3 f_{2,6} f_{0,-2}}{f_{0,2}}
        + \order{D^{-1}}
    ],
\end{align}
which are pushed toward infinity in the large-$D$ limit.
Generically, the number of such saddles that are pushed toward infinity increases as higher-order terms are included in the action \eqref{eq:LD_phase_wres},\footnote{The prefactor can also contain higher-order corrections, which affect the positions of saddles.} but in the limit $D\to\infty$, they do not appear in the leading large-$D$ saddle-point structure.

%%%%%%%%%%%%%%%%%%%%%%%%%%%%%%%%%%%%%%%
%%%%%%%%%%%%%%%%%%%%%%%%%%%%%%%%%%%%%%%
\section{Discussion and Conclusion}
\label{sec:conclusion}
%%%%%%%%%%%%%%%%%%%%%%%%%%%%%%%%%%%%%%%
%%%%%%%%%%%%%%%%%%%%%%%%%%%%%%%%%%%%%%%

The Lorentzian path integral method with Picard-Lefschetz theory provides an efficient way to identify the favored scenario for the creation of our universe, based on explicitly specified boundary conditions. For the quantum creation of the finite-size universe from nothing, it has been shown that the tunneling proposal is preferred in 4D GR~\cite{Feldbrugge:2017kzv, Honda:2024aro}. On the other hand, we previously found that the Hartle-Hawking no-boundary scenario can be realized in the 2D JT gravity and that the resulting universe is linearly stable~\cite{Honda:2024hdr}. To reveal the dimensionality dependence of these results, we have investigated Lorentzian quantum cosmology in $D=3, 4$, and $5$, as well as in the large-$D$ limit, and have comprehensively identified favored scenarios under Dirichlet-type boundary conditions (see \tabref{tab:summary} for the summary of the results).

As our main result, we have shown that, for the creation from nothing ($0=q_0<q_1$), the tunneling proposal remains robust in $D\ge 3$ closed de Sitter spacetimes. In all cases, the requirement that the contour be analytically deformed onto the appropriate steepest descent path identifies the favored saddle as the one with $\im{N}>0$, corresponding to the tunneling scenario. These results follow from the negativity of the effective kinetic term in the Einstein-Hilbert action, as can be seen, e.g., in Eq.~\eqref{eq:4DHam}. 
To make the integral absolutely convergent, this negativity requires the anti-Wick rotation, which corresponds to $\im{N}>0$. Applying Picard-Lefschetz theory to the Lorentzian path integral of the Einstein-Hilbert action favors the tunneling saddle, irrespective of the spacetime dimensionality.

We have further shown that similar Stokes phenomena can arise in the Euclidean-to-Euclidean transition ($q_0 < q_1 < q_{\rm crit}$) in the cases $D=3$, $4$, and $5$. In all such cases, there exists an ambiguity in the choice of the contributing tunneling saddle and its associated integration path. However, through a resurgence analysis based on the Borel-Pad\'{e} resummation, we have explicitly demonstrated in the 3D and the 4D cases that this ambiguity in the thimble selection precisely corresponds to the Borel ambiguity, and 
the amplitudes have no ambiguities. The analogous thimble structure in the 5D case suggests that the same mechanism also applies there. This analysis generalizes our previous work \cite{Honda:2024aro}, in which resurgence was applied to the case $q_0=q_1$ in 4D GR, and illustrates that the resurgence technique is useful in a broader class of Lorentzian quantum cosmological settings.

Furthermore, in the large-$D$ limit, we have shown how the transition is controlled by the large-$D$ expansion and truncation of the effective potential $\widetilde{U}_D$ in the Hamiltonian. After suitable rescaling of the physical constants and truncating the expansion at the first sub-leading order, $\widetilde{U}_D$ is effectively governed by its constant, which depends on the spatial curvature and the cosmological constant: if this constant is positive, the classical transition can occur, while a negative value instead results in the tunneling transition. 
Including higher-order corrections in the $1/D$ expansion not only shifts the leading-order saddles but also gives rise to additional saddles. Some of these remain at finite locations as $D\to\infty$, while the others are pushed toward infinity in the complex $N$-plane and therefore do not appear in the leading large-$D$ saddle-point structure. This suggests a connection between the more complicated saddle-point structures found in lower dimensions and the simpler structure appearing in the large-$D$ expansion, while also suggesting that a resummation of the $1/D$ expansion may be important for understanding this relation in more detail.

We now comment on the linear stability of each scenario discussed above. In Lorentzian quantum cosmology, the instability of the realized tunneling scenario has often been regarded as a serious issue. In 4D GR, it has been suggested that, when linear test-field perturbations are introduced around the background associated with the tunneling scenario, the corresponding matter contribution to the total amplitude takes an inverse-Gaussian form. As a result, higher-energy modes become more strongly excited, leading to an unstable universe (see Refs.~\cite{Feldbrugge:2017fcc,DiazDorronsoro:2017hti, Feldbrugge:2017mbc, Feldbrugge:2018gin, DiazDorronsoro:2018wro, Halliwell:2018ejl, Janssen:2019sex, Vilenkin:2018dch, Vilenkin:2018oja, Bojowald:2018gdt, DiTucci:2018fdg, DiTucci:2019dji, DiTucci:2019bui, Lehners:2021jmv, Matsui:2022lfj, Matsui:2024bfn,Ailiga:2024nkz,Yamada:2025rld}). This can be understood as follows. In the gravitational path integral, where the kinetic term is negative, the anti-Wick rotation ensures the convergence of the path integral, as discussed above. In contrast, the path integral for an ordinary matter action (with a positive kinetic term) converges under the usual Wick rotation ($\im{N}<0$), and diverges under the anti-Wick rotation. Owing to this mismatch, perturbative instabilities arise in the tunneling proposal.

In the cases analyzed in this work, namely $D=3, 4$, and $5$, as well as the large-$D$ limit, the tunneling scenario is found to be likewise favored. 
This implies that the tunneling-like transition is a universal feature in the quantum genesis of the universe under Einstein gravity regardless of dimensions, although we do not explicitly analyze the thimble structure in other dimensions. 
Furthermore, while we have not directly investigated perturbative instabilities, the structure of the kinetic term discussed above is common to all the cases considered here. We therefore expect the perturbative instability to persist in the same manner. This suggests that varying the spacetime dimensionality alone offers limited scope for obtaining a stable universe-creation scenario.

On the other hand, as shown previously in Ref.~\cite{Honda:2024hdr}, the no-boundary proposal is realized as a perturbatively stable creation-from-nothing scenario in 2D JT gravity, under the additional boundary condition that the final dilaton value is smaller than its initial value. These results imply that, in order to realize a stable creation of the universe under the Dirichlet-type boundary conditions, it is essential to introduce additional gravitational degrees of freedom or nontrivial couplings, rather than merely increasing the spacetime dimensionality.
We leave for future work a systematic investigation of the Lefschetz-thimble structures of Lorentzian path integrals in $D$-dimensional extended theories of gravity, including dilaton gravity. This includes determining whether the tunneling or no-boundary proposal more naturally describes the quantum creation of the universe, as well as carefully analyzing the associated metric perturbations.

\section*{Acknowledgments}
%%%%%%%%%%%%%%%%%%%%%%%%%%%%%%%%%%%%%
We would like to thank Akihiro Ishibashi, Taiga Miyachi, Vikramaditya Mondal, Shinji Mukohyama, Ryo Namba, Jun Nishimura, Daiki Saito, Kotaro Shinmyo, and Hiromasa Tajima for useful discussions.
This work is supported by JSPS Grant-in-Aid for Scientific Research Numbers JP22H01222 (MH), 23K13100 (HM), 26K17143 (KN), JP23KJ1162 (KO), and 26H00403 (TT). 
M.~H.~is supported by JST CREST Grant Number JPMJCR24I3, JSPS Grant-in-Aid for Transformative Research Areas (A) ``Extreme Universe'' JP21H05190 [D01], and the Royal Society grants ICA/R2/242058 and IEC/R3/243103.
T.T.'s work was partly supported by the 34th (FY 2024) Academic research grant (Natural Science) No.~9284 from DAIKO FOUNDATION.

%%%%%%%%%%%%%%%%%%%%%%%%%%%%%%%%%%%%%%%%%%
\clearpage
\appendix

\section{Transition from a finite-size universe in five dimensions \label{sec:5D_fin}}

\begin{figure}[tbp]%
    \centering%
    \includegraphics[keepaspectratio, width=0.5\linewidth]{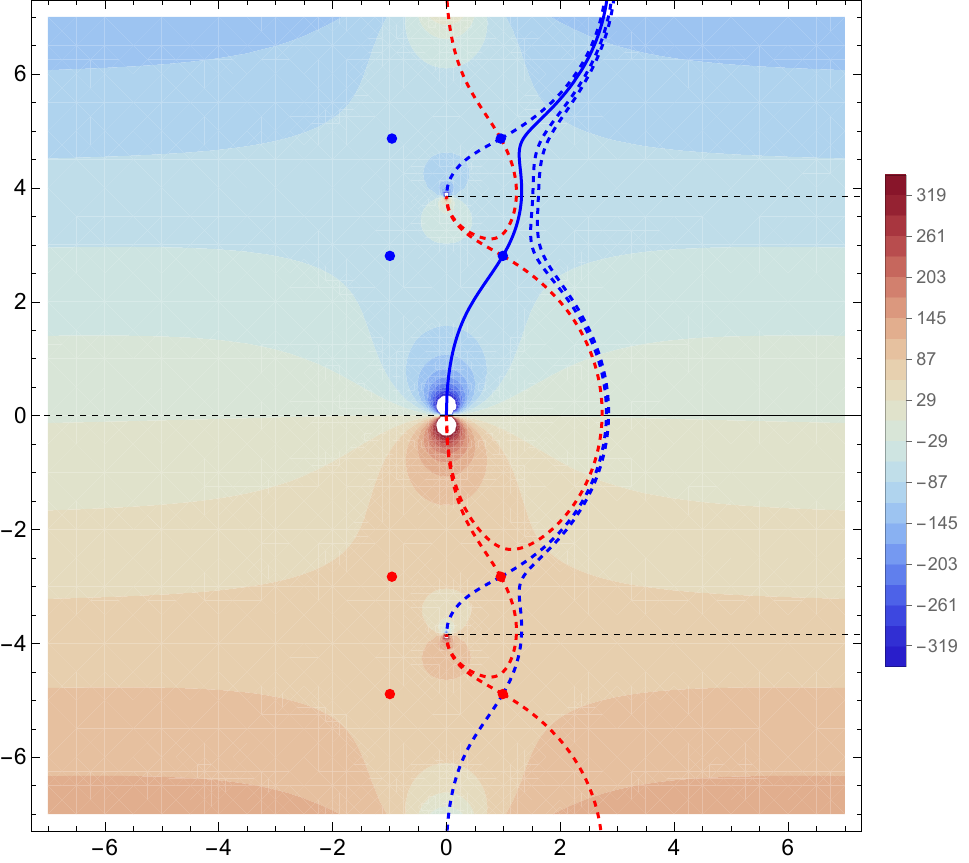}%
    \label{fig:5dINQ_thimbles_fromfint}%
  \caption{Thimbles in the 5D Euclidean-to-Lorentzian case with $q_0>0$. }
  \label{fig:5dINQ_fin}
\end{figure}%

In this appendix, we illustrate some results for cases (ii) Euclidean-to-Lorentzian and (iii) Euclidean-to-Euclidean, beginning from a finite-size 5D universe $0<q_0<q_{\rm crit}$. As mentioned below Eq.~\eqref{eq:saddle_constants_5D}, saddles in the semiclassical limit $\hbar\rightarrow 0$ have no degeneracy in these cases. 

\figref{fig:5dINQ_fin} illustrates the contour plot and the thimbles for the Euclidean-to-Lorentzian case with $q_0 = k/\Lambda$ and $q_1 = 7k/\Lambda$. As in \figref{fig:5dINQ_thimbles}, we find that a single tunneling saddle also unambiguously contributes in the $q_0\neq 0$ case.

\begin{figure}[htbp]%
  \begin{minipage}[t]{0.5\linewidth}%
    \centering%
    \includegraphics[keepaspectratio, width=\linewidth]{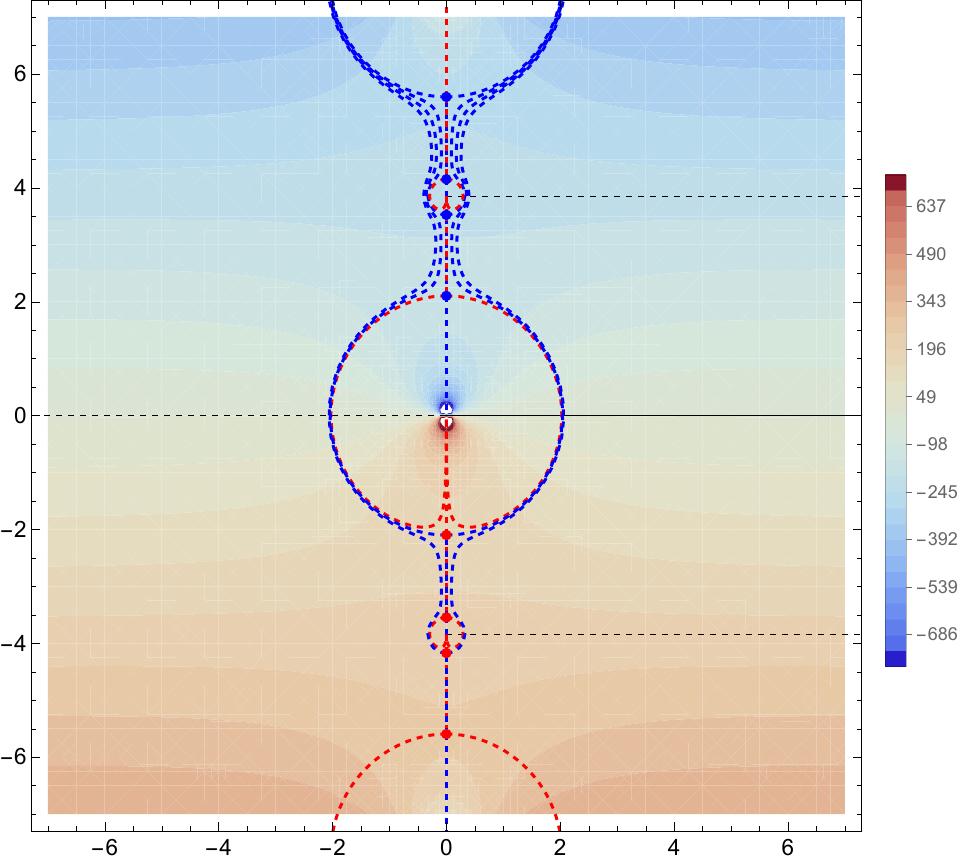}%
    \subcaption{Thimbles}%
    \label{fig:5dq_thimbles}%
  \end{minipage}%
  \\[5pt]
  \begin{minipage}[t]{0.5\linewidth}%
    \centering%
    \includegraphics[keepaspectratio, width=\linewidth]{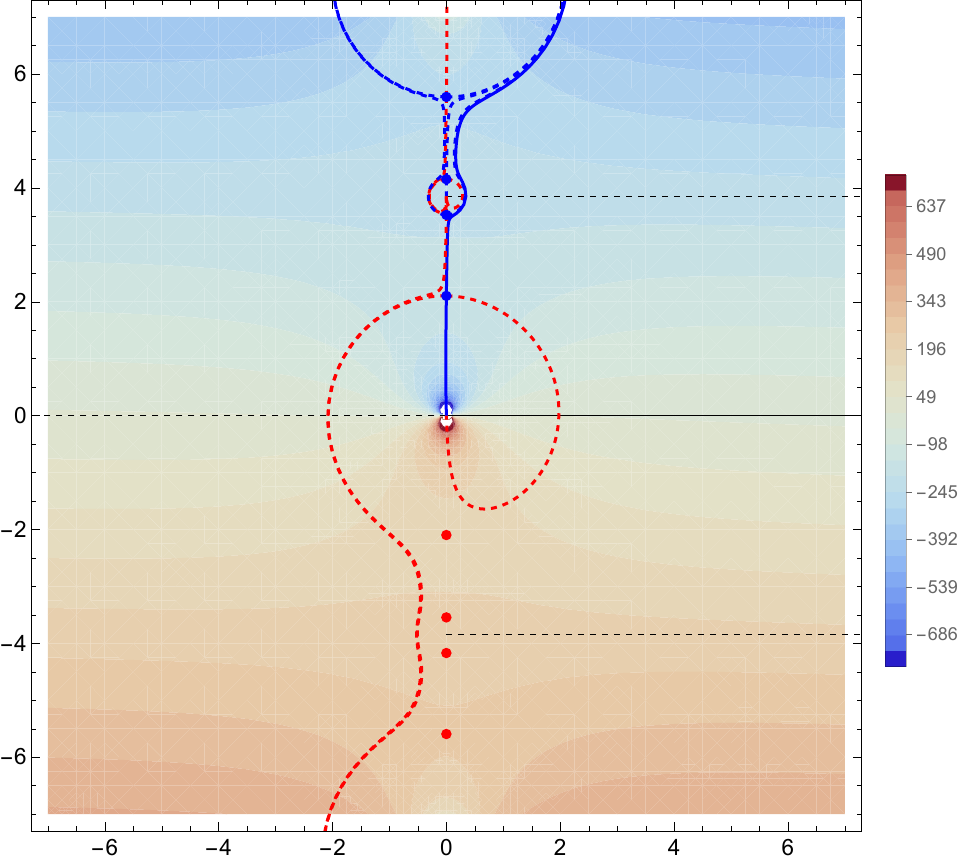}%
    \subcaption{$ \hbar \rightarrow \hbar \e^{-i\pi/120}$
    }%
    \label{fig:5dq_RM}%
  \end{minipage}%
  \begin{minipage}[t]{0.5\linewidth}%
    \centering%
    \includegraphics[keepaspectratio, width=\linewidth]{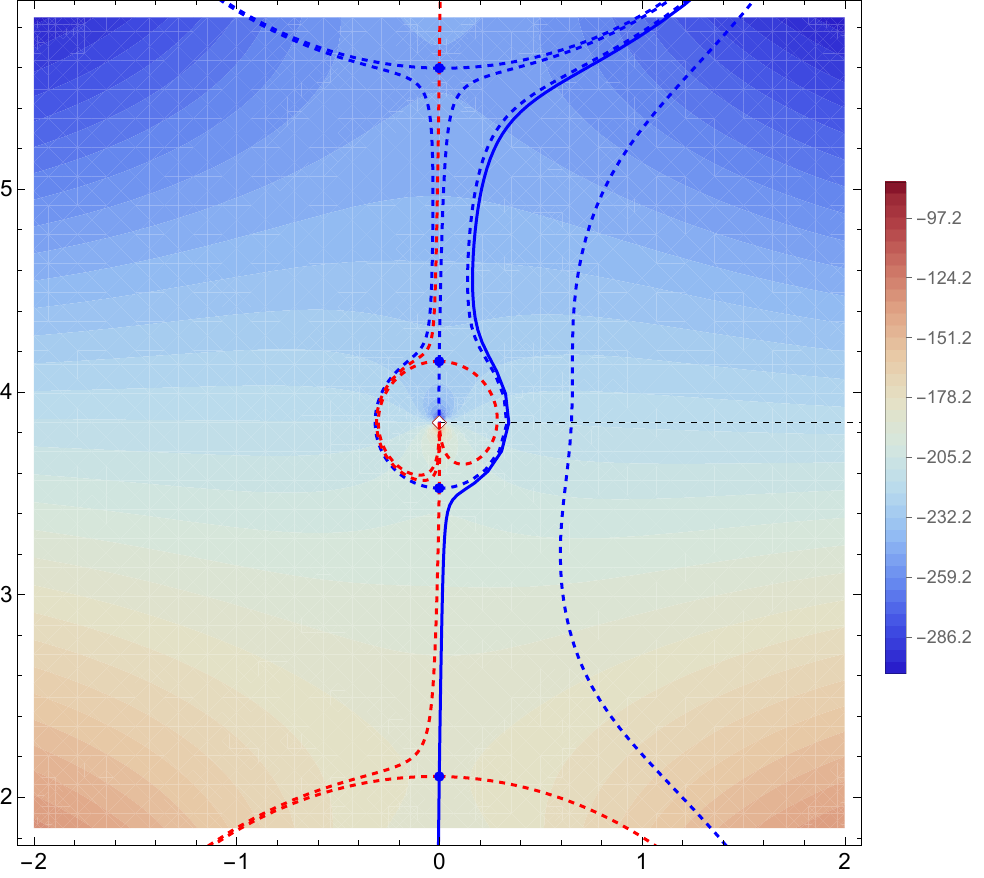}%
    \subcaption{$ \hbar \rightarrow \hbar \e^{-i\pi/120} $ (around $N=i\pi \sqrt{\frac{3}{2\Lambda T^2}}$)
    }%
    \label{fig:5dq_RM_focus}%
  \end{minipage}%
  \\[5pt]
  \begin{minipage}[t]{0.5\linewidth}%
    \centering%
    \includegraphics[keepaspectratio, width=\linewidth]{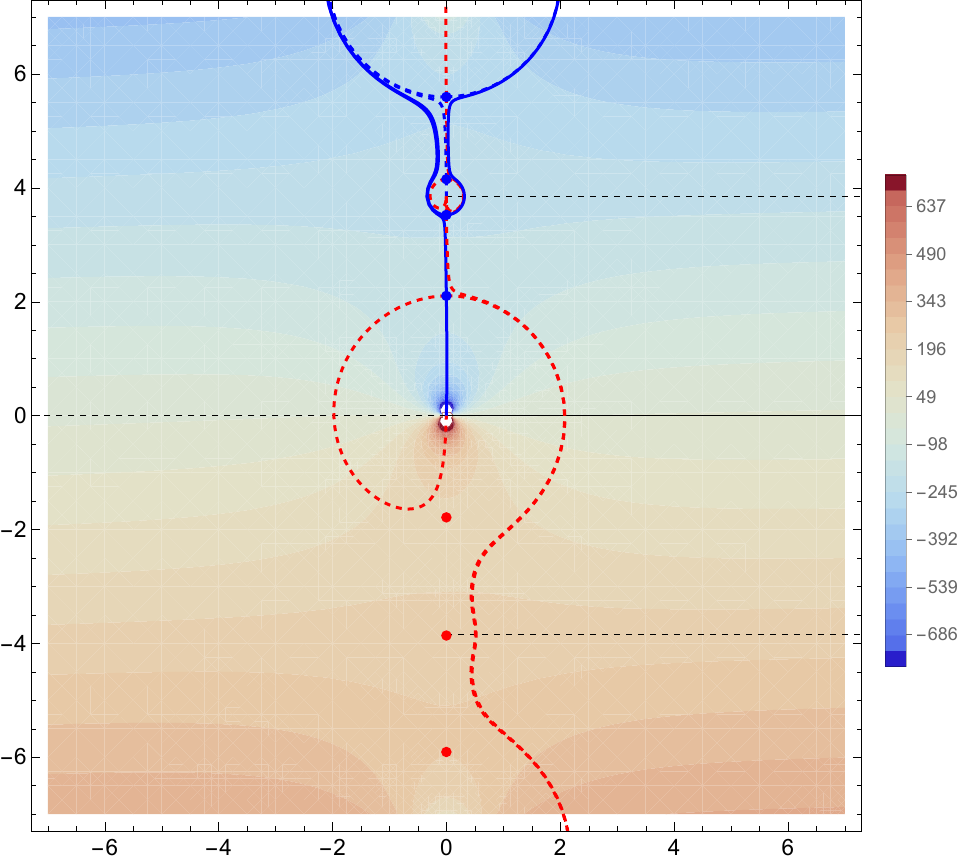}%
    \subcaption{$ \hbar \rightarrow \hbar \e^{+i\pi/120}$}%
    \label{fig:5dq_RP}%
  \end{minipage}%
  \begin{minipage}[t]{0.5\linewidth}%
    \centering%
    \includegraphics[keepaspectratio, width=\linewidth]{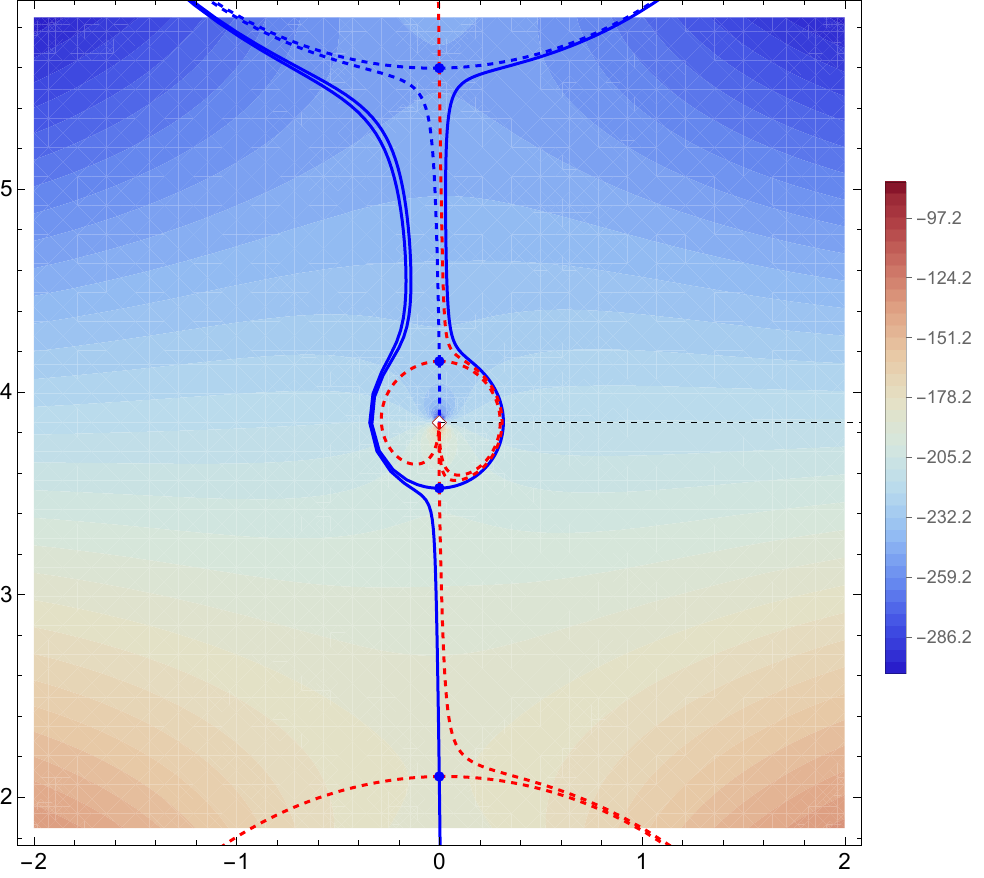}%
    \subcaption{$ \hbar \rightarrow \hbar \e^{+i\pi/120}$ (around $N=i\pi \sqrt{\frac{3}{2\Lambda T^2}}$)}%
    \label{fig:5dq_RP_focus}%
  \end{minipage}%
  \caption{Thimbles in the 5D Euclidean-to-Euclidean case with $q_0>0$.
  }
  \label{fig:5dq}
\end{figure}

\figref{fig:5dq} shows the Euclidean-to-Euclidean case with $q_0 = k/2\Lambda$ and $q_1 = 5k/\Lambda$.
As illustrated in FIG.~\ref{fig:5dq_thimbles}, the relevant thimbles are degenerate and lie on a Stokes line: several steepest-descent paths intersect the original contour with equal weight. This degeneracy can be resolved by rotating $\hbar$ toward complex value $\hbar \rightarrow \hbar \e^{i\Delta \theta}$
as shown in FIGs.~\ref{fig:5dq_RM} and~\ref{fig:5dq_RP}. For a small negative phase rotation,
$\hbar \rightarrow \hbar e^{-i|\Delta\theta|}$, only a single tunneling saddle contributes (FIGs.~\ref{fig:5dq_RM} and \ref{fig:5dq_RM_focus}), whereas for a small positive phase rotation,
$\hbar \rightarrow \hbar e^{+i|\Delta\theta|}$, two tunneling saddles with
$\im{N_{i,n}} >0$ contribute (FIGs.~\ref{fig:5dq_RP} and \ref{fig:5dq_RP_focus}). The deformation of the integration contour as a function of $\Delta\theta$ is thus discontinuous at $\Delta\theta = 0$, which is characteristic of the Stokes phenomenon. Since the amplitude \eqref{eq:5Dphase} has a structure analogous to that of the 3D case \eqref{eq:3dphase}, the same type of resurgent analysis using Borel--Pad\'e resummation can be applied to show that the computed amplitude varies continuously as $\Delta \theta$ is changed.

%%%%%%%%%%%%%%%%%%%%%%%%%%%%%%%%%%
\clearpage
\bibliography{Refs}
\bibliographystyle{JHEP}
%%%%%%%%%%%%%%%%%%%%%%%%%%%%%%%%%%

\end{document}